\documentclass[10pt,aps,prd,nofootinbib,longbibliography,showpacs,showkeys]{revtex4-1}
\usepackage{amsmath,bm,amssymb,amsfonts,bbold,latexsym}
\usepackage{pstricks,pst-node,pst-plot,pst-3d,pst-3dplot,pst-text,pst-eps}
\usepackage{graphicx}
\usepackage{hyperref}
\providecommand{\U}[1]{\protect\rule{.1in}{.1in}}

\newcommand{\bmath}{\begin{displaymath}}
\newcommand{\emath}{\end{displaymath}}
\newcommand{\bite}{\begin{itemize}}
\newcommand{\eite}{\end{itemize}}

\newcommand{\tr}{{\rm Tr}}
\renewcommand{\P}{\mathcal{P}}
\newcommand{\eps}{\varepsilon}
\newcommand{\drop}[1]{}

\newcommand{\one}{{\mathbb{1}}}
\newcommand{\half}{{\textstyle \frac{1}{2}}}

\newcommand{\om}{{\omega}}

\newcommand{\bn}{\mathbf{n}}
\newcommand{\bLambda}{\mathbf{\Lambda}}
\newcommand{\bDelta}{\mathbf{\Delta}}

\newcommand{\phm}{\phantom{-}}
\newcommand{\bel}[1]{\begin{equation}\label{#1}}
\newcommand{\bal}[1]{\begin{eqnarray}\label{#1}}
\newcommand{\ee}{\end{equation}}
\newcommand{\ea}{\end{eqnarray}}
\newcommand{\equ}[1]{~Eq.(\ref{#1})}

\newcommand{\Tr}{{\rm Tr}}

\newcommand{\fig}[1]{~Fig.~\ref{#1}}
\renewcommand{\b}[1]{\relax{\bar{#1}}}
\newcommand{\tud}[3]{{{#1}^{#2}_{{\hspace{.5em}{#3}}}}}

\newcommand{\tuu}[3]{{{#1}^{#2 #3}}}
\newcommand{\tdd}[3]{{{#1}_{#2 #3}}}
\newcommand{\SLC}{\textrm{$SL(2,\mathbb{C})\,$}}
\newcommand{\vev}[1]{\langle #1\rangle}

\newcommand{\sect}[1]{~Sect.\ref{#1}}

\begin{document}
\date{\today}
\preprint{HEP-TH}
\title[Causal Lattice]{Causal Space-Times on a Null Lattice}
\author{Martin Schaden}
\affiliation{Department of Physics, Rutgers University, 101 Warren Street, Newark NJ 07102}

\begin{abstract}
I investigate a discrete model of quantum gravity on a causal null-lattice with \SLC structure group. The description is geometric and foliates in a causal and physically transparent manner. The general observables of this model are constructed from local Lorentz symmetry considerations only. For smooth configurations, the local lattice actions reduce to the Hilbert-Palatini action, a cosmological term and the three topological terms of dimension four of Pontyagin, Euler and Nieh-Yan. Consistency conditions for a topologically hypercubic complex with null 4-simplexes are derived and a topological lattice theory that enforces these non-local constraints is constructed.  The lattice integration measure is derived from an \SLC-invariant integration measure by localization of the non-local structure group. This measure is unique up to a density that depends on the local 4-volume. It can be expressed in terms of manifestly coordinate invariant geometrical quantities. The density provides an invariant regularization of the lattice integration measure that suppresses configurations with small local 4-volumes. Amplitudes conditioned on geodesic distances between local observables have a physical interpretation and may have a smooth ultraviolet limit. Numerical studies on small lattices in the unphysical strong coupling regime of large imaginary cosmological constant suggest that this model of triangulated causal manifolds is finite.   Two topologically different triangulations of space-time are discussed: a single, causally connected universe and a duoverse with two causally disjoint connected components. In the duoverse, two hypercubic sublattices are causally disjoint but the local curvature depends on fields of both sublattices.  This may simulate effects of dark matter in the continuum limit.       
\end{abstract}





\startpage{1}
\endpage{120}
\maketitle

\section{Introduction}
The first order Hilbert-Palatini formulation\cite{Palatini:1919ph} of classical general relativity (GR) is equivalent to Einstein\rq{}s in the absence of torsion. Astronomical observations currently cannot distinguish between the two formulations. However, in the presence of fermionic matter it seems natural to include an a priori independent \SLC-connection that describes the parallel transport of spinors in curved space-time.  Including the cosmological term, the Hilbert-Palatini action on the four-dimensional Lorentzian manifold $M$ is given by the  differential volume form,  is\cite{Ashtekar:2004eh},
\bel{HP}
S_{\rm HP}= \frac{1}{l_P^2}\int_M e^a \wedge e^b \wedge [\frac{\Lambda}{6}e^c\wedge e^d-R^{cd}(\om)] \eps_{abcd}\ .
\ee
The length scale $l_P=\sqrt{32\pi G\hbar/c^3}$ here is proportional to the Planck length. Phenomenologically\cite{Barrow:2011zp} the cosmological constant  $\Lambda$  is positive and incredibly small in natural units\footnote{In natural units $\hbar=c=l_P=1$. The Minkowski metric $\eta=\text{diag}(1,\dots,1,-1)$ with "mostly positive" signature $(+++-)$ is used throughout and Einstein's summation convention for repeated \emph{diagonal} indices is adopted. The symbol $\eps({abcd})=\eps_{abcd}=
-\eps^{abcd}$ gives the sign of the permutation of its arguments with $\eps(1234)=1$. Following common conventions, lower- (upper-) case letters from the beginning of the Latin alphabet denote Lie-algebra indices of $so(3,1)$ and $sl(2,C)$ respectively. Latin indices from the middle of the alphabet are used for tensors. Greek indices will be reserved for labeling parametrization invariant (lattice) objects.}, with $\lambda=\Lambda l_P^2=32\pi \Lambda\ell_\text{Planck}^2\sim 1.7\times10^{-119}$. From a semi-classical point of view this small cosmological constant is related to quantum fluctuations of the total 4-volume of the currently observable universe (see Appendix~\ref{cosconstant}). The  $e^a$ in \equ{HP} are the  Einstein-Cartan co-frame 1-forms,
\bel{frameoneform}
e^a= e^a_k dx^k,
\ee
and $R^{ab}(\om)$ is the (dimensionless) $so(3,1)$ curvature 2-form,
\bel{F}
\tuu{R}{a}{b}(\om)=d\tuu{\om}{a}{b}+\tud{\om}{a}{c}\wedge\tuu{\om}{c}{b}
\ee
with  connection 1-form $\tuu{\om}{a}{b}=-\tuu{\om}{b}{a}$.  The Hilbert-Palatini action of \equ{HP} does not depend on the frame and is defined even if the co-frame is not invertible everywhere.

This first order formulation in terms of co-frames differs from one in terms of frames in that it is polynomial in all fields and depends on the \emph{signed} invariant volume element. The Lagrangian of \equ{HP} is proportional to $\text{det}(e)$ rather than $|\text{det}(e)|$ and changes sign under improper (local) Lorentz transformations. The action of\equ{HP} classically is equivalent to the Einstein-Hilbert action only for orientable manifolds with $\text{det}(e)> 0$ everywhere. The Hilbert-Palatini action thus includes information on the local orientation of the manifold not provided by the metric. For a quantum theory of orientable manifolds, the co-frames of the discretized model  must all have the same orientation.  More important for a discrete version of GR is that forms are geometrical quantities that do not depend on the parametrization of the manifold. Contrary to the co-frame $e^a_k(x)$ and the connection $\omega^{ab}_k(x)$, the 1-forms $e^a(x)$ and $\omega^{ab}(x)$ are geometrical objects. In the first order formulation, the coupling to matter may also be expressed in terms of forms only (see Appendix~\ref{Matteractions}).   The remaining local internal symmetry of such an oriented model is  $SO(3,1)$ and causality further restricts this to the connected component $SO^+(3,1)\sim$\SLC.       
 
In the absence of torsion, the Hilbert-Palatini action of \equ{HP} is (twice) the real part of the action for self-dual connections and the classical phase space may be restricted accordingly\cite{Ashtekar:1986,Ashtekar:1987,Jacobson:1988yy}. This has been the starting point of most Hamiltonian approaches to quantum gravity\cite{Ashtekar:1987,Jacobson:1988yy,Smolin:2003rk,Ashtekar:2004eh}. 

It is difficult to regulate Hamiltonian continuum theories. A lattice formulation in terms of geometrical quantities potentially could overcome these difficulties. 
However, in dynamic triangulations based on Regge's simplicial description\cite{Regge:1961px,Loll:1998aj} of GR, a crumpling instability leads to non-causal configurations\cite{Ambjorn:1991pq,Agishtein:1992xx,Bialas:1996eh,Catterall:1997xj}.
Ambj{\o}rn, Jurkiewicz and Loll\cite{Ambjorn:2004qm,Ambjorn:2007jm} realized that a restriction to causal manifolds may stabilize this model in some regions of the coupling space. They solved the causality constraints\cite{Teitelboim:1983fh,Teitelboim:1983fk} with a particular dynamic triangulation.
    
Here I elaborate on a recently proposed\cite{Schaden:2015yea} causal lattice formulation of \equ{HP}.  The discrete counterparts of \emph{forms} provide a geometric triangulation of space-time.  To accommodate the conventional coupling to matter, the co-frame 1-form will be treated as distinct from the $\SLC$ connection. The co-frame thus will \emph{not} be considered a component of a single gauge connection as in\cite{MacDowell:1977ma,Mansouri:1977f, Witten:1988wi,Catterall:2009nz}.

The resulting  model is a hybrid between a conventional lattice gauge theory and the rigid simplicial approach of Regge\cite{Regge:1961px,Loll:1998aj}, Ambj{\o}rn and Loll\cite{Ambjorn:2004qm,Ambjorn:2007jm,Ambjorn:2009ts,Ambjorn:2013apa}. 
It differs from ordinary lattice gauge theory in that the geometry of the lattice depends on the field configuration on it, but unlike the Ambj{\o}rn-Loll approach, the basic variables are \SLC-variant spinors. The lattice retains a fixed (hypercubic) coordination but its cells vary in shape and size.  Distances between nodes of the lattice in particular depend on the configuration of lattice variables.  We consider triangulations for which the separations between neighboring events are light-like. Such a null-lattice does not of itself introduce a length scale and an ultraviolet  cutoff is required to ensure a minimal coarseness of the triangulation. The latter breaks scale invariance dynamically but allows one to compute finite distances on a finite lattice. Measured  quantities are related to lattice amplitudes with certain geometrical characteristics. The critical limit of this model thus differs from that of ordinary lattice gauge theories where the lattice geometry is assumed fixed from the outset.  
\vspace{1em}

The present model is based on the same principle as the Global Positioning System\cite{Fang:1986, Chaffee:1991,Fang:1992} (GPS): \emph{The intersection of forward light-cones from four spatially separate events is a later event.}
\vspace{1em} 

In the GPS, the (spatial) space-time separations of the emitted signals are determined using atomic clocks on satellites with known orbits. These satellites furthermore emit a string of signals (events). In the idealized lattice setting, the emitting event is itself specified by the intersection of four previously emitted signals. The intersection of light cones in this sense defines an event of the causal null lattice.  The resulting network of events with light-like separation in principle could be traced all the way back to the Big Bang, the single event from which all others derive causally. 

The analogy with the GPS implies that exactly 4 events must lie on the backward light cone of every event and that every event must illuminate 4 others. The coordination of the network thus has a fixed value of $2d$ in $d$ space-time dimensions and is not arbitrary. We will use the nodes of a hypercubic lattice to label the intersection events and indicate their causal relation.       

The use of light-like links solves the causality problem. However, instead of reality constraints\cite{Ashtekar:1987,Jacobson:1988yy}, certain conditions have to be met to ensure  that a configuration gives a triangulated manifold. These consistency conditions are derived in \sect{MC} and a local Topological Lattice Theory (TLT) which enforces them is constructed in \sect{secTLT}. The model thus represents causal manifolds by simplicial complexes. Its point set  of events preserves causal order, a feature this model shares with causal set theory\cite{Sorkin:2007bd,Bombelli:1987,Brightwell:2015laa}.  However, local Lorentz invariance is imposed by averaging over equivalent configurations rather than emergent and the network is designed to  represent triangulated causal manifolds \emph{only}. In the strong coupling limit of \sect{SClimit} the model becomes a random measure of spatial lengths that satisfy certain constraints. 
 
\section{The lattice}
\label{thelattice}
We will consider a  four-dimensional topological lattice $\bLambda$ with hypercubic coordination.  This lattice labels events and gives causal relations between them, but does not of itself specify a manifold. Topologically hypercubic implies that the adjacencies of events are those of a hypercubic lattice but, of itself, does not specify distances or angles between neighboring events. It does imply that each event has 8 neighbors and that links between neighboring events can be oriented. 
\begin{figure}[h]
\centering
\begin{minipage}[b]{0.5\textwidth}
\centering
\includegraphics[width=0.5\textwidth]{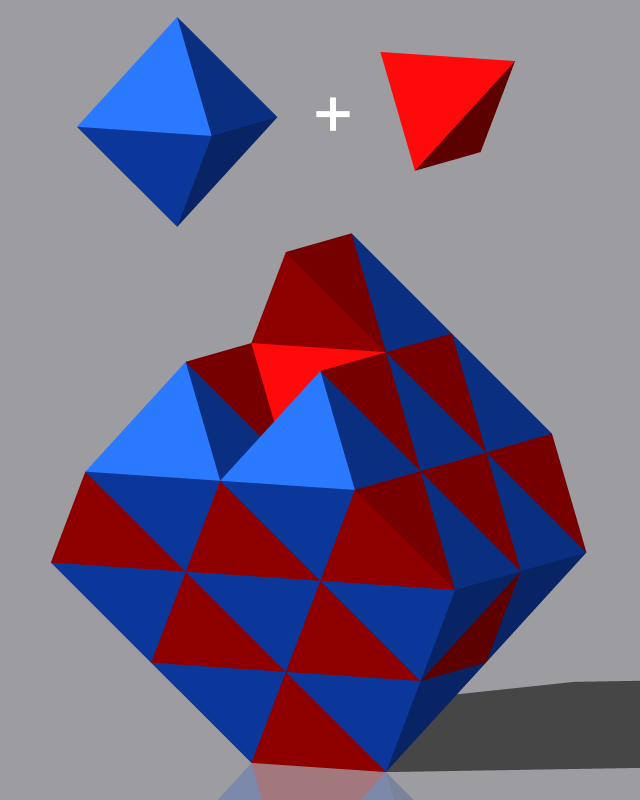}
\\a)
\label{1a}
\end{minipage}%
\begin{minipage}[b]{0.5\textwidth}
\centering
\includegraphics[width=1.0\textwidth]{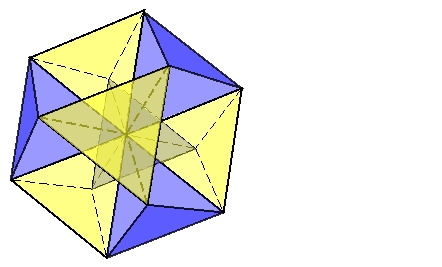}
b)
\label{1b}
\end{minipage}
\caption{\small a) [Wikipedia File:HC P1-P3.png] Tetrahedral-octahedral  triangulation of a spatial hyperplane. Each node has coordination 12.   b) the immediate neighborhood of a node on the hyperplane.  The four B-tetrahedrons with a common node are shown in blue and the four F-tetrahedrons are shown in yellow.  Each of these two sets of tetrahedrons includes all nodes of the spatial hyperplane although only their union with the interstitial octahedrons is space-filling. Every edge is shared by one F- and one B-tetrahedron and every vertex is common to four B- and four F- tetrahedrons. Note that the 8 faces of the octahedral voids coincide with those of four B- and four F-tetrahedrons.  For an general 3-dimensional Cauchy surface the tetrahedrons in general will be neither equal nor regular.}
\label{tetra}
\end{figure}

By analogy with the GPS,  the coordination of the null-lattice in 3+1 dimensions is 8, with 4 future and 4 past events associated with each node.  This may be realized by a lattice with hypercubic structure. That this is the only possibility can be seen by examining flat Minkowski space-time. To this end consider first a spatial hyperplane $P_0$ of Minkowski space. The intersection of light-cones from a tetrahedron of adjacent nodes in $P_0$ defines a single node in a later hyperplane. To conserve the number of nodes, each node of a spatial hyperplane therefore must be common to 4, otherwise disjoint tetrahedrons. One thus is led to the triangulation of $P_0$ by a regular tetrahedral-octahedral honeycomb. The nodes of the hyperplane $P_0$ (at "time" $n_4=0$)\footnote{In Minkowski space the spatial hyperplanes correspond to global time, but a spatial hypersurface with constant depth $n_4$ of Lorentzian space-time in general \emph{will not} be one to constant cosmic time.} thus can be labeled by even integers,
\bel{spatialhp}
P_0=\{(2 n_1, 2 n_2, 2 n_3,0);\  n_1,n_2,n_3\in\mathbb{Z} \text{ with }  n_1+n_2+n_3=0\mod 2\}\ .
\ee
This tetrahedral-octahedral tesselation of flat 3-dimensional space with coordination 12  is shown in \fig{tetra}a). The 12 spatial neighbors of a node form 4 disjoint sets of 3 that neighbor each other. Together with the common node, these are the vertices of 4 (spatial) tetrahedrons. We will call this a set of F-tetrahedrons (or "forward" tetrahedrons). In \fig{tetra}b) they are colored yellow. The complementary set of 4  (blue in \fig{tetra}b)) B-tetrahedrons (or "backward" tetrahedrons) have the same common node and every edge of a B-tetrahedron also is the edge of  an F-tetrahedron, but  B- and F- tetrahedrons do not share a face. The vicinity of the node $(0,0,0,0)$ of the lattice $\bLambda$ includes the following sets of nodes,
\begin{align}\label{neighborhood}
\begin{smallmatrix}(-1,\phm 1,\phm 1,-1)\\(\phm 1,-1,\phm 1,-1)\\(\phm 1,\phm 1,-1,-1)\\(-1,-1,-1,-1)\end{smallmatrix}
\begin{smallmatrix}
\xrightarrow{\text{illumination}}\\\xleftarrow{\text{intersection}}
\end{smallmatrix}& \overbrace{\left.\begin{smallmatrix}
(0,\phm 0,\phm 0,\phm 0)&(-2,\phm 2,\phm 0,\phm 0)&(-2,\phm 0,\phm 2,\phm 0)&(0,\phm 2,\phm 2,\phm 0)\\(2,-2,\phm 0,\phm 0)&(\phm 0,\phm 0,\phm 0,\phm 0)&(\phm 0,-2,\phm 2,\phm 0)&(2,\phm 0,\phm 2,\phm 0)
\\(2,\phm 0,-2,\phm 0)&(\phm 0,\phm 2,-2,\phm 0)&(\phm 0,\phm 0,\phm 0,\phm 0)&(2,\phm 2,\phm 0,\phm 0)\\(0,-2,-2,\phm 0)&(-2,\phm 0,-2,\phm 0)&(-2,-2,\phm 0,\phm 0)&(0,\phm 0,\phm 0,\phm 0)
\end{smallmatrix}\ \right\}}^{\text{B-tetrahedrons}}\text{F-tetrahedrons}\nonumber\\
&\hspace{5em}\text{illumination}\uparrow\downarrow\text{intersection}\nonumber\\
&\hspace{.3em}\begin{smallmatrix}(1,-1,-1,\phm 1)&(-1,\phm 1,-1,\phm 1)&(-1,-1,\phm 1,\phm 1)&(1,\phm 1,\phm 1,\phm 1)\end{smallmatrix}\ .
\end{align}

If the four nodes of a column/row label the vertices of a B/F- tetrahedron, \equ{neighborhood} may be interpreted as follows: the \emph{forward} light cones\footnote{To represent Minkowski space by an integer lattice the speed of light may conveniently be taken as $c=1/\sqrt{3}$.} of the B-tetrahedron at depth $n_4=-1$ illuminate the four F-tetrahedrons (rows) at depth $n_4=0$. The events of the selected B-tetrahedron of the $n_4=-1$ hyperplane thus are on the \emph{backward} light cone of the $(0,0,0,0)$ node. Each node at $n_4=-1$ is at the intersection of four \emph{backward} light cones based at the vertices of an F-tetrahedron in the $n_4=0$ hyperplane. The rows/columns of the matrix of vertices  on the $n_4=0$ hyperplane label the vertices of  F/B- tetrahedrons with the common node $(0,0,0,0)$. The forward light cones from the vertices of a B-tetrahedron in the $n_4=0$ hyperplane intersect at a vertex of the $n_4=1$ hyperplane and the corresponding B-tetrahedron is illuminated by this vertex. The four vertices obtained as intersections of light cones from four B-tetrahedrons that include the $(0,0,0,0)$ node form the F-tetrahedron of the $n_4=1$ hyperplane in \equ{neighborhood}. Its vertices are on the forward light cone of $(0,0,0,0)$.   

Although but a small local section of the whole lattice, \equ{neighborhood} illustrates the general construction:
\begin{itemize}
\item A spatial Cauchy surface is triangulated by B-tetrahedrons with no common edge. Every vertex of this triangulation is common to four B-tetrahedrons.
\item Vertices of the next spatial hypersurface correspond to the intersection of four forward light cones emanating from the nodes of a B-tetrahedron of the earlier Cauchy surface.
\item The vertices obtained from four B-tetrahedrons with a common node form an F-tetrahedron of the later spatial hypersurface. They lie on the forward light cone of the node that is common to the B-tetrahedrons.  Four of these F-tetrahedrons have one node in common but do not share an edge.
\item The corresponding B-tetrahedrons of the later hyperplane share an edge with each of six surrounding F-tetrahedrons. 
\item the procedure is repeated to obtain the spatial hypersurface of the next depth. One thus arrives at a foliated causal description of space-time.
\end{itemize}    

\begin{figure}[h]
\includegraphics[width=3in, angle=-90]{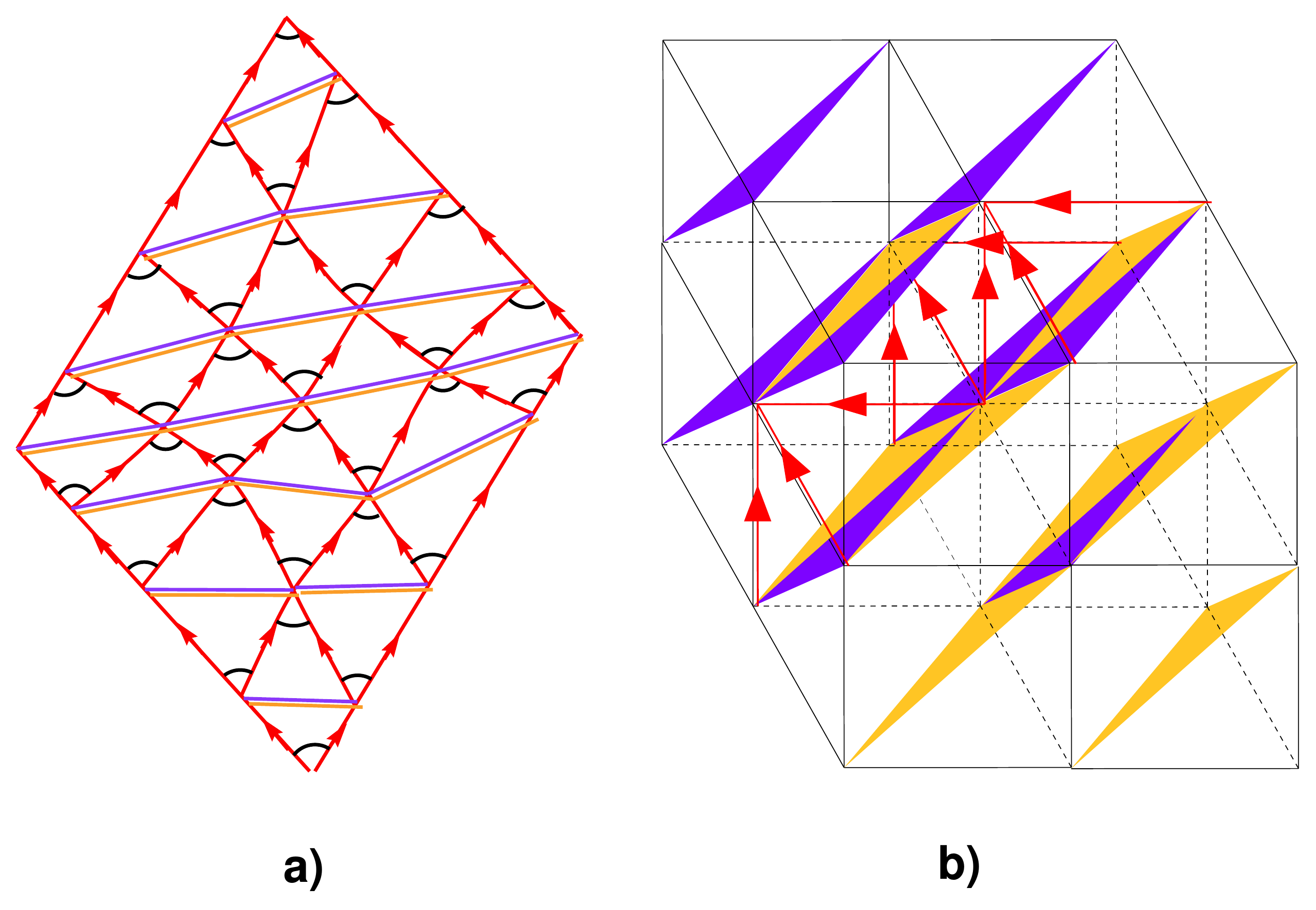}
\caption{\small Topological lattices in $1+1$ and $2+1$~space-time dimensions. Forward null vectors are in red and the corresponding backward (B) and forward (F) spatial simplexes are shown in blue and yellow respectively.  a) In 1+1 dimensions  these spatial simplexes are 1-dimensional and coincide. Forward and backward light-cones are triangles with indicated apexes. Events and their causal relation correspond to nodes and links on this   topologically square null-lattice.  b) In $2+1$~space-time dimensions the topological null-lattice is cubic and the 2-dimensional spatial hypersurfaces are hexagonally triangulated with coordination~6. [As discussed in the text, the construction of a null-lattice in $3+1$ dimensions is analogous, with hypercubes replacing cubes and tetrahedrons replacing the triangles. The tetrahedral-octahedral  triangulation of 3-dimensional spatial hypersurfaces in \fig{tetra} has coordination~12.] The 3(4) vertices of blue B-triangles (tetrahedrons) are the apexes of  future light cones (with rays shown in red). Their intersection corresponds to the node of a  yellow F-triangle (tetrahedron) on a future Cauchy surface. The subset of F- or B-triangles (tetrahedrons) both triangulate a 2(3)-dimensional Cauchy surface.  The triangles (tetrahedrons)  of a causal 2(3)+1 dimensional Lorentzian manifold in general are not regular and the lattice is deformed as in a).}
\label{3Dtriangulation}
\end{figure}

\fig{3Dtriangulation} illustrates the analogous construction in 1+1 and  2+1 space-time dimensions. The hypercubes in 2+1 dimensions are  three-dimensional cubes and the tetrahedrons of a spatial hypersurface become triangles of a two-dimensional, hexagonally triangulated surface.   Note that the F- and B- triangles individually suffice to determine the (hexagonally) triangulated two-dimensional surface.  In 1+1 dimensional space-time the cubes reduce to squares and the (F/B) triangles become line segments that coincide. Note that the number of independent \emph{spatial} lengths is the number of geometrical degrees of freedom in any dimension, that is $1,3$ and $6$ per node in $d=2,3$ and $4$ dimensions respectively.     

We triangulated flat Minkowski space, but the construction can be generalized to causal Lorentzian manifolds. Future events again are at the intersection of (four) forward light-cones from a spatial  B-tetrahedron of earlier events, although these B-tetrahedrons generally are not regular. Four spatial B-tetrahedrons with a common event define the events of a future F-tetrahedron. The construction fails if the four light cones emanating from the vertices of a B-tetrahedron do not intersect. With only a finite number of nodes this can occur near curvature singularities such as the singular world line of a black hole or the singular surface of a cosmic string. In the present construction such singular defects can only occur in the critical limit where the number of nodes becomes arbitrary large while the total 4-volume remains  finite.      

The main challenge will be to ensure that field configurations describe causally triangulated \emph{manifolds}. Since each space-like edge of  a B-tetrahedron is shared with an F-tetrahedron, the endpoints of every spatial edge lie on the \emph{backward} light cone of a unique future event \emph{and} on the \emph{forward} light cone of a unique past event.  One can glue the edges of F- and B-tetrahedons to a pure  3-complex only if these (spatial) lengths coincide for the respective past and future 4-simplexes.  The Topological Lattice Theories (TLT's) constructed in \sect{MC} and Appendix~\ref{coframeMC}  ensure this consistency condition is satisfied and allow one to uniquely construct the triangulated causal manifold. 

\equ{neighborhood} implies that events of the piecewise linear oriented manifold may be labeled by integer quartets of a hypercubic lattice,
\bel{evenlat}
\bLambda=\{\bn = \sum_{\mu=1}^4 n_\mu \bDelta_\mu; \  n_\mu\in\mathbb{Z}\}\ ,
\ee
with forward lattice displacements,
\bel{deltas}
\bDelta_1=(1,-1,-1,1),\ \bDelta_2=(-1,1,-1,1),\ \bDelta_3=(-1,-1,1,1),\ \bDelta_4=(1,1,1,1)\ .
\ee
The labels of 12 spatially neighboring events of a node on the same hyperplane are found by adding the (12) differences of these displacements to the label of the node. This completes the construction of the topologically hypercubic lattice $\bLambda$ that labels events of the causal universe.

In an alternate scenario connections and co-frames are placed on separate links. In addition to $\bLambda$ we in this case consider the lattice with "time-conjugate" forward displacements, 
\bel{dualb}
\bar\bDelta_1=(1,-1,-1,-1),\ \bar\bDelta_2=(-1,1,-1,-1),\ \bar\bDelta_3=(-1,-1,1,-1),\ \bar\bDelta_4=(1,1,1,-1)\ . 
\ee
By construction these satisfy,
\begin{align}\label{sumdual}
\bDelta_\mu-\bDelta_\nu+\bar\bDelta_\nu-\bar\bDelta_\mu&=0 \text{ for all } \ \mu,\nu\ \nonumber\\
\bDelta_\mu+\bDelta_\nu+\bar\bDelta_\rho+\bar\bDelta_\sigma&=0 \text{ for }\ \ \{\mu,\nu,\rho,\sigma\}=\{1,2,3,4\}\ .
\end{align}
The displacements of \equ{dualb} and \equ{deltas} are oppositely oriented\footnote{The lattice displacements $\{\bDelta_\mu\}$ ($\{\bar\bDelta_\mu\}$) form left(right)-handed systems.} and related by the orthogonal matrix $O$ with  $\det{ O}=-1$,
\bel{orthogonal}
\bar\bDelta_\mu=\sum_\nu O_{\mu\nu}\bDelta_\nu \ \ \text{with } O=\frac{1}{2}{\left(\begin{array}{rrrr} 1 & -1 & -1 & -1 \\-1 & 1 & -1 & -1 \\-1 & -1 & 1 & -1 \\-1 & -1 & -1 & 1 \\\end{array}\right)}\ .
\ee

The conjugate forward displacements of \equ{dualb} generate a hypercubic lattice with nodes labeled by,
\bel{duallattice}
\bar\bLambda=\{\bn = \sum_{\mu=1}^4 n_\mu \bar\bDelta_\mu; \  n_\mu\in\mathbb{Z}\}\ .
\ee
By \equ{sumdual}, the labels of all 'even' nodes of $\bLambda$ and $\bar\bLambda$ coincide and the corresponding events are identified, with 
\bel{latintersection}
\bLambda\cap\bar\bLambda=\{\bn = \sum_{\mu=1}^4 n_\mu \bar\bDelta_\mu; \  \sum^4_{\mu=1} n_\mu=0\mod 2\}\ .
\ee
However, the union $\bLambda\cup \bar\bLambda$ leaves 'odd' nodes of $\bar \bLambda$ out in the cold and generally leads to an unstable model. This can be avoided by including a second, translated copy $\bLambda'$ of $\bLambda$ that coincides with $\bar\bLambda$ at odd nodes,
\bel{oddlat}
\bLambda'=\{\bn =\bar\bDelta_4+ \sum_{\mu=1}^4 n_\mu \bDelta_\mu; \  n_\mu\in\mathbb{Z}\}\ .
\ee
The combined lattice,
\bel{lattice}
\bLambda_d=\bLambda\cup\bLambda'\cup\bar\bLambda= \bLambda\cup\bLambda'\ ,
\ee
is the union of the hypercubic lattice $\bLambda$ with vertices of the $\bLambda'$ lattice at the centers of its cells. This body-centered hypercubic $C_4^2$ lattice, the 16-cell honeycomb of 4-dimensional Euclidean space,  is the 4-dimensional analogue of a body-centered cubic lattice. The Euclidean distance to each of the 24 nearest neighbors of a vertex on this lattice  is 2 in the present parameterization and unit 3-spheres whose centers coincide with vertices of $\bLambda_d$ have the densest packing in 4-dimensional Euclidean space\cite{Musin:2003}. The hypercubic sub-lattices $\bLambda$ and $\bLambda'$ are in dual positions, i.e. $\bLambda$ could be interpreted as the dual hypercubic lattice to $\bLambda'$. $\bar\bLambda\subset\bLambda_d$ is a hypercubic sub-lattice of $\bLambda_d$ whose even and odd nodes coincide with nodes of $\bLambda$ and $\bLambda'$ respectively.    

Inclusion of the dual hypercubic lattice leads to a more symmetric and appealing construction at the price of "doubling" the number of causally connected components. Locally, the two components, $\bLambda$ and $\bLambda'$, of the "duoverse" $\bLambda_d$ are causally disjoint although the space-time geometry of each connected component will be determined  by the field content on both. \drop{This scenario potentially could describe dark matter: one causal component of this "duoverse" affects the curvature of the other but does not otherwise interact with the other. Depending on the global arrangement, the arrow of time could be reversed in the two components of the duoverse and a preponderance of matter in one may correspond to a preponderance of anti-matter in the other.  One might further speculate that $\bLambda$ and its dual $\bLambda'$ are globally connected to a non-orientable manifold much like the locally two sides of a M{\"o}bius strip (or the 4-dimensional analog of a Klein bottle)  with dark and ordinary matter occupying locally oppositely oriented "sides". The simplest of these global arrangements is obtained by identifying the initial (Big Bang) events of the two causal components of the duoverse.}   From the point of view of one of its causally connected components, the duoverse $\bLambda_d$ resembles a single universe constructed on $\bLambda$ only. The (gravitational) field content of both scenarios in fact is the same.       

\section{Fields and Observables}
\label{Observables}
To obtain the Lorentzian lattice model with causal dynamics, consider first the topologically hypercubic lattice $\bLambda$ of the universe. For the duoverse a similar construction holds on the hypercubic $\bar\bLambda$ sub-lattice.

Links are ordered pairs of \emph{neighboring} nodes of $\bLambda$.  We adopt standard conventions and use $[\bn,\bn'] \ \text{or  } [\bn,\bn+\bDelta_\mu]\equiv [\bn,\bn+\mu]\equiv[\bn,\mu]$ to identify links of $\bLambda$ - whichever notation is more convenient. $\bLambda$ is naturally oriented by the displacements of \equ{deltas}.  Links $[\bn+\mu, \bn]\equiv[\bn+\mu,-\mu]$ are said to be reversed.   
Sometimes more general  paths $[\bn_0, \bn_1,\bn_2,\dots]$ have to be considered. They are given by a contiguous set of oriented links $[\bn_i,\bn_{i+1}]\in\bLambda$.

The variables $E^{AA'}_\mu(\bn)$ of this lattice formulation are a finite lattice version of the continuum 1-form $e^{AA'}_k(\bn) dx^k$ in spinorial representation\cite{Penrose:1973}.  The differential $dx^k$ is represented by a contravariant vector $l^k_\mu$ that approximates the "displacement" to the node $\bn+\mu$ along the geodesic ${\cal C}[\bn,\bn+\mu]$ between the two events, 
\bel{E} 
E^{AA'}_\mu(\bn):= e^{AA'}_k(\bn) l^k_\mu(\bn)\sim \left. e^{AA'}_k\frac{dx^k}{d\lambda}\right|_\bn \frac{\Delta\lambda}{l_P}\ .
\ee 
The contravariant displacement vector $l^k_\mu$ is uniquely defined using the affine parametrization of a geodesic with $x(\lambda=0)\equiv x(\bn)$ and $x(\lambda=1)\equiv x(\bn+\mu)$ and $\Delta\lambda=1$. In natural units it is the tangent $u^k_\mu(\lambda=0)=u^k_\mu(\bn)$  to the so affine parameterized geodesic ${\cal C}[\bn,\bn+\mu]$ joining the two events\footnote{This geometric interpretation of $E^{AA'}_\mu(\bn)$ is owed\cite{communT} to T. Jacobson.},  
\bel{disp}
 l^k_\mu(\bn)=\frac{1}{l_P}\left.\frac{dx^k}{d\lambda}\right|_{{\cal C}[\bn,\bn+\mu]}\hspace{-3em}(\lambda=0):=\frac{1}{l_P} u^k_\mu(\bn)\ .
\ee
The dimension of the displacement vector has here been absorbed by $l_P$ so that all  variables of the lattice model are dimensionless. We effectively are measuring length in units of $l_P=\sqrt{32\pi \hbar G/c^3}\approx 1.6\times 10^{-34}m$,  time in units of $l_P/c$ and energy in units of $\hbar c/l_P$. $l_P$ here is a dimensional unit on the same footing as $\hbar$ and $c$. The only  gravitational coupling of this lattice model is the dimensionless cosmological constant $\lambda=\Lambda l_P^2$. It plays a r{\^o}le  analogous to that of the (in 3+1 dimensional space-time) dimensionless couplings $\alpha=e^2/(\hbar c)$ and $g^2$ of the electro-weak and strong interactions. 

The displacement vector $l^k_\mu(\bn)$ of \equ{disp} for given events $\bn$ and $\bn+\mu$ is a (uniquely defined) contravariant vector and its contraction with the co-frame $e^{AA'}_k(\bn)$ to  $E_\mu(\bn)$ in \equ{E} is a scalar. A smooth change of coordinates in the vicinity of the event $\bn$ in particular transforms  $l^k_\mu(\bn)$ \emph{and} the co-frame $e_k(\bn)$, but \emph{not} $E_\mu(\bn)$. For lack of a better name, we refer to $\{E_\mu(\bn),\mu=1,\dots,4\}$ as a lat-frame at the node $\bn$, even though this "frame" is parametrization invariant. 

In Appendix~\ref{Matteractions} first order actions for spinors, scalars and gauge fields are written in terms of (diffeomorphism invariant) forms.  Displacements and co-frames thus never appear separately and the corresponding actions for the discretized model can be written in terms of lat-frames only. This is desirable, since the diffeomorphism group does not act on the triangulation of a continuous manifold by a finite number of points. Apart from an internal \SLC symmetry, the present lattice model thus is a dynamical triangulation that preserves coordination number.  

The matrices $E_\mu(\bn)$ transform homogeneously under local \SLC transformations,
\bel{transformations}
E_\mu(\bn)\longrightarrow g(\bn) E_\mu(\bn) g^\dagger(\bn)\ \ \text{ with   } g(\bn)\in\SLC \ ,
\ee
and the \SLC invariant inner product on this space of anti-hermitian matrices  is,
\bel{sp}
X\cdot Y:=-\half\Tr\eps X^T \eps Y=-\half X^{AB'}\eps_{AB}Y^{BA'}\eps_{A'B'} \in\mathbb{R}
\ee
where $X^T=-X^*$ denotes the transpose of the anti-hermitian matrix $X=-X^\dagger$ and $\eps=\left(\begin{smallmatrix} 0&1\\-1&0\end{smallmatrix}\right)$ is the \SLC  invariant tensor.  

It sometimes is more convenient to consider lat-frames as $so^+(3,1)$ vectors. The relation to their spinorial representation is provided by an orthogonal basis of anti-hermitian $2\times 2$ matrices\footnote{We in particular consider a basis in which the $\vec \sigma=(\sigma_1,\sigma_2,\sigma_3) $-matrices are traceless anti-hermitian generators of an $su(2)$ algebra and $\sigma_4=i\one$.}, 
\bel{sigbasis}
\{\sigma^{AB'}_a, a=1,\dots,4; \sigma_a=-\sigma_a^\dagger\ \text{with } \ \Tr \,\sigma^a \sigma_b=2\delta^a_b , \text{ where}\  \sigma^a_{A'B}:=\eps_{BC}\sigma^{CC'}_a\eps_{C'A' }\}.
\ee
The spinor and vector components of lat-frames are related by,
\bel{changebasis}
E^{AB'}_\mu(\bn)=\sum_{a=1}^4\sigma_a^{AB'} E^a_\mu(\bn);\ E^a_\mu=\half \Tr \sigma^a  E_\mu\ .
\ee
and the scalar product of \equ{sp} in the vector representation is,
\bel{spv}
X\cdot Y=X^a\eta_{ab}X^b\ ,
\ee
with $\eta_{ab}=\text{diag}(1,1,1,-1)$ \ .

As discussed previously, to ensure causality and in analogy with the GPS,  all forward displacements are chosen  light-like, 
\bel{nullvecs}
E_\mu(\bn)\cdot E_\mu(\bn)=g_{jk}(\bn)l^j_\mu(\bn)l^k_\mu(\bn)=0 \ \text{ with }\ -i\Tr E_\mu>0\ .
\ee

The same causal manifold may be triangulated in a number of ways, but its triangulation by a topological null-lattice is unique up to the placement of events on the boundary of the manifold. [For a more detailed discussion of a conic manifold see Appendix~\ref{SLR}.] 

Up to \SLC transformations, a null lat-frame represents the 6 spatial distances $\ell_{\mu\nu}(\bn)>0$ between four events on the forward light-cone of $\bn$,
 \bel{lengths}
\ell^2_{\mu\nu}(\bn)=(E_\mu-E_\nu)\cdot (E_\mu-E_\nu)=-2 E_\mu(\bn)\cdot E_\nu(\bn)=-2g_{ik}(\bn)l^i_\mu(\bn)l^k_\nu(\bn)\ge 0\ . 
\ee
These lengths are coordinate invariant \emph{geometrical} quantities. Modulo \SLC transformation, a null lat-frame thus describes $6$ geometric degrees of freedom. 

In general, lat-frames are \emph{not} proportional to the co-frames in any coordinate system\footnote{The displacement matrices $l^k_\mu(\bn)$ generally do not  correspond to a smooth coordinate transformation $\partial x^k/\partial x'^\mu(\bn)$ at more than one node.}.  However,  they may be identified with (forward) null co-frames in Minkowski space-time\footnote{The null lat-frames for instance can be taken proportional to co-frames of the system $S^\prime$ related to Minkowski space with metric $g_{ik}=\text{diag}(1,1,1,-1)$ by the linear coordinate transformation $4x^{\prime 1}=-x^1+x^2+x^3+x^4/\sqrt{3},\  4x^{\prime 2}=x^1-x^2+x^3+x^4/\sqrt{3},\  4x^{\prime 3}=x^1+x^2-x^3+x^4/\sqrt{3}, 4x^{\prime 4}=-x^1-x^2-x^3+x^4/\sqrt{3} $. This transformation preserves orientation and, as required by \equ{lengths}, scalar products of the lat-frames are negative semi-definite. The metric in  $S^\prime$ is $g^\prime_{\mu\nu}= E^\text{Mink}_\mu\cdot E^\text{Mink}_\nu=-4$ for $\mu\neq \nu$ and vanishes for $\mu=\nu$.}, for instance, 
\bel{Minkframe}
E^\text{Mink}_1=i\left(\begin{smallmatrix}\sqrt{3}+1 & -1+i \\ -1-i & \sqrt{3}-1 \\\end{smallmatrix}\right),\ E^\text{Mink}_2=i\left(\begin{smallmatrix}\sqrt{3}+1 & 1-i \\ 1+i & \sqrt{3}-1 \\\end{smallmatrix}\right),\  E^\text{Mink}_3=i\left(\begin{smallmatrix}\sqrt{3}-1 & 1+i \\ 1-i & \sqrt{3}+1 \\\end{smallmatrix}\right),\  E^\text{Mink}_4=i\left(\begin{smallmatrix}\sqrt{3}-1 & -1-i \\ -1+i & \sqrt{3}+1 \\\end{smallmatrix}\right) .
\ee 
\SLC-transformed and scaled null lat-frames give an equivalent description of  discretized Minkowski space-time.  Note that the lattice displacement vectors in $S^\prime$ are  $l^k_\mu=a \delta^k_\mu$. The \emph{global} lattice constant $a$ gives the coarseness of the discretization (in units of $l_P$).  This reflects the possibility of choosing a local inertial system that describes the immediate neighborhood of  an event. The resulting piecewise linear \emph{approximation} to a smooth manifold can be justified for a sufficiently "fine" lattice, but fails to reproduce curvature singularities. These singularities may only be recovered in the critical limit where the number of events becomes infinite.

\equ{sp} and \equ{nullvecs} imply that null lat-frames are singular matrices and may be represented by complex bosonic 2-component  spinors $\xi^A_k,\, (A=1,2)$,-
\bel{spinors}
E_\mu^{A B'}(\bn)= i (\xi_\mu\otimes\xi^*_\mu)_\bn^{AB'}=i\xi_\mu^A(\bn)\xi^{* B'}_\mu(\bn)\ ,
\ee
where $\xi^*_\mu$ denotes the complex conjugate spinor. A primed upper case Latin index indicates that the spinor transforms with the complex conjugate representation of \SLC.
There is no summation over the repeated Greek index\footnote{Only \emph{diagonally related}, repeated indices, i.e. $_A\!\!\nearrow^A$  and $^B\!\!\searrow_B$, are summed.} $\mu$ in \equ{spinors}. Components of the conjugate $\chi^c$ of any spinor  $\chi$ are,
\bel{conjugate}
\chi^c_A:=\chi^B \eps_{BA}\ .
\ee
$\xi_\mu(\bn)$ and its conjugate $\xi^c_\mu(\bn)$ thus transform inversely under  \SLC,
\bel{trafoxi}
	\xi^A_\mu(\bn)\longrightarrow g^A_{\ B}(\bn)\xi^B_\mu(\bn)\ \Leftrightarrow\   \xi^c_{\mu\, A}(\bn)\longrightarrow \xi^c_{\mu\, B}(\bn)g^{-1 B}_{\ \ \ \ A}(\bn),\text{  for } g(\bn)\in \SLC\ . 
\ee 

A lat-frame in addition is invariant under local $U^4(1)$ phase transformations of the spinors,
\bel{U1s}
\xi_\mu(\bn)\rightarrow\xi'_\mu(\bn)= e^{-i\psi_\mu(\bn)} \xi_\mu(\bn)\text{   and    }   \xi^*_\mu(\bn)\rightarrow\xi'^*_\mu(\bn)= e^{i\psi_\mu(\bn)} \xi^*_\mu(\bn),\ \ \psi_\mu(\bn)\in\mathbb{R}\  .
\ee
 
A spinor may be compared to another by parallel transport along links of the lattice to a common node. The parallel transport of a spinor from  $\bn+\Delta_\mu$ to $\bn$  is given by matrix $U_\mu(\bn)\equiv U[\bn,\mu]\equiv U[\bn,\bn+\Delta_\mu]\in\SLC$. Nodes of the lattice thus are  associated with four spinors and its oriented links $[\bn,\mu]$ with parallel transport matrices.  On a (periodic) hypercubic null-lattice with $N$ nodes there are altogether $4N$  spinors and  $4N$ transport matrices. Under the  \SLC   structure group the transport matrices transform as ($g(\bn)\in\SLC$),
\bel{Utrafo}
U^A_{\mu B}(\bn)\rightarrow g^A_{\ C}(\bn) U^C_{\mu D}(\bn) g^{-1 D}_{\ \ \ \ B}
(\bn+\mu)\ .
\ee

Although consistent with \equ{Utrafo}, the reversed lattice link will \emph{not} be associated with the inverse-, but rather with the \emph{hermitian conjugate} transport matrix,
\bel{reverse}
U[\bn,\bn^\prime]=U^\dagger[\bn^\prime,\bn]\ .
\ee
Since the representation of $\SLC$ is not unitary, \equ{reverse} eliminates closed (Wilson) loops of \SLC transport matrices\footnote{The identification $U[\bn,\bn^\prime]=U^{-1}[\bn^\prime,\bn]$  would include closed (Wilson) loops of the form $\Tr U[\bn_1,\bn_2] U[\bn_2,\bn_3]\dots U[\bn_r,\bn_1]$ as basic \SLC invariants.} from the \SLC-invariant  observables one can construct for this lattice. Eqs.~(\ref{trafoxi}),~(\ref{Utrafo})~and~(\ref{reverse}) imply that the basic \SLC invariants of this model are contiguous strings of transport matrices bookended by spinors,
\bel{invariants}
I_{\mu\nu}^{(r)}[\bn_0,\bn_1,\dots,\bn_r]:=\xi^c_\mu(\bn_0) U[\bn_0,\bn_1] U[\bn_1,\bn_2]\dots U[\bn_{r-1},\bn_r]\xi_\nu(\bn_r)\ ,
\ee
where $[\bn_0,\bn_1,\dots,\bn_r]$ is a continuous chain of forward oriented links. For reference we separately list and name the three \emph{shortest} invariants,
\begin{subequations}
\label{shortinv}
\begin{align}
\label{f}
f_{\mu\nu}(\bn)&:=\xi^c_{\mu A}(\bn) \xi^A_\nu(\bn)\equiv \xi^c_\mu(\bn)\xi_\nu(\bn)=:I_{ij}^{(0)}[\bn]\ ,\\
\label{psi}
\psi_{\mu\nu\rho}(\bn)&:=\xi^c_{\mu A}(\bn)U^A_{\rho B}(\bn)\xi^B_\nu(\bn+\rho)\equiv\xi^c_\mu(\bn) U[\bn,\bn_1] \xi_\nu(\bn_1)=:I_{\mu\nu}^{(1)}[\bn,\bn+\rho]\ , \\ 
\label{chi}
\chi_{\mu\nu\rho\sigma}(\bn)&:=\xi^c_{\mu A}(\bn)U^A_{\rho B}(\bn)U^B_{\sigma C}(\bn+\rho)\xi^C_\nu(\bn+\rho+\sigma)\nonumber\\
&\hspace{3em}\equiv \xi^c_\mu(\bn) U[\bn,\bn_1]U[\bn_1,\bn_2] \xi_\nu(\bn_2)=:I_{\mu\nu}^{(2)}[\bn,\bn+\rho,\bn+\rho+\sigma]\ .
\end{align}
\end{subequations}

The invariants of Eqs.~(\ref{invariants})~and~(\ref{shortinv}) in general are complex and \emph{physical} observables necessarily depend on the corresponding complex conjugate invariants as well. The path of the latter is reversed and they are bookended by  hermitian conjugate spinors.  Invariance under the local $U^4(1)$ phase transformations of \equ{U1s} implies that \emph{physical} observables locally conserve four separate spinor numbers. They thus depend only on the null lat-frames of \equ{spinors}.   \sect{secTLT} exploits the $U^4(1)$ symmetry to construct a local TLT that  constrains the space of lattice spinor configurations to those that correspond to triangulated causal manifolds.

The magnitude of the symplectic form $f_{\mu\nu}(\bn)=-f_{\nu\mu}(\bn)$ on the space of spinors at site $\bn$ is the spatial length $\ell^2_{\mu\nu}(\bn)$ of \equ{lengths}\drop{\footnote{It is tempting to interpret the spinors $\xi_\mu(\bn)$ as (not normalized) single-particle amplitudes for a fermion at site $\bn$ to occupy state $\mu=1,\dots,4$ and $U[\bn,\bn^\prime]$ as a (non-unitary) transition amplitude from site $\bn^\prime$ to site $\bn$. In this interpretation $f_{\mu\nu}(\bn)$ is the amplitude for two identical fermions to occupy single-particle states $\mu$ and $\nu$ at the same space-time point and  $\ell^2_{\mu\nu}$ is proportional to the probability that a pair of fermions resides in these states at site $\bn$.}},
\bel{lengths2}
\ell^2_{\mu\nu}(\bn)=\Tr \eps E^T_\mu(\bn)\eps E_\nu(\bn)=|f_{\mu\nu}(\bn)|^2\ge 0\ ,
\ee  
where we used  Eqs.~(\ref{spinors})~and~(\ref{f}). Note that the length $\ell^2_{\mu\nu}$ vanishes only if the two spinors $\xi_\mu$ and $\xi_\nu$ are linearly dependent.

 In four dimensions the 6 complex variables of the anti-symmetric matrix $f_{\mu\nu}(\bn)$ are constrained by the fact that its Pfaffian vanishes,
\bel{Pfaffian}
\text{Pf}(f(\bn)):=\frac{1}{8}\sum_{\mu\nu\rho\sigma}\eps(\mu\nu\rho\sigma) f_{\mu\nu}(\bn) f_{\rho\sigma}(\bn)=f_{12}(\bn)f_{34}(\bn)+f_{13}(\bn)f_{42}(\bn)+f_{14}(\bn)f_{23}(\bn)=0\ .
\ee
The matrix $f(\bn)$ thus depends on 10 real parameters only. Four of these are overall phases of the spinors\cite{Penrose:1973} (see \equ{U1s}). The remaining 6 real degrees of freedom are the spatial lengths $\ell_{\mu\nu}(\bn)\ge 0$ of \equ{lengths2}.

The vanishing Pfaffian implies that the magnitudes of the three complex numbers in  \equ{Pfaffian} satisfy triangle inequalities, 
\bel{tineq}
0\leq a(\bn)\leq b(\bn)+c(\bn),\ \ \ 0\leq b(\bn)\leq c(\bn)+a(\bn),\ \ \ 0\leq c(\bn)\leq a(\bn)+b(\bn)\ ,
\ee
with,
\begin{align}\label{defineabc}
a(\bn):=&\ell_{12}(\bn)\ell_{34}(\bn)=|f_{12}(\bn)f_{34}(\bn)|\nonumber\\
b(\bn):=&\ell_{13}(\bn)\ell_{24}(\bn)=|f_{13}(\bn)f_{42}(\bn)|\nonumber\\
c(\bn):=&\ell_{14}(\bn)\ell_{23}(\bn)=|f_{14}(\bn)f_{23}(\bn)|\ .
\end{align}
For non-negative $a,b$, and $c$, the triangle inequalities of \equ{tineq} can be combined to the single inequality, 
\bel{detgineq}
0\ge \left. a^4+b^4+c^4-2 a^2 b^2-2 b^2 c^2-2 c^2 a^2\right|_{\bn}=\det{\ell^2_{\mu\nu}(\bn)}\ .
\ee
The triangle inequalities of \equ{tineq} thus are equivalent to requiring that the local four volume $V(\bn)=\sqrt{-\det{\ell^2_{\mu\nu}(\bn)}}$ be real.   
The fact that causality restricts the spatial lengths $\ell_{\mu\nu}(\bn)$ so simply is one reason for considering null lat-frames. The converse holds as well: given six non-vanishing positive lengths $\{\ell_{\nu\mu}=\ell_{\mu\nu}>0,1\leq \mu<\nu\leq 4\}$ with $\det{\ell^2_{\mu\nu}}\leq 0$, an anti-symmetric complex matrix $f_{\mu\nu}$ with $|f_{\mu\nu}|=\ell_{\mu\nu}$ can be constructed whose Pfaffian vanishes. The spinors then are reconstructed as follows.

The vanishing Pfaffian of \equ{Pfaffian} implies that  $f_{\mu\nu}(\bn)$ and its dual $ f^{\mu\nu}(\bn):=\half\sum_{\rho\sigma}\epsilon(\mu\nu\rho\sigma) f_{\rho\sigma}(\bn)$ are antisymmetric $4\times 4$ matrices of rank 2. The null space of $ f^{\mu\nu}(\bn)$ thus is spanned by two linearly independent vectors $\xi_\mu^A$,
\bel{xis}
\sum_\nu  f^{\mu\nu}(\bn) \xi_\nu^A(\bn)=0 , \ \text{for} \ A=1,2 \ .  
\ee
Up to normalization these define $f_{\mu\nu}(\bn)$ itself. The fact that any 4-dimensional  anti-symmetric matrix with vanishing Pfaffian is represented by a set of spinors is exploited in \sect{MC} and Appendix~\ref{Reconstruction}.

\section{Field Assignments and Lattice Actions}
\label{Fields}
The model is completed by assigning this field content to links and nodes of the lattice and constructing the lattice action and integration measure. We will consider two distinct scenarios:
\subsection{The  Universe}
\label{Universe}
Only the hypercubic $\bLambda$ lattice of \equ{evenlat} is utilized  in this case. Each of its oriented links $[\bn,\mu]$ is associated with an \SLC transport matrix $U_\mu(\bn)$. The reversed link is associated with the hermitian conjugate transport matrix $U^\dagger_\mu(\bn)$  as in \equ{reverse}.  

Each node $\bn\in\bLambda$ furthermore  is associated with a null lat-frame $\{E_\mu(\bn),\;\mu=1,\dots, 4\}$ that gives the light-like displacements in the \emph{forward} direction from $\bn$  to the events $\{\bn+\bDelta_\mu,\mu=1,\dots,4\}$.

\begin{figure}[h]
\includegraphics[width=1.7in,angle=270,clip=true,trim=0 0 2.1in 0]{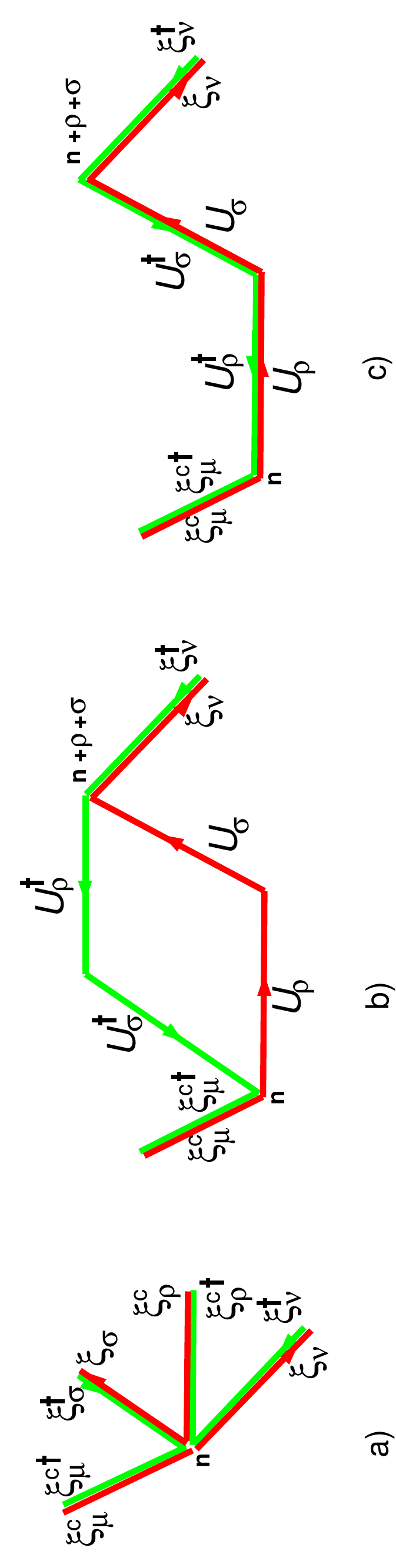}
\caption{\small Quasi-local contributions to the invariant action on a null-lattice in $d=3+1$ dimensions. a) The cosmological term $\mathcal{V}_{\mu\nu\rho\sigma}(\bn)$  at site $\bn$ of \equ{Vform}. Spinors $\xi_\mu(\bn)$ are shown as red segments along lattice direction $\mu$, hermitian conjugate spinors $\xi^\dagger_\mu$ are depicted in green. Color flows from conjugate spinors $\xi^c_\mu$ defined by\equ{conjugate} to spinors $\xi_\nu$ and from $\xi^\dagger_\nu$ to $\xi^{c\,\dagger}_\mu$.  Two directed line segments of the same color joined at $\bn$ form the \SLC-invariant  (red) $f_{\mu\nu}(\bn)$ of\equ{f} (or its complex conjugate (green)). The contribution to the cosmological term of  site $\bn$ is the product of these invariants with the sign determined by the permutation of $(\mu\nu\rho\sigma)$. b) The ${\cal P}_{\mu\nu\rho\sigma}(\bn)$ curvature contribution to \equ{HPaction1}. The depiction of spinors at sites $\bn$ and $\bn+\rho+\sigma$ is as in a), but these now are parallel transported along links in $\rho$- and $\sigma$- direction by \SLC transport matrices as shown. The (red) spinors $\xi_\mu$ are parallel transported by the \SLC-matrices $U$, whereas their (green) hermitian conjugates counterparts $\xi^\dagger_\mu$ are transported by the hermitian conjugate \SLC-matrices $U^\dagger\neq U^{-1}$. The contribution to the lattice Hilbert-Palatini action is given by the product of these two overlaps with a sign determined by the permutation of $(\mu\nu\rho\sigma)$.  c) The ${\cal Q}_{\mu\nu\rho\sigma}$ topological term of \equ{topHPaction}. In the continuum limit his term does not affect  classical equations of motion and could be included in the lattice action with a purely imaginary coefficient. Similar to the cosmological contribution a), the corresponding  path does not enclose a plaquette and does not measure local curvature.}
\label{action4}
\end{figure}

The lattice action is constructed from purely imaginary \SLC  invariant lattice 4-forms with real coefficients. The two most local candidates for the lattice action are,
\begin{subequations}
\label{HPactions}
\begin{align}
\label{Vform} 
\mathcal{V}_{\mu\nu\rho\sigma}(\bn):=&\Tr E_\mu(\bn)\eps E^T_\nu(\bn)\eps E_\rho(\bn)\eps E^T_\sigma(\bn)\eps=2i\eps(\mu\nu\rho\sigma)\det[E]:=2i\eps(\mu\nu\rho\sigma)V(\bn)\nonumber\\
=&f^*_{\mu\nu}(\bn)f_{\nu\rho}(\bn)f^*_{\rho\sigma}(\bn)f_{\sigma\mu}(\bn)\ ,\\
\label{HPaction1}
{\cal P}_{\mu\nu\rho\sigma}(\bn):=&\chi_{\mu\nu\rho\sigma}(\bn)\chi^*_{\mu\nu\sigma\rho}(\bn)=\Tr \eps E^T_\mu(\bn)\eps U_\rho(\bn)U_\sigma(\bn+\rho) E_\nu(\bn+\rho+\sigma) U^\dagger_\rho(\bn+\sigma) U^\dagger_\sigma(\bn) \nonumber\\
=&\xi^c_{\mu A}(\bn)U^A_{\rho B}(\bn)U^B_{\sigma C}(\bn+\rho)\xi^C_\nu(\bn+\rho+\sigma)\xi^{*\dot C}_\nu(\bn+\sigma+\rho) U_{\rho\dot C}^{*\ \dot B}(\bn+\sigma)U_{\sigma \dot B}^{*\ \dot A}(\bn)\xi^{*c}_{\mu\dot A}(\bn)\ ,
\end{align}
\end{subequations}
These contributions to the action are depicted graphically in Figs.~\ref{action4}a)~and~\ref{action4}b) respectively.

The other two quasi-local observables, 
\begin{subequations}
\label{HPdiscard}
\begin{align}
\label{guv} 
\ell^2_{\mu\nu}(\bn):=&\Tr \eps E^T_\mu(\bn)\eps E_\nu(\bn)=|f_{\mu\nu}(\bn)|^2=\ell^2_{\nu\mu}(\bn)\\
\label{topHPaction}
{\cal Q}_{\mu\nu\rho\sigma}(\bn):=&|\chi_{\mu\nu\rho\sigma}(\bn)|^2=\Tr \eps E^T_\mu(\bn)\eps U_\rho(\bn)U_\sigma(\bn+\rho) E_\nu(\bn+\rho+\sigma) U^\dagger_\sigma(\bn+\rho) U^\dagger_\rho(\bn)\nonumber\\
=&\xi^c_{\mu A}(\bn)U^A_{\rho B}(\bn)U^B_{\sigma C}(\bn+\rho)\xi^C_\nu(\bn+\rho+\sigma)\xi^{*\dot C}_\nu(\bn+\sigma+\rho) U^{*\ \dot B}_{\sigma\dot C}(\bn+\rho) U^{* \ \dot A}_{\rho \dot B}(\bn)\xi^{*c}_{\mu\dot A}(\bn)\ ,
\end{align}
\end{subequations}
are real. The first of these invariants are the spatial lengths of \equ{lengths2} and is symmetric in its indices. The totally anti-symmetric part of ${\cal Q}_{\mu\nu\rho\sigma}(\bn)$ is the lattice analog of a topological contribution to the continuum action. This term could be (and its continuum analog often is\cite{Holst:1995pc,Ashtekar:1986,Kaul:2011va}) included but will not be considered in the following.  We furthermore do not investigate other, less local, contributions to the lattice action. The invariant (real) lattice action\footnote{The weight $e^{iS^{\text{inv.}}_\text{HP}}$ of a configuration is a pure phase.} 
for a causally connected universe thus is, 
\bel{LatHP}
S^{\text{uni.}}_\text{HP}= i\sum_{\mu\nu\rho\sigma}\eps(\mu\nu\rho\sigma)\sum_{\bn\in\bLambda} [\P_{\mu\nu\rho\sigma}(\bn)+\frac{\lambda}{12}\mathcal{V}_{\mu\nu\rho\sigma}(\bn)]\ ,
\ee 
where $\lambda$ is the dimensionless cosmological constant. We have absorbed the coefficient (and dimensionality) of the curvature term in the normalization of the spinors (or equivalently, of the lat-frame $E_\mu$). On a hypercubic lattice the HP-action takes the same form whether the lat-frame is light-like or not, but the relation between events of  $\bLambda$ in general would be acausal. Null lat-frames assure causality, but only a subset of such lattice configurations corresponds to triangulated manifolds (see \sect{MC}).  
    
\subsection{The Duoverse} 
\label{Duoverse}
\begin{figure}[h]
\includegraphics[width=4.0in]{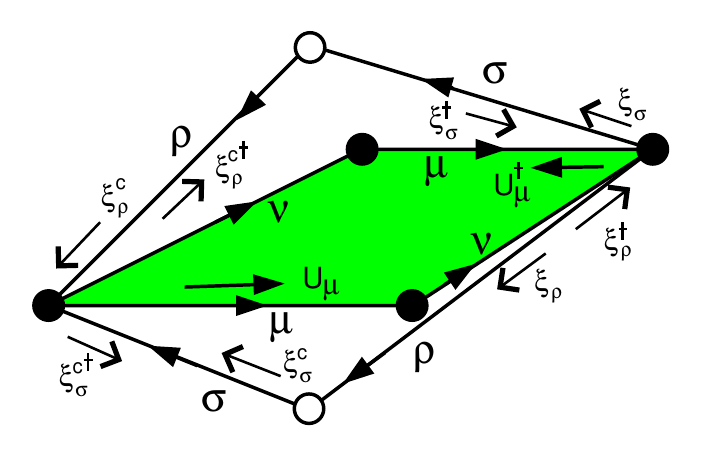}
\caption{\small Plaquettes of the duoverse.  A $\mu\nu$-plaquette of events (solid nodes) in $\bar\bLambda$ and the corresponding $\rho\sigma$-plaquette of $\bLambda$ ($\bLambda'$). Solid dots ($\bullet$) denote active events of $\bar\bLambda$. Passive events of $\bLambda$ ($\bLambda'$) are shown as empty circles ($\circ$).  Note that spinors $\xi$ ($\xi^c$) reside on outgoing (incoming) passive links whereas \SLC transport matrices occupy active links. Links traversed in reverse direction are associated with hermitian conjugate variables. The local \SLC structure group acts at active sites of $\bar\bLambda$ only.}
\label{plaquettes}
\end{figure}

The duoverse is obtained by placing the field content on the lattice $\bLambda_d$ of \equ{lattice}. In some ways this scenario is more restrictive and appealing than the universe. In analogy with optics, nodes  $\bn\in\bar\bLambda$ of the hypercubic sub-lattice $\bar\bLambda\subset\bLambda_d$ will be referred to as "active" nodes. All other nodes  $\bar\bn\in \bLambda_d\backslash\bar\bLambda$ of $\bLambda_d$ are "passive" or inert in this construction.  In Table~\ref{assignments} and \fig{plaquettes}, active nodes are denoted by solid dots and passive ones by open circles. Half the nodes of $\bLambda$ and $\bLambda'$ are passive. Active links are between two active nodes whereas passive links are between an active and a passive node. The hypercubic sub-lattices $\bLambda$ and $\bLambda'$ are composed of passive links only. Depending on whether the initial or final node is active, \emph{oriented} passive links  are of two kinds $[\bn,\bar\bn]$ and $[\bar\bn,\bn]$. There are no links between two passive nodes in this construction.  \SLC transport matrices $U_{\mu}(\bn)$ reside on oriented active links $[\bn,\bn'=\bn+\bar\bDelta_\mu]\in\bar\bLambda$  only. The reversed active link being associated with the hermitian conjugate transport matrix as in \equ{reverse} and Table~\ref{assignments}. \fig{plaquettes} depicts these assignments for a $\rho\sigma$-plaquette of $\bLambda$ (or $\bLambda'$) and a $\mu\nu$-plaquette of $\bar\bLambda$ that have two active nodes in common.

\begin{table}[h]
\begin{equation*}
\begin{array}{c|c|c}
& \text{\ \ \ oriented \ \ } & \text{\ \ reversed\ \ }\\
\hline
\text{active links}&&\\
{\bn}\;\bullet\xrightarrow{\ \ \bar\bDelta\mu\ \ \ }\bullet\;\bn' & U_{\mu}(\bn) & U^\dagger_{\mu}(\bn)\\
\hline
\text{passive links}&&\\
{\bn}\;\bullet\xrightarrow{\ \ \bDelta\mu\ \ \ }\circ\;\bar\bn &\xi_\mu(\bn)&\xi^\dagger_\mu(\bn)\\
\bar\bn\;\circ\xrightarrow{\ \ \bDelta\mu\ \ \ }\bullet\;{\bn} &\xi^c_\mu(\bn)&\xi^{c\dagger}_\mu(\bn)
\end{array}
\end{equation*}
\caption{Links and field assignments of the duoverse. Solid dots ($\bullet$) denote active nodes of the sublattice $\bar\bLambda$, whereas open circles ($\circ$) are passive nodes of $\bLambda$ (or $\bLambda'$). Assignments depend on whether a closed path traverses a link in the direction of its orientation or reverse to it. Passive ($\circ$) nodes act like mirrors: they are never crossed and reverse the direction of a path. A path on a passive link thus always retraces itself. Active nodes on the other hand do not reverse the direction of a path.  Physical observables of the duoverse are associated with closed lattice loops.} 
\label{assignments}
\end{table}

The number of degrees of freedom of  the universe and of the duoverse are the same, since they have the same number of active nodes and active links. In both arrangements each active link is associated with an \SLC transport matrix and each active node with four forward null vectors. The structure group \SLC acts at active nodes only. The passive nodes of the duoverse carry no additional dynamical information.   In the next section the triangulated manifold is reconstructed from outgoing passive links of $\bLambda$ ($\bLambda'$) only.  In the duoverse forward and backward light cones of active nodes are glued at passive nodes, whereas they are glued at active nodes in the universe. The universe and duoverse thus differ only in the manner the complex is formed from the basic simplexes.  

A path on $\bLambda$  either traverses a link in the direction of its orientation or in the reverse direction. In the duovers one considers closed loops on the lattice whose direction reverses at \emph{passive} nodes $\bar\bn\in \bLambda_d\backslash \bar\bLambda$. Passive nodes (open circles ($\circ$) in Table~\ref{assignments} and \fig{plaquettes}) are never crossed and reverse the direction of a path. Paths on the other hand do not reverse direction at active nodes.   

The four shortest loops on $\bLambda_d$ of this kind are,
\begin{align}\label{actionloops}
{\cal C}_a[\bn;\mu\nu\rho\sigma]&:=[\bn+\bDelta_\mu,\bn,\bn-\bDelta_\nu,\bn,\bn+\bDelta_\rho,\bn,\bn-\bDelta_\sigma,\bn,\bn+\bDelta_\mu]\ ;\ \text{sgn}[{\cal C}_a]=\eps(\mu\nu\rho\sigma)\ ,\nonumber\\ 
{\cal C}_b[\bar\bn;\mu\nu\rho\sigma]&:=[\bar\bn,\bar\bn+\bDelta_\sigma,\bar\bn+\bDelta_\sigma+\bar\bDelta_\mu, 
\bar\bn-\bDelta_\rho,\bar\bn,\bar\bn-\bDelta_\rho, \bar\bn+\bDelta_\sigma+\bar\bDelta_\nu,\bar\bn+\bDelta_\sigma,\bar\bn]\ ;\ \text{sgn}[{\cal C}_b]=\eps(\mu\nu\rho\sigma)\ ,\nonumber\\
{\cal C}_c[\bn;\mu\nu]&:=[\bn+\bDelta_\mu,\bn,\bn-\bDelta_\nu,\bn,\bn+\bDelta_\mu]\ ,\nonumber\\ 
{\cal C}_d[\bar\bn;\mu\nu\rho\sigma]&:=[\bar\bn,\bar\bn+\bDelta_\sigma,\bar\bn+\bDelta_\sigma+\bar\bDelta_\mu, 
\bar\bn-\bDelta_\rho,\bar\bn,\bar\bn-\bDelta_\rho, \bar\bn-\bDelta_\rho-\bar\bDelta_\nu,\bar\bn+\bDelta_\sigma,\bar\bn]\ ;\ \text{sgn}[{\cal C}_d]=\eps(\mu\nu\rho\sigma)\ .
\end{align}
Using the assignments of Table.~\ref{assignments},  these closed loops correspond precisely to the densities of Eqs.~(\ref{HPactions})~and~(\ref{HPdiscard}).  Reflection at passive nodes is required by the local $U^4(1)$ symmetry and ensures that spinors combine with complex conjugate spinors to anti-hermitian (null) lat-frames. Introducing the sign of a loop as in \equ{actionloops}, the lattice action of \equ{LatHP} for the duoverse can be succinctly written,
\bel{LatHPduo}
iS^{\text{duo.}}_\text{HP}=\sum_{{\cal C}_b}\text{sgn}[{\cal C}_b] \P({\cal C}_b)-\frac{\lambda}{12}\sum_{{\cal C}_a}\text{sgn}[{\cal C}_a]\mathcal{V}({\cal C}_a)\ ,
\ee 
where the sums extend over the elementary loops of \equ{actionloops} on $\bLambda_d$. Expressions for the densities $\mathcal{V}({\cal C}_a[\bn;\mu\nu\rho\sigma])=\mathcal{V}_{\mu\nu\rho\sigma}(\bn)$ and $\P({\cal C}_b[\bar\bn;\mu\nu\rho\sigma])=\P_{\sigma\rho\mu\nu}(\bar\bn+\bDelta_\sigma)$   are given by  \equ{Vform} and \equ{HPaction1} respectively.  The imaginary parts of the expressions in \equ{topHPaction} and \equ{guv} vanish.  ${\cal C}_d$-loops with real coefficients give purely imaginary contributions to the lattice action.  

Field assignments and reflection rules thus dictate the form of the lattice action of the duoverse.   Every configuration of the duoverse corresponds to two causally disjoint manifolds whose events are labeled by the hypercubic sub-lattices $\bLambda$ and $\bLambda'$. The events on each of these manifolds are causally related to events of the same manifold, but not to events on the other. Parallel transport on spatial links between one sub-manifold and the other is possible, but the matter fields of each causal component do not interact. Matter and energy of one causally connected component of the duoverse thus influences the curvature of the other, but there apparently is no causal communication between the two manifolds.   \footnotetext{Lyrics to the song "The Makings of You" by Curtis Mayfield: Add a little sugar, honeysuckle/
And a great big expression of happiness/
Boy, you couldn't miss/
With a dozen roses/
Such will astound you/
The joy of children laughing around you/
These are the makings of you/
It is true, the makings of you/
The righteous way to go/
Little one would know/
Or believe if I told them so/
You're second to none/
The love of all mankind/
Should reflect some sign of these words/
I've tried to recite/
They're close but not quite/
Almost impossible to do/
Reciting the makings of you/}    
 
\section{Makings of{\protect\footnotemark[\value{footnote}]} Causal Manifolds}
\label{MC}
Although causal manifolds are triangulated by null lat-frames  as described in \sect{thelattice}, not every lattice configuration corresponds to a  triangulated manifold. For a causal simplicial complex the vertices of every B-tetrahedron must lie on the backward light cone of a future event. This condition is the same for uni- and duo-verses, since the two manifolds corresponding to $\bLambda$ and $\bLambda'$ lattices can be reconstructed independently. For definiteness we here consider the reconstruction of the $\bLambda$ manifold.

\subsection{Consistency Conditions of the Simplicial Complex}
\label{complex}
\begin{figure}[h]
\includegraphics[width=3in]{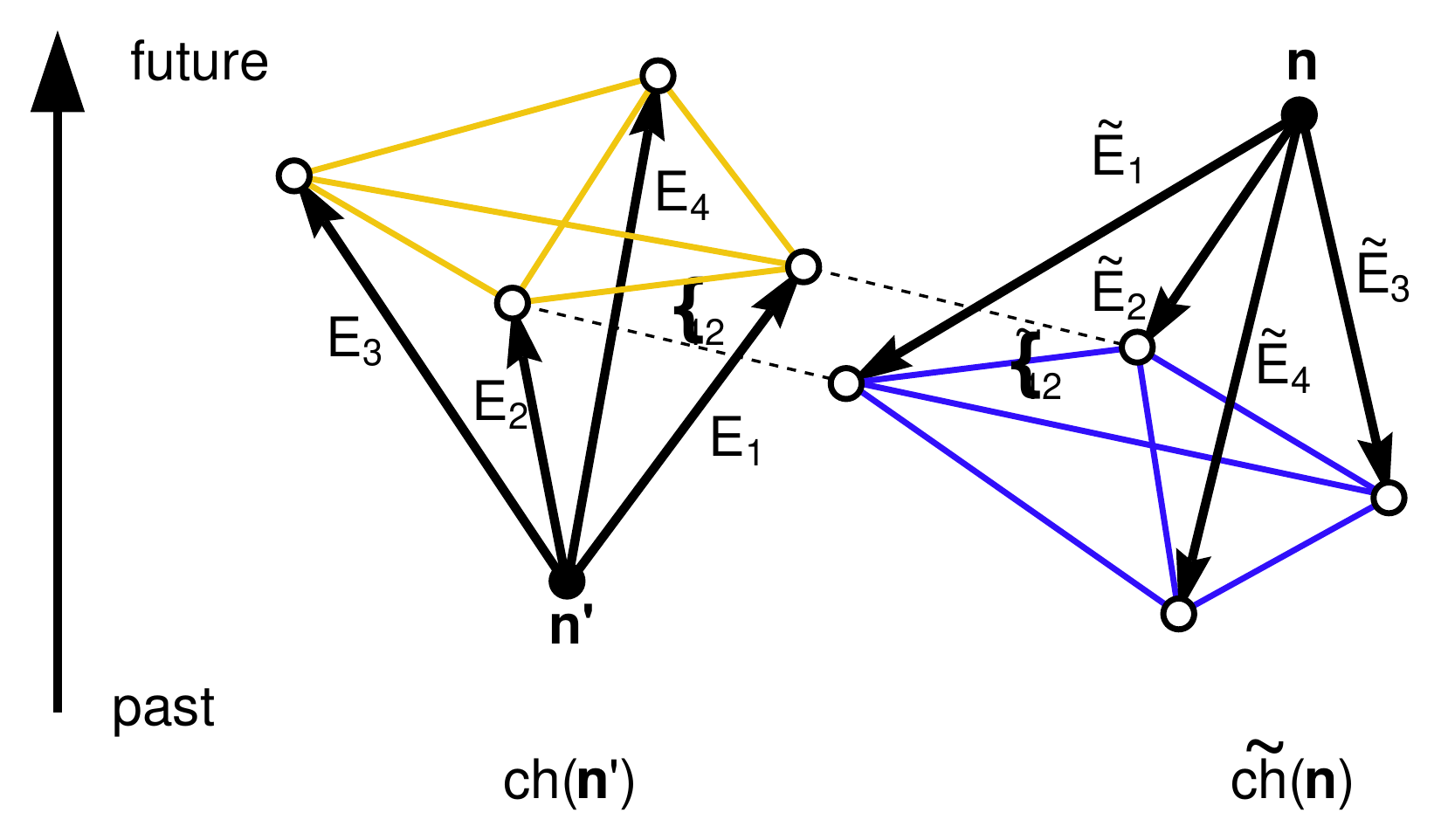}
\caption{Color online. Schematic depiction of the elementary null-simplexes $ch(\bn')$ and $\widetilde{ch}(\bn)$ forming the simplicial complex of the duoverse with passive ($\circ$) and active ($\bullet$) nodes. In the universe all nodes are active.  The four arrows of each simplex represent displacements to nodes on the forward ($ch(\bn')$) and backward ($\widetilde{ch}(\bn)$) light cone of the active nodes $\bn'$ and $\bn$. They are specified by the forward and backward lat-frames $E_\mu(\bn')$ and $\widetilde E_\mu(\bn)$ respectively. These illuminate F- and B-tetrahedrons whose 6 spatial lengths are shown in yellow and blue. Dashed lines connect nodes that are to be identified in the pure simplicial complex. The corresponding common edge of the F-and B-tetrahedron has spatial length $\ell_{12}(\bn-\Delta_1-\Delta_2)=\widetilde \ell_{12}(\bn)$. This is the geometrical content of \equ{symmc}. The four nodes $(\bn,\bn-\Delta_1,\bn-\Delta_2,\bn=\bn-\Delta_1-\Delta_2$ form a plaquette of the topologically hypercubic lattice with $\ell_{12}(\bn-\Delta_1-\Delta_2)=\widetilde\ell_{12}(\bn)$ on the diagonal.  Note that the backward null lat-frames $\widetilde E_\mu$ are redundant (and not included as variables of the model): if the 6 spatial lengths $\widetilde \ell_{\mu\nu}$ of the B-tetrahedron satisfy the inequalities of \equ{inequalities}, the relative location of the apex  $\bn$ of the oriented backward null-simplex is uniquely determined.        
}  
\label{simplices}
\end{figure}

Necessary and sufficient  conditions for constructing the triangulated manifold are obtained by considering the 4-simplexes of the pure complex. The complex is composed of two types of simplexes (which could be viewed as being local charts of an atlas),  $ch(\bn)$ and  $\widetilde{ch}(\bn)$, 
\begin{align}\label{charts}
ch(\bn)=\{\bn,\bn+\mu;\mu=1,\dots 4\}&\ \ \text{and}\ \ \ \widetilde{ch}(\bn)=\{\bn,\bn-\mu;\mu=1,\dots 4\}\ .
\end{align}
As schematically indicated in \fig{simplices}, these simplexes are composed of an apex  event and four causally related events on its forward or backward light cone respectively. We will refer to these special Minkowski simplexes with four light-like and six spatial edges as forward (backward) null-simplexes\footnote{The corresponding 2-dimensional road atlas would be composed of charts that include just three cities, with one-way roads connecting a central city to the other two.  The central city of any chart also is the central city of one other chart with reversed one-way roads (these two charts could be labeled past and future). In the universe every city is a central city on some chart. In the atlas of the duoverse peripheral cities (passive nodes) never are central ones on any chart. Every city is at the intersection of two one-way roads in both atlases.}.

Four of the faces of such a null-simplex of Minkowski space are tetrahedrons with 3 spatial and 3 null edges, the edges of the fifth (F- or B-) tetrahedron are all spatial.  Any two of these simplexes  (charts) have at most two events (and thus at most one length) in common.  The non-trivial intersections of such simplexes are (with $\mu\neq\nu$),
\begin{align}\label{overlap}
ch(\bn)&\cap ch(\bn+\mu)&=&\{\bn+\mu\}&&\nonumber\\
ch(\bn)&\cap \widetilde{ch}(\bn+\mu)&=&\{\bn,\bn+\mu\}, &&\text{with  } \ell^2_{\mu\mu}(\bn)=\widetilde\ell^2_{\mu\mu}(\bn+\mu)=0\nonumber\\
ch(\bn)&\cap ch(\bn+\mu-\nu)&=&\{\bn+\mu\}&&\nonumber\\
ch(\bn)&\cap \widetilde{ch}(\bn)&=&\{\bn\}&&\nonumber\\
ch(\bn)&\cap\widetilde{ch}(\bn+2\mu)&=&\{\bn+\mu\}&&\nonumber\\ 
ch(\bn)&\cap \widetilde{ch}(\bn+\mu+\nu)&=&\{\bn+\mu,\bn+\nu\},&&\text{with  } \ell^2_{\mu\nu}(\bn)=\widetilde\ell^2_{\mu\nu}(\bn+\mu+\nu)\ .
\end{align}
Simplexes (charts) of the duoverse include a single active node ($\bn$) and four passive ones.The first two intersections of \equ{overlap} can be ignored in this case because they do not occur in the simplicial complex of the duoverse (i.e. are not in the atlas).

 Introducing the \emph{backward} null lat-frame $\{\widetilde E_\mu(\bn), \mu=1,\dots,4\}$ on the backward light cone of an event $\bn$ in the same manner as the \emph{forward} null lat-frame $\{ E_\mu(\bn), \mu=1,\dots,4\}$ was defined on the forward light cone of $\bn$, 
the simplexes (charts) can be consistently glued at their common overlap only if 
\bel{symmc}
\ell^2_{\mu\nu}(\bn'):=-2E_\mu(\bn')\cdot  E_\nu(\bn')=-2\widetilde E_\mu(\bn)\cdot  \widetilde E_\nu(\bn):=\widetilde\ell^2_{\mu\nu}(\bn)\ ,\  \text{with } \bn'=\bn-\mu-\nu \ \text{for all  } \mu,\nu\ \text{ and } \bn\ .
\ee

\begin{figure}[h]
\includegraphics[width=2in,angle=-90]{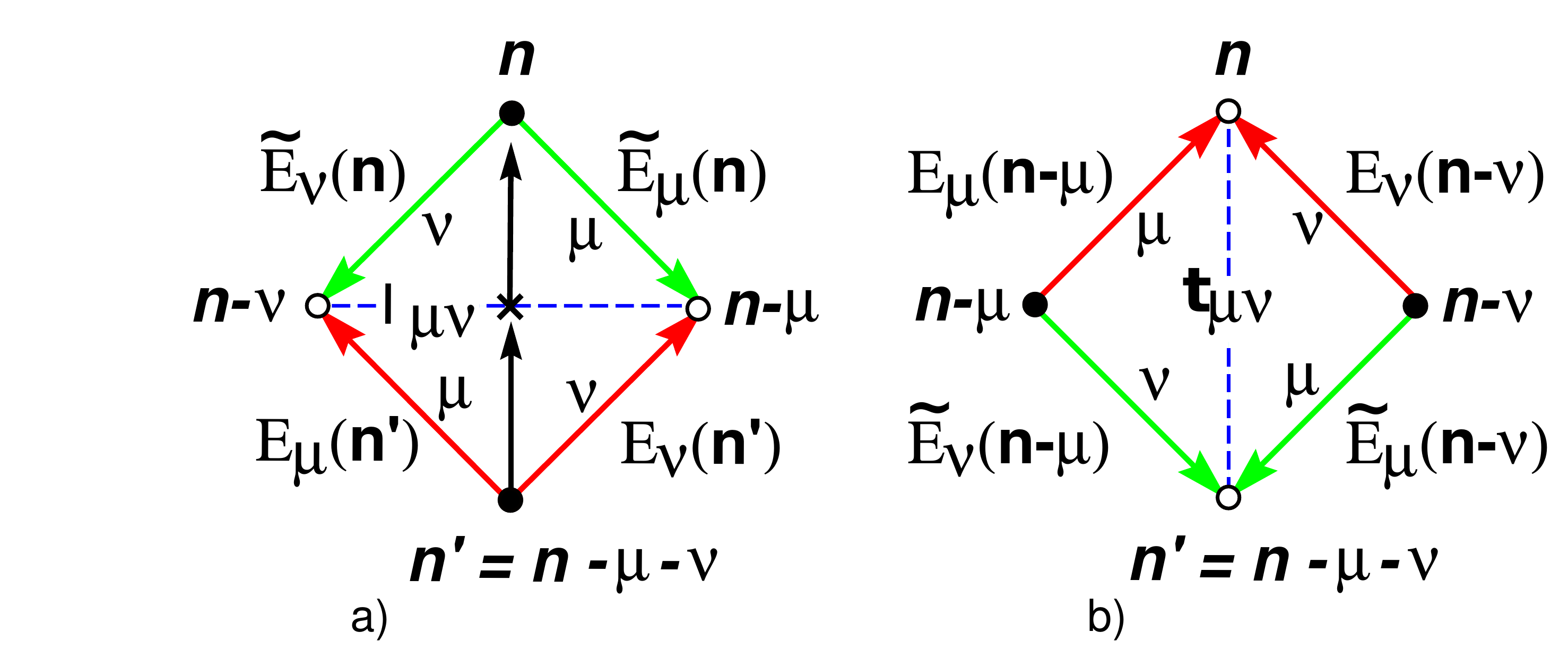}
\caption{Color online. Constraints of a triangulated manifold. Plaquettes of $\bLambda$ in the \emph{duoverse} are of type a), with two spatially separate passive ($\circ$) nodes and two time-like separate active ($\bullet$) nodes or of type b), with two spatially separate active nodes and two time-like separate passive ones. Forward (backward) null lat-frames of the configuration are shown as red (green) arrows that point away from an active vertex and toward a passive one.  In the \emph{universe} all nodes are active.  To construct a consistent simplicial complex, the spatial separation $\ell_{\mu\nu}$ between the events labeled  by the nodes $\bn-\mu$ and $\bn-\nu$  must be the same for local inertial systems at $\bn$ and $\bn-\mu-\nu$. The time-like separation $t_{\mu\nu}$ between events represented by passive nodes $\bn$ and $\bn^\prime$ also is the same for local inertial systems at the active nodes $\bn-\mu$ and $\bn-\nu$.  In a) the center-event of a plaquette at the midpoint of the spatial diagonal is marked by a cross ($\times$) and its time-like separation to the two active nodes indicated by solid (black) arrows (see \sect{PT}). }  
\label{plaquette}
\end{figure}

The same conclusion is reached by examining the typical  $\mu\nu$-plaquette of the $\bLambda$ lattice shown in \fig{plaquette}. If the lattice configuration is to represent a triangulated manifold, the events labeled by $\bn-\mu$ and $\bn-\nu$ need to be on the backward  light cone of  $\bn$ \emph{and} on the forward light cone of the event labeled by $\bn'=\bn-\mu-\nu$. Note that $\bn-\mu=\bn'+\nu$ and $\bn-\nu=\bn'+\mu$  are the only nodes of the lattice in the intersection of these light cones.  

Including the case $\mu=\nu$, \equ{symmc} altogether imposes 10 independent \SLC-invariant constraints per site on the backward lat-frames.  

Equality of the time-like separation $t^2_{\mu\nu}$ between events $\bn$ and $\bn^\prime$ (see \fig{plaquette}b ) as viewed from inertial systems associated with $\bn-\mu$ and $\bn-\nu$ similarly requires that,
\bel{antic}
0=E_\nu(\bn-\nu)\cdot  \widetilde E_\mu(\bn-\nu)- E_\mu(\bn-\mu)\cdot  \widetilde E_\nu(\bn-\mu)\ \text{for all  } \mu,\nu\ \text{ and } \bn\ .
\ee
These are 6 additional independent linear constraints per site. The altogether 16 constraints per site of \equ{symmc} and \equ{antic} offset  the 16 degrees of freedom per site of the $\widetilde E$'s. No \emph{additional} dynamical degrees of freedom have been introduced and one thus should be able to  construct a TLT that constrains lattice configurations to those representing triangulated causal \emph{manifolds}. 

\equ{symmc} as well as \equ{antic} are invariant under local \SLC transformations of all the lat-frames, but \equ{symmc} is invariant under \emph{independent} \SLC transformation of the backward and forward lat-frames at each active site. For a given configuration of forward null vectors $\{E_\mu(\bn); [\bn,\mu]\in\bLambda\}$ the constraints~(\ref{symmc}) imply that, 
\bel{Bcond}
\{ 0<\widetilde\ell_{\mu\nu}(\bn)=\ell^2_{\mu\nu}(\bn-\mu-\nu)=-2 E_\mu(\bn-\mu-\nu)\cdot E_\nu(\bn-\mu-\nu), \ \text{for } 1\leq\mu<\nu\leq 4\} \ ,
\ee
are the spatial lengths of the 6 edges of a B-tetrahedron whose four vertices are on the backward light-cone of the event $\bn$.   

The six additional constraints of \equ{antic} on the other hand  apparently only serve to localize the additional \SLC-invariance $\widetilde E_\mu(\bn)\rightarrow g(\bn)\widetilde E_\mu(\bn) g^\dagger(\bn)$ of the solution to \equ{symmc}. 
The ten \SLC-invariant constraints of \equ{symmc} in fact suffice to (uniquely) construct the whole complex. The geometry of this complex does not depend on  the additional  \SLC degrees of freedom of the backward lat-frames.   

As in the GPS \cite{Fang:1986, Chaffee:1991,Fang:1992}, the apex $\bn$ can be  reconstructed from the vertices of the B-tetrahedron of $\widetilde{ch}(\bn)$.  The discussion of light-cones in \sect{Fields} implies that this is possible only if the 6 spatial lengths $\widetilde\ell_{\mu\nu}(\bn)=\ell_{\mu\nu}(\bn-\mu-\nu)>0$ of \equ{Bcond} satisfy the inequalities,
\begin{align}\label{inequalities}
\widetilde\ell_{12}(\bn)\widetilde\ell_{34}(\bn)&\leq \widetilde\ell_{13}(\bn)\widetilde\ell_{24}(\bn)+\widetilde\ell_{14}(\bn)\widetilde\ell_{23}(\bn)\ ,\nonumber\\
\widetilde\ell_{13}(\bn)\widetilde\ell_{24}(\bn)&\leq
\widetilde\ell_{12}(\bn)\widetilde\ell_{34}(\bn)+\widetilde\ell_{14}(\bn)\widetilde\ell_{23}(\bn)\ ,\nonumber\\
\widetilde\ell_{14}(\bn)\widetilde\ell_{23}(\bn)&\leq
\widetilde\ell_{12}(\bn)\widetilde\ell_{34}(\bn)+\widetilde\ell_{13}(\bn)\widetilde\ell_{24}(\bn)\ ,\ \text{for all} \ \bn\ .
\end{align}
Since all the lengths are positive, we again can combine the triangle inequalities of \equ{inequalities} to the single requirement that,
\bel{dettgineq}
\det{\widetilde\ell_{\mu\nu}(\bn)}\leq 0\ ,\ \text{for all} \ \bn\ .
\ee
The inequalities of \equ{inequalities} are \SLC-invariant and select configurations of null lat-frames that represent triangulated causal manifolds. Due to \equ{dettgineq} the consistency conditions are the requirement that the 4-volume of every backward simplex $\widetilde V(\bn):=\sqrt{-\det{\widetilde \ell^2_{\mu\nu}(\bn)}}$ be real.

These quasi-local but non-linear consistency conditions on the lattice configuration not only are reasonable and necessary but also are sufficient to reconstruct the oriented \emph{backward} lat-frame up to $\SLC$ transformation from the set of  lengths $\{\widetilde\ell_{\mu\nu}(\bn)\}$ (see Appendix~\ref{Reconstruction}). As outlined in \sect{thelattice}, one then can reconstruct the simplicial complex of the triangulated manifold uniquely. The conditions of \equ{dettgineq} in some sense are "integrability conditions" of the causal lattice.   

\subsection{The Manifold TLT}
\label{secTLT}
We here construct a local TLT whose partition function vanishes when \equ{inequalities} is violated and is a (non-vanishing) constant otherwise. The TLT partition function evidently is proportional to the product of Heaviside functions that enforce the inequalities of\equ{inequalities}. We wish to  have a local integral representation of it. Many such representations exist.  In Appendix~\ref{coframeMC} an equivariant BRST construction is used to obtain a local TLT in terms of the lat-frames that ensures \equ{symmc}. Unfortunately this local TLT depends explicitly on the backward lat-frames and more than doubles the number of lattice variables. One also can simply enforce the single inequality of \equ{dettgineq}, which is of rather high order in the lattice variables. 

Here we construct a simpler TLT based on the spinor formulation of the lattice model. It requires just two new variables (spinor phases) per site to enforce \equ{inequalities} with a local lattice action of the same scaling dimension as the cosmological term.     

Consider therefore the representation of backward null lat-frames $\widetilde E_\mu^{A \dot B}(\bn)$ by spinors $\widetilde \xi^A_\mu(\bn)$,
\bel{backE}
\widetilde E_\mu^{AB'}(\bn)=- i (\widetilde\xi_\mu\otimes\widetilde\xi^*_\mu)^{AB'}=-i\widetilde\xi_\mu^A(\bn)\widetilde \xi^{*B'}_\mu(\bn)\ .
\ee
The only difference to representation of forward lat-frames by spinors in\equ{spinors} is the minus sign in \equ{backE}. It implies that $-i\Tr \widetilde E_\mu (\bn) =-(|\widetilde \xi_\mu^1(\bn)|^2+|\widetilde \xi_\mu^2(\bn)|^2)<0$. $\widetilde E_\mu(\bn)$ thus is on the backward light cone.

Events labeled by $\bn-\mu$ and $\bn-\nu$  are spatially separated by,
\bel{sluv}
\widetilde \ell_{\mu\nu}(\bn)=|\widetilde\xi^A_\mu(\bn)\eps_{AB}\widetilde\xi^B_\nu(\bn)|\ , 
\ee
and \equ{symmc} demands that,
\bel{manconsp}
\widetilde \ell^2_{\mu\nu}(\bn)=\ell^2_{\mu\nu}(\bn-\mu-\nu)=| f_{\mu\nu}(\bn-\mu-\nu) |^2\ .
\ee
A configuration of forward spinors $\{\xi_\mu(\bn)\}$ thus is compatible with a  triangulated manifold only if real phases $\varphi_{\mu\nu}(\bn)=\varphi_{\nu\mu}(\bn)$ and backward spinors $\widetilde\xi_\mu (\bn)$ exist for which,
\bel{manspinor}
\widetilde\xi^A_\mu(\bn)\eps_{AB}\widetilde\xi^B_\nu(\bn)=\widetilde f_{\mu\nu}(\bn):=e^{i\varphi_{\mu\nu}(\bn)} f_{\mu\nu}(\bn-\mu-\nu) \ \text{for all  }\bn,\mu,\nu \ .
\ee
The skew-symmetric matrix $\widetilde f_{\mu\nu}(\bn)$ defined by the right-hand-side of  \equ{manspinor} is described by spinors only if its Pfaffian vanishes\footnote{If $\text{Pf}[\widetilde f(\bn)]=0$, $\widetilde f_{\mu\nu}(\bn)=\gamma(X_\mu Y_\nu-X_\nu Y_\mu)$ where the two linearly independent vectors $X_\mu$ and $Y_\mu$ span the two-dimensional kernel of the dual tensor $\widetilde f^{\mu\nu}=\sum_{\rho,\sigma}\half \eps(\mu\nu\rho\sigma) \widetilde f_{\rho\sigma}(\bn)$ and $\gamma$ is a complex number. A possible set of spinors then is $\widetilde\xi_\mu(\bn)=(\gamma X_\mu,Y_\mu)$.}. A configuration therefore  represents a triangulated manifold only if a set of phases $\{\varphi_{\mu\nu}(\bn)=\varphi_{\nu\mu}(\bn)\in \mathbb{R}\}$ exists  for which 
\bel{Manspin2}
0=\text{Pf}(\widetilde f(\bn))=\widetilde f_{12}(\bn)\widetilde f_{34}(\bn)+\widetilde f_{13}(\bn)\widetilde f_{42}(\bn)+\widetilde f_{14}(\bn)\widetilde f_{23}(\bn)\ \text{for all}\ \bn\ ,
\ee
where $\widetilde f_{\mu\nu}(\bn)$ is given by \equ{manspinor}. Note that this formulation of the consistency condition assures the existence of  backward spinors $\widetilde\xi_\mu(\bn)$ without explicitly constructing them.

The phases  $\{\varphi_{\mu\nu}(\bn)\}$ can be found and \equ{Manspin2} satisfied only if the magnitudes of the three complex numbers,
\begin{align}\label{defabc}
\widetilde a(\bn)&:=|f_{12}(\bn-\bDelta_1-\bDelta_2) f_{34}(\bn-\bDelta_3-\bDelta_4)|=\widetilde\ell_{12}(\bn)\widetilde\ell_{34}(\bn)\nonumber\\
\widetilde b(\bn)&:=|f_{13}(\bn-\bDelta_1-\bDelta_3) f_{42}(\bn-\bDelta_2-\bDelta_4)|=\widetilde\ell_{13}(\bn)\widetilde\ell_{24}(\bn)\nonumber\\
\widetilde c(\bn)&:=|f_{14}(\bn-\bDelta_1-\bDelta_4) f_{23}(\bn-\bDelta_2-\bDelta_3)|=\widetilde\ell_{14}(\bn)\widetilde\ell_{23}(\bn)\ ,
\end{align}
form the sides of a triangle. To construct  the pure simplicial complex of a triangulated causal manifold, the spatial lengths $\widetilde \ell_{\mu\nu}(\bn)=|f_{\mu\nu}(\bn-\bDelta_\mu-\bDelta_\nu)|=\ell_{\mu\nu}(\bn-\bDelta_\mu-\bDelta_\nu) $ thus must satisfy the inequalities of \equ{inequalities}.

Suppose that for a configuration of forward spinors a set of phases $\varphi_{\mu\nu}$ can be found for which \equ{Manspin2} is satisfied. We first demonstrate that a physically equivalent configuration of forward spinors exists in this case that satisfies \equ{Manspin2} with vanishing phases. 

The $U^4(1)$ invariance of physical observables (and lat-frames) of \equ{U1s} implies the equivalence of  phases,
\bel{equivphase}
\varphi_{\mu\nu}(\bn)\equiv \varphi^\prime_{\mu\nu}(\bn)=\varphi_{\mu\nu}(\bn)-\psi_\mu(\bn-\mu-\nu)-\psi_\nu(\bn-\mu-\nu)\ .
\ee

For $\text{Pf}[\widetilde f(\bn)]=0$ the overall phase is irrelevant and  we need only construct a representative set of spinors for which the phases of the three terms of the Pfaffian coincide, that is 
\bel{Manspin3}
\varphi^\prime_{12}(\bn)+\varphi^\prime_{34}(\bn)=\varphi^\prime_{13}(\bn)+\varphi^\prime_{42}(\bn)=\varphi^\prime_{14}(\bn)+\varphi^\prime_{23}(\bn)\ .
\ee
 Given the  initial set of phases $\{\varphi_{\mu\nu}(\bn)\}$ for which  \equ{Manspin2} is satisfied, the $2N$ linear constraints of \equ{Manspin3} on the $4N$ phases $\psi_\mu(\bn)$ in general allow for an infinite set of solutions. To have a more manageable set we restrict to   $\psi_1(\bn)=-\psi_2(\bn)$ and  $\psi_3(\bn)=-\psi_4(\bn)$.  Defining $\psi_\pm(\bn):=\psi_1(\bn)\pm\psi_3(\bn)=-\psi_2(\bn)\mp\psi_4(\bn)$, the conditions of \equ{Manspin3}  then decouple into two equations,
\begin{align}\label{solvePhases}
\psi_+(\bn-\bDelta_2-\bDelta_4)-\psi_+(\bn-\bDelta_1-\bDelta_3)&=\varphi_{12}(\bn)+\varphi_{34}(\bn)-\varphi_{13}(\bn)-\varphi_{24}(\bn)\hspace{3em}\mod{2\pi}\nonumber\\
\psi_-(\bn-\bDelta_2-\bDelta_3)-\psi_-(\bn-\bDelta_1-\bDelta_4)&=\varphi_{12}(\bn)+\varphi_{34}(\bn)-\varphi_{14}(\bn)-\varphi_{23}(\bn)\hspace{3em}\mod{2\pi}\ ,
\end{align}
that relate the  $\psi_\pm$ angles at the endpoints of two diagonals of a hypercube. With appropriate boundary conditions \equ{solvePhases} uniquely determines the angles $\psi_\pm(\bn), \mod{2\pi}$. The angles $\psi_1(\bn)=(\psi_+(\bn)+\psi_-(\bn))/2$ and $\psi_2(\bn)=(\psi_+(\bn)-\psi_-(\bn))/2$ then are known modulo~$\pi$ and the new spinors   $\{\xi^\prime_\mu(\bn),\mu=1,\dots,4\}$ are determined up to sign.      

Let us therefore attempt to construct the TLT that enforces the manifold condition by imposing $\text{Pf}[\widetilde f(\bn)]=0$ at all nodes $\bn$ as a $U^2(1)$ gauge condition on the spinors. This "gauge" conditions has a solution only if the corresponding configuration of forward null lat-frames represents a triangulated manifold\footnote{The forward null lat-frames do not depend on the spinor phases and are $U^4(1)$ invariants.}.

We introduce ghosts $c_\pm$ and a corresponding BRST-doublet of complex anti-ghosts $\bar c,\bar c^*$ as well as complex Lagrange multiplier fields $\eta,\eta^*$. The  nilpotent BRST-variation of the spinor phases $\psi_\pm$  and of these additional fields is,
\begin{align}\label{BRSTMansp}
s \psi_\pm(\bn)&=c_\pm(\bn) ,\
s c_\pm(\bn)=0 &s \bar c(\bn)&=\eta(\bn)  ,\
s \eta(\bn)=0 &s \bar c^*(\bn)&=\eta^*(\bn)  ,\
s \eta^*(\bn)=0 \ ,
\end{align} 
with a trivial extension to other fields. We therefore consider the TLT with the BRST-exact Lagrangian,
\begin{align}\label{L_TLT}
\mathcal{L}_{TLT}(\bn)&=s [\bar c^*(\bn) \text{Pf}[f'(\bn)]+\bar c(\bn)\text{Pf}[f'^*(\bn)]+i \frac{\alpha}{2} (\eta^*(\bn) \bar c(\bn)+\eta(\bn) \bar c^*(\bn))]\nonumber\\ 
&=\eta^*(\bn) \text{Pf}[f'(\bn)] +\eta(\bn) \text{Pf}[f'(\bn)]+ i\alpha \eta(\bn)\eta^*(\bn)+\\
&\hspace{3em}
+i\left(\begin{matrix} \bar c(\bn) & \bar c^*(\bn)\end{matrix}\right)\cdot\left(\begin{matrix} f'^*_{13}(\bn)f'^*_{42}(\bn) & f'^*_{14}(\bn)f'^*_{32}(\bn)\\ f'_{13}(\bn)f'_{42}(\bn)& f'_{14}(\bn)f'_{32}(\bn)\end{matrix}\right)\cdot\left(\begin{matrix}c_+(\bn-\bDelta_1-\bDelta_3)-c_+(\bn-\bDelta_2-\bDelta_4) \\ c_-(\bn-\bDelta_1-\bDelta_4)-c_-(\bn-\bDelta_2-\bDelta_3)\end{matrix}\right)\ , \nonumber
\end{align}
where we have introduced a gauge-parameter $\alpha\ge 0$. $f'(\bn)$ in \equ{L_TLT} depends on the spinorial phases and is the skew-symmetric matrix with components,
\begin{align}\label{fbar}
f'_{12}(\bn)&=-f'_{21}(\bn):=f_{12}(\bn-\bDelta_1-\bDelta_2)\nonumber\\
f'_{34}(\bn)&=-f'_{43}(\bn):=f_{34}(\bn-\bDelta_3-\bDelta_4)\nonumber\\
f'_{13}(\bn)&=-f'_{31}(\bn):=e^{-i\psi_+(\bn-\bDelta_1-\bDelta_3)}f_{13}(\bn-\bDelta_1-\bDelta_3)\nonumber\\
f'_{24}(\bn)&=-f'_{42}(\bn):=e^{+i\psi_+(\bn-\bDelta_2-\bDelta_4)}f_{24}(\bn-\bDelta_2-\bDelta_4)\nonumber\\
f'_{14}(\bn)&=-f'_{41}(\bn):=e^{-i\psi_-(\bn-\bDelta_1-\bDelta_4)}f_{14}(\bn-\bDelta_1-\bDelta_4)\nonumber\\
f'_{23}(\bn)&=-f'_{32}(\bn):=e^{+i\psi_-(\bn-\bDelta_2-\bDelta_3)}f_{23}(\bn-\bDelta_2-\bDelta_3)\ .
\end{align}
The partition function of this TLT is,
\begin{align}\label{UTLT}
Z^{TLT}_\alpha[\xi]&\propto \prod_\bn\int dc_+(\bn) dc_-(\bn) d\bar c(\bn) d\bar c^*(\bn) \int_{S_1\times S_1} \hspace{-1.8em}d\psi_+(\bn) d\psi_-(\bn)\int_{\mathbb{R}^2} d\eta(\bn) d\eta^*(\bn)  e^{i \mathcal{L}_{TLT}(\bn)}\nonumber\\
&\propto \prod_\bn \int_{S_1\times S_1} \hspace{-1.8em}d\psi_+(\bn) d\psi_-(\bn) \widetilde V(\bn) \;e^{-\frac{1}{\alpha}\left|\text{Pf}[f'(\bn)]\right|^2}\ .
\end{align}
The last expression is obtained by integrating over ghost and Lagrange multiplier fields and 
\bel{defVt}
\widetilde V(\bn):=\half\text{Im} f'_{13}(\bn)f'^*_{32}(\bn)f'_{24}(\bn)f'^*_{41}(\bn)={\textstyle  {\frac i 4}}\det{\left(\begin{matrix} f'^*_{13}(\bn)f'^*_{42}(\bn) & f'^*_{14}(\bn)f'^*_{32}(\bn)\\ f'_{13}(\bn)f'_{42}(\bn)& f'_{14}(\bn)f'_{32}(\bn)\end{matrix}\right)}\ .
\ee

In the limit $\alpha\rightarrow 0^+$, only spinor configurations that satisfy $\text{Pf}[f'(\bn)]\sim 0$  contribute significantly to the integral of \equ{UTLT}. This is the condition we wish to enforce, but one can show  \cite{Witten:1988,Baulieu:1996rp, Schaden:1998hz}  that\footnote{Viewing  $\text{Pf}[f'(.)]$ as a smooth 2N-dimensional real vector field on $T_{2N}=(S_1\times S_1)^N$ with coordinates $(\psi_+(.),\psi_-(.))$,  this is a consequence of the Poincar{\'e}-Hopf theorem\cite{Birmingham:1991ty}.}  $Z^{TLT}_\alpha[\xi]$ is proportional to the Euler characteristic  of the 2N-dimensional torus $T_{2N}=(S_1\times S_1)^N$.  As defined in \equ{UTLT} $Z^{TLT}_\alpha[\xi]$ therefore  vanishes\cite{Neuberger:1986xz,Schaden:1998hz} for \emph{any}  $\alpha>0$ and spinor configuration $\{\xi\}$. This TLT thus \emph{fails} to constrain to configurations that satisfy the consistency condition. 

Fortunately this can be rectified. We already know that any skew-symmetric complex matrix with vanishing Pfaffian may be written in terms of spinors. Configurations that  contribute significantly to the integral of \equ{UTLT}  in the limit $\alpha\rightarrow 0^+$ thus can be written in terms of \emph{backward} spinors as,
\bel{Maninverse}
 f'_{\mu\nu}(\bn)=\widetilde\xi^A_\mu(\bn)\eps_{AB}\widetilde \xi^B_\nu(\bn)\ . 
\ee
For $\text{Pf}[f'(\bn)]=0$,  $\widetilde V(\bn)$, given by \equ{defVt}, therefore is just the (signed) four dimensional volume element,
\bel{barV0}
\widetilde V(\bn)=- \textstyle{\frac{i}{2}} \Tr \widetilde E_3(\bn)\eps \widetilde E^T_2(\bn) \eps\widetilde E_4(\bn)\eps \widetilde E_1^T(\bn)=\det[\widetilde E(\bn)]\ ,
\ee
of the backward null lat-frames given by\equ{backE}.
The lengths $\widetilde\ell_{\mu\nu}(\bn)=|f'_{\mu\nu}(\bn)|$ themselves do not depend on spinor phases. Positive and negative contributions to $Z^{TLT}_{0^+}[\xi]$ in \equ{UTLT} thus arise from different signs of the 4-volumes $\widetilde V(\bn)$ of configurations with the same set of spatial lengths $\{\widetilde\ell_{\mu\nu}(\bn)\}$.  Non-singular solutions to $\text{Pf}[f'(\bn)]=0$ in fact come in pairs with $\widetilde V(\bn)$ of opposite sign and this leads to $Z^{TLT}_{0^+}[\xi]=0$. 

\emph{Orientable} manifolds on the other hand ought to be triangulated by spinor configurations for which $\widetilde V(\bn)>0$ at all sites $\bn$. It thus is tempting to modify the TLT to,
\bel{tildeZ}
\widetilde Z^{TLT}[\xi]= \lim_{\alpha\rightarrow 0^+} \prod_\bn \alpha^{-1}\int_{S_1\times S_1} \hspace{-1.8em}d\psi_+(\bn) d\psi_-(\bn) \widetilde V(\bn) \Theta[\widetilde V(\bn)]\;e^{-\frac{1}{\alpha}\left|\text{Pf}[f'(\bn)]\right|^2}\ ,
\ee
where $\Theta[x]$ is the Heaviside distribution. The question arises  whether $\widetilde Z^{TLT}$ is a topological integral that does not depend on generic variations of the spinor configuration. This generally will not be true if the number of (pairs of) solutions depends on the configuration. 

We can assure ourselves that this is not the case by explicit evaluation of \equ{tildeZ}.  Note that $\widetilde V(\bn)$, defined in \equ{defVt}, depends on the following differences of the angles in \equ{solvePhases} only,
\begin{align}\label{newphases}
\theta(\bn)&=\half(\Psi_+(\bn-\bDelta_2-\bDelta_4)-\Psi_+(\bn-\bDelta_1-\bDelta_3)+ \Psi_-(\bn-\bDelta_2-\bDelta_3)-\Psi_-(\bn-\bDelta_1-\bDelta_4))\nonumber\\
\varphi(\bn)&=\Psi_+(\bn-\bDelta_2-\bDelta_4)-\Psi_+(\bn-\bDelta_1-\bDelta_3)- \Psi_-(\bn-\bDelta_2-\bDelta_3)+\Psi_-(\bn-\bDelta_1-\bDelta_4)\ .
\end{align}
The partition function of \equ{tildeZ} thus factorizes in the form,
\bel{tildeZ1}
\widetilde Z^{TLT}[\xi]= \prod_\bn \Upsilon[f_{12}(\bn-\bDelta_1-\bDelta_2) f_{34}(\bn-\bDelta_3-\bDelta_4),f_{13}(\bn-\bDelta_1-\bDelta_3) f_{42}(\bn-\bDelta_2-\bDelta_4),f_{14}(\bn-\bDelta_1-\bDelta_4) f_{23}(\bn-\bDelta_2-\bDelta_3)]\ ,
\ee 
with,
\bel{Idef}
\Upsilon[a,b,c]:=\lim_{\alpha\rightarrow 0^+} \frac{1}{\pi\alpha}\int_0^{2\pi} \hspace{-1em}d\theta \int_0^{2\pi}\hspace{-1em}d\varphi\ \text{Im} (b c^* e^{i\varphi}) \Theta[ \text{Im} (b c^* e^{i\varphi})] \exp\left\{-\frac{1}{\alpha}\left|a+b\; e^{i(\theta+\varphi/2)}+c\; e^{i(\theta-\varphi/2)}\right|^2\right\}\ .
\ee

\begin{figure}[h]
\includegraphics[width=5in]{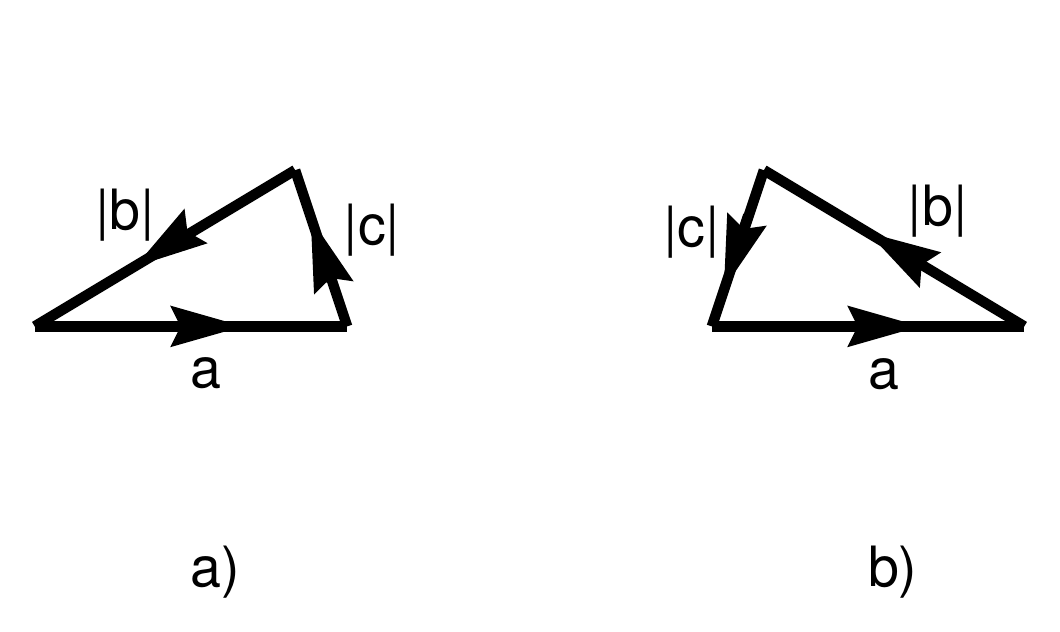}
\caption{\small The pair of generic extrema with $a+b+c=0$ for given magnitudes of the complex numbers $b$ and $c$. The magnitudes $|a|,|b|,|c|$ satisfy the triangle inequalities of \equ{ineq}. Both solutions coincide only if $|a|=||b|\pm|c||$.}
\label{absextrema}
\end{figure}

For $\alpha\rightarrow  0^+$ the two-dimensional integral of \equ{Idef} may be evaluated semiclassically.  The exponent  vanishes only at (absolute) extrema that satisfy the  triangle inequalities, 
\bel{ineq}
|a|\leq |b|+|c|\ \wedge\  |b|\leq |c|+|a| \ \wedge\ |c|\leq |a|+|b| \ ,
\ee 
If \equ{ineq} holds, these extrema correspond to angles $\bar\theta,\bar\varphi$ that satisfy,
\bel{absextremum} 
|a|^2=|b|^2+|c|^2+2\text{Re}\;b c^* e^{i\bar\varphi}\ ,\  e^{2 i\bar\theta}=\frac{a(b^*e^{-i\bar\varphi/2}+c^* e^{i\bar\varphi/2})}{a^*(b e^{i\bar\varphi/2}+c e^{-i\bar\varphi/2})}\ . 
\ee
As \fig{absextrema} illustrates, the exponent generically vanishes at pairs of  absolute minima\footnote{If $a$ is taken to be real, the two extrema correspond to letting $(b,c)\rightarrow(b^*,c^*)$.}, with angles $\bar\varphi$ and $\bar\varphi'$ related by $b c^* e^{i\bar\varphi}=b^* c\,  e^{-i\bar\varphi'}$.  Both extrema satisfy \equ{absextremum}, but give the opposite sign for $\widetilde V(\bn)=\text{Im}\;  b c^* e^{i\bar\varphi}$.
The Gaussian integral over fluctuations around these absolute minima furthermore are the same and proportional to $\alpha/|\text{Im } b c^* e^{i\bar\varphi}|$  The total contribution to the partition function for $\alpha\rightarrow 0^+$ thus gives the vanishing Euler characteristic of the  2-torus if one sums over \emph{both} minima. Taking only solutions with \emph{positive} volume in\equ{Idef} on the other hand leads to a non-vanishing partition function that generically does not depend on $a, b$ and $c$ as long as these satisfy the inequalities of \equ{ineq}. $\Upsilon[a,b,c]$ on the other hand vanishes exponentially in the limit $\alpha\rightarrow 0+$ when \equ{ineq} is violated. $\Upsilon[a,b,c]$ thus is an integral representation of the distribution, 
\bel{upsilon}
\Upsilon(a,b,c)=\Theta(|a|+|b|-|c|)\  \Theta(|b|+|c|-|a|)\ \Theta(|c|+|a|-|b|)=\Theta(2 a^2 b^2+2 b^2 c^2+2 c^2 a^2-a^4-b^4-c^4)\ ,
\ee
Where $\Theta(x)$ is the Heaviside function. It is a constant if $a,b,c$ satisfy triangle inequalities and vanishes otherwise. 

The TLT partition function of \equ{tildeZ} thus constrains the space of configurations to those satisfying \equ{inequalities} that correspond to triangulated causal manifolds. Note that $\widetilde Z^{TLT}$ is a topological lattice integral in the limit $\alpha\rightarrow 0^+$ only.  For $\alpha>0$, the integral \emph{does} depend on the configuration. However, the limit $\alpha\rightarrow 0^+$ appears to be smooth and well-defined for $|a|^2>\alpha$. Regularization of the model in \sect{Regularization} ensures that this bound is satisfied.

\section{Partial Localization to an $SU(2)$-structure group}
\label{Localization}
The action of\equ{LatHP} is real and the weight $e^{i S_\text{HP}}$ of a configuration therefore oscillatory. Although $S_\text{HP}$ is unbounded,  it is conceivable that configurations with very large classical action do not contribute to the integral, as in ordinary Fresnel integrals. 

Even if this is the case, the Lorentzian  model with the lattice action of \equ{LatHP} is not well defined because its \SLC structure group is not compact.  The infinite volume of this group formally cancels in the expectation value of invariant physical observables, but it prevents one from defining a finite generating function even for a lattice with a finite number of sites. Contrary to ordinary lattice gauge theories with compact structure group, the \SLC structure group of this model has to be localized to a compact subgroup for the generating  function to be defined.  The localization to an $SU(2)$-invariant lattice theory (where only the subgroup of local spatial rotations remains free)  in fact is unique. The following  partial gauge fixing of lattice configurations in this sense does not suffer of a Gribov \cite{Gribov:1978ab, Singer:1978} ambiguity.

Consider the local $SU(2)$-invariant Morse function constructed from the spinors of a single active node  $\bn$,
\bel{Morsefunction}
W_\xi[g(\bn)]=\sum_{\mu} w_\mu(\bn) \xi^\dagger_\mu(\bn) g^\dagger(\bn) g(\bn)\xi_{\mu}(\bn)=\sum_\mu w_\mu(\bn)\tau^{(g)}_\mu(\bn)\ ,
\ee
where $\{w_\mu(\bn)\}$ is a set of non-negative weights.  This Morse function is proportional to an average over the \emph{positive} time components $\tau^{(g)}_\mu=-i\half\Tr g(\bn)E_\mu(\bn) g^\dagger(\bn)$ of the \SLC-transformed null lat-frame. It is bounded below and invariant under the $SU(2)$ subgroup of spatial rotations. Considered as a function of $g(\bn)\in\SLC$ for a given spinor configuration,  $W_\xi[g(\bn)]$ thus is a function on the coset space $\SLC/SU(2)$ only. Decomposing $g\in\SLC$ into a hermitian and a unitary part, both of unit determinant,  
\bel{decomp}
g=u h, \text{with } h=h^\dagger=e^{-i\vec v\cdot\vec\sigma},\ \text{and } u\in SU(2)\ ,
\ee
critical points of $W_\xi$ are characterized by,
\bel{gaugecond}
0=-i\sum_{\mu}w_\mu(\bn) \xi^\dagger_\mu(\bn) \,\vec\sigma\,\xi_{\mu}(\bn)\ .
\ee
The $3\times 3$ Hessian matrix $H$ of this Morse function is strictly positive and  proportional to the identity,
\bel{Hessian}
H(\bn)=\one \sum_\mu w_\mu(\bn) \tau^{(g)}_\mu(\bn)=\one W_\xi[g(\bn)]\ .
\ee
$\det[H(\bn)]$ vanishes only if  $W_\xi[g(\bn)]=0$, that is, when the null lat-frame is singular\footnote{The invariant regularization of the theory in \sect{Regularization} implies that $\det{E(\bn)}:=\det {(E^a_\mu(\bn))}>0$ for all $\bn$.}. The non-trivial solution to \equ{gaugecond} thus is unique  (modulo spatial $SU(2)$ rotations) and the Euler characteristic  $\chi(\SLC/SU(2))=\chi(H_3)=1$. Note that this partial gauge eliminates one of the Lorentz components of a forward null lat-frame as a dynamical variable.

Several partial localizations of this kind are of special interest. They differ only in the choice of weights $w_\mu(\bn)\ge 0$ in \equ{Morsefunction}.
\begin{itemize}
\item{The most practical partial localization uses just two linearly independent\footnotemark[\value{footnote}]  null vectors at each node. Setting $w_1(\bn)=w_2(\bn)=0$ and $w_3(\bn)=w_4(\bn)>0$, the gauge condition of \equ{gaugecond} implies that $\vec E_4(\bn)=-\vec E_3(\bn)$ and $\tau_3(\bn)=\tau_4(\bn)=|\vec E_3(\bn)|$, i.e. the corresponding two events appear simultaneous in this local inertial system.    A spinorial solution to this condition is,
\bel{gauge2}
\xi_4^1(\bn)= -\xi_3^{2*}(\bn),\  \  \ \xi_4^2(\bn)=\xi_3^{1*}(\bn)\ .   
\ee
It is unique up to $U(1)$ phase transformations of the spinors. Note that \equ{gauge2} is compatible with the choice made in \sect{MC} to solve the consistency constraints with $\psi_3(\bn)=-\psi_4(\bn)$. Due to its simplicity, the partial localization of \equ{gauge2} will be used in much of the following.}

\item{It may be desirable to choose equal weights for three of the four components, i.e. $w_4(\bn)=0$ and $w_1(\bn)=w_2(\bn)=w_3(\bn)>0$. The spatial components of three null vectors here form a triangle whose side-lengths are the temporal components of these null vectors.  However, an explicit solution to \equ{gaugecond} in terms of the spinors is quite involved.}

\item{Sometimes a more equitable gauge condition is preferable that weights all components equally $w_1(\bn)=w_2(\bn)=w_3(\bn)=w_4(\bn)$ . \equ{gaugecond} then implies that the spatial parts of the four null-vectors of the lat-frame sum to zero. This gauge condition is used to reconstruct the backward lat-frame  in Appendix~\ref{Reconstruction}.}

\item{The weights $w_\mu(\bn)$ in \equ{Morsefunction} can be any positive \SLC-invariants of the fields. A judicious choice of these invariants may sometimes  be advantageous (as 't Hooft gauges are in spontaneously broken gauge theories). In the next section we have occasion to consider path-dependent gauges where the weights $w_\mu(\bn)$ depend on $\bn$ in a manner that simplifies the computation of a path-dependent quantity, such as the proper time. } 
\end{itemize}   

The partial localization of \SLC to the SU(2) subgroup at each site of the lattice may also be achieved by an equivariant BRST construction \cite{Birmingham:1991ty, Schaden:1998hz}. Since the gauge condition here is ultra-local, the construction here is rather trivial, but it illustrates the general principle. The nilpotent BRST variation of the spinors, ghosts ($\vec c$ and $\vec\omega$) and auxiliary fields ($\vec b$ and $\vec {\bar c}$) is,
\begin{align}\label{BRSTslc}
s\xi_\mu &=\half(\vec c+\vec\omega)\cdot\vec\sigma \xi_\mu\ ,& s\xi^{\dagger}_\mu &= \half\xi^\dagger_\mu\vec\sigma\cdot(\vec c-\vec\omega)\nonumber\\
s\vec c&=\vec\omega\times \vec c\ ,&s\vec\omega&=\half\vec c\times\vec c+\half\vec\omega\times\vec\omega \nonumber\\
s \vec{\bar c}&=\vec b+\vec\omega\times\vec{\bar c}\ , &s\vec b&=\vec \omega\times \vec b+\half \vec{\bar c}\times(\vec c\times\vec c) \ ,
\end{align}
Note that the $\vec\omega$-ghosts here generate $SU(2)$ transformations. The partial localization then is implemented by extending the \SLC-invariant lattice action by the $SU(2)$-invariant  and BRST-exact part,
\begin{align}\label{SGF}
S_{GF}&=s\sum_\bn  \vec{\bar c}(\bn) \cdot \left(\frac{i\gamma}{2}\Big( \vec b(\bn) +\kappa\,\vec{\bar c}(\bn)\times \vec c(\bn)\Big)+\sum_\mu w_\mu(\bn)\xi_\mu^\dagger(\bn)\vec\sigma\xi_\mu(\bn)\right)\nonumber\\
&=\sum_\bn  \Big[\vec b(\bn)\cdot \left(i\gamma\kappa\, \vec{\bar c}(\bn)\times\vec c(\bn)+ \sum_\mu w_\mu(\bn)\xi_\mu^\dagger(\bn)\vec\sigma\xi_\mu(\bn)\right)+\vec{\bar c}(\bn)\cdot\vec c(\bn)  \sum_\mu w_\mu(\bn)\xi_\mu^\dagger(\bn)\one\xi_\mu(\bn)+\nonumber\\
&\hspace{7em}+\frac{i\gamma}{2}\,\vec b(\bn)\cdot\vec b(\bn) -\frac{i\gamma}{4} (\vec{\bar c}(\bn)\times\vec{\bar c}(\bn))\cdot (\vec c(\bn)\times\vec c(\bn))\Big]\nonumber\\
&=i\sum_\bn \Big[\frac{\gamma}{2}\left(\vec b(\bn)+\kappa\, \vec{\bar c}(\bn)\times\vec c(\bn)-\frac{i}{\gamma} \sum_\mu w_\mu(\bn)\xi_\mu^\dagger(\bn)\vec\sigma\xi_\mu(\bn)\right)^2+\frac{1}{2\gamma}\left(\sum_\mu w_\mu(\bn)\xi_\mu^\dagger(\bn)\vec\sigma\xi_\mu(\bn)\right)^2\ +\nonumber\\
&\hspace{3em}+i\kappa\, (\vec{\bar c}(\bn)\times\vec c(\bn))\cdot  \sum_\mu w_\mu(\bn)\xi_\mu^\dagger(\bn)\vec\sigma\xi_\mu(\bn)-i\vec{\bar c}(\bn)\cdot\vec c(\bn)  \sum_\mu w_\mu(\bn)\xi_\mu^\dagger(\bn)\one\xi_\mu(\bn)\ +\nonumber\\
&\hspace{6em} +\gamma \frac{\kappa^2-1}{4} (\vec{\bar c}(\bn)\times\vec{\bar c}(\bn))\cdot (\vec c(\bn)\times\vec c(\bn))\Big]
\end{align} 
where $\kappa$ and $\gamma\geq 0$ are gauge parameters. Observables and the extended lattice action, do not depend on the $\vec \omega$ ghosts that generate local SU(2) variations. The Grassmann integrals of $\vec \omega$ therefore can be saturated and the equivariant BRST variation $s_e$ is obtained by formally setting $\vec\omega=0$ in \equ{BRSTslc},
\begin{align}\label{eBRST}
s_e\xi_\mu &=\half\vec c\cdot\vec\sigma \xi_\mu\ ,& s_e\xi^{\dagger}_\mu &= \half\xi^\dagger_\mu\vec\sigma\cdot\vec c\nonumber\\
s_e\vec c&=0\ ,&&\nonumber\\
s_e \vec{\bar c}&=\vec b\ , &s_e\vec b&=\half \vec{\bar c}\times(\vec c\times\vec c)  \ .  
\end{align}  
The equivariant BRST variation $s_e$ is nilpotent on SU(2)-invariant functionals only: formally $s^2_e$ generates $SU(2)$ variations with gauge parameters $\half \vec c(\bn)\times\vec c(\bn)$.  Note that the quartic ghost interaction of \equ{SGF} vanishes for $\kappa=\pm 1$ or $\gamma=0$ only. Since $S_{GF}$ is ultra-local, this part of the action can be absorbed in the integration measure at each site. The fact that the expectation of \SLC-invariant observables does not depend on the gauge parameters implies that the limit $\gamma\rightarrow 0^+$ is smooth. One recovers the gauge condition of \equ{gaugecond} in this limit and the Grassmann integral over the ghost fields provides the factor $\det[H(\bn)]$ in the local bosonic measure of the spinors, where the Hessian $H(\bn)$ is given by \equ{Hessian}.  $S_\text{GF}$ shows that \equ{gaugecond} is just one of many ways to localize $\SLC$ to the  compact $SU(2)$ subgroup.  

\section{Proper times of causal paths}
\label{PT}
A basic quantity of interest is the geodesic distance between two nodes of this lattice. Contrary to ordinary lattice gauge theory, this distance depends on the lattice configuration and is not specified a priori. The proper time of a causal (time-like) path between two nodes depends on the configuration and correlation functions will have to be conditional on this proper time.  

In Sect.~\ref{MC} we saw that  lattice configurations which satisfy the inequalities of \equ{inequalities} can be interpreted as triangulated causal manifolds (or of two disjoint causal manifolds in the case of the duoverse). Here we define geodesic distances on these manifolds and in particular define the proper time of a causal path between two nodes for any consistent lattice configuration.

Two nodes $\bn$ and $\bn'$ of the $\bLambda$- (or of the $\bLambda'$-) lattice are (uniquely) related by,
\bel{diffnodes}
\bn'=\bn+\sum_{\mu} h_\mu \bDelta_\mu\ .
\ee
The four integers $\{h_\mu\}$ give the following causal relations between the nodes $\bn$ and $\bn'$,
\begin{align}\label{PFP}
h_\mu&\geq 0, \ \text{for all } \mu=1,\dots,4\ ,& \bn'\geq\bn;&\ \text{$\bn'$ is in the future of $\bn$}\nonumber\\
h_\mu&\leq 0, \ \text{for all } \mu=1,\dots,4\ ,& \bn'\leq\bn;&\ \text{$\bn'$ is in the past of $\bn$}\nonumber\\
&\text{otherwise}\ ,& \bn'\not\sim\bn;&\ \ \text{$\bn'$  and $\bn$ are not causally related}\ .
\end{align}

A lattice path $\mathfrak{P}_L$ of depth $L$  from node $\bn_0$ to node $\bn_L$  is  a list of contiguous nodes,
\bel{nullpaths}
\mathfrak{P}_L=[\bn_0,\bn_1,\dots,\bn_L]=\{\bn_0,\bn_i=\bn_{i-1}\pm\bDelta_{\mu_{i-1}},\ \text{for } i=1,\dots,L\}\ .
\ee 
 Since the separation between any two adjacent events of the null lattice is light-like, the proper time of a path like $\mathfrak{P}_L$  vanishes. To allow for paths that are time-like, additional events of the manifold must be considered. It is natural to include events that bisect the spatial diagonals of plaquettes. In \fig{plaquettes} these are marked by a cross ($\times$). These additional (centered) events of a plaquette based at the active node $\bn$ will be denoted by $\widehat\bn$. Note that the plaquette is uniquely determined by the events on the path before and after $\widehat\bn$.  The temporal and spatial separation of these center-events to other events of the same plaquette are readily computed in terms of the null lat-frames \emph{of a single inertial system}.  

Consider a $\mu\nu$-plaquette based at the active node $\bn$ with "center" $\widehat\bn$. Viewed from an inertial system with origin at the active node $\bn$, the event proper time to the event $\hat\bn$ is $\Delta s=\sqrt{-\half E_\mu(\bn)\cdot E_\nu(\bn)}=\half \ell_{\mu\nu}(\bn)>0$. Due to \equ{symmc} the proper time between $\widehat\bn$ and $\bn'=\bn+\mu+\nu$ on the same plaquette. determined in the inertial system at $\bn'$, is the same $\Delta s'=\sqrt{-\half \widetilde E_\mu(\bn)\cdot \widetilde E_\nu(\bn)}=\half \ell_{\mu\nu}(\bn)=\Delta s$ . Note that the proper time $2\Delta s=\ell_{\mu\nu}$ between $\bn$ and $\bn'$ is \emph{largest} for a path that passes through the central node $\widehat\bn$ of the diagonal. The spatial separation of the event at $\widehat\bn$ from  the events labeled by $\bn+\mu$ and $\bn+\nu$ also\footnote{Although the \SLC invariant square of the distances have opposite sign, proper times and spatial lengths  here are both defined positive.}  is $\half\ell_{\mu\nu}(\bn)$. 

\begin{figure}[h]
\includegraphics[width=6in]{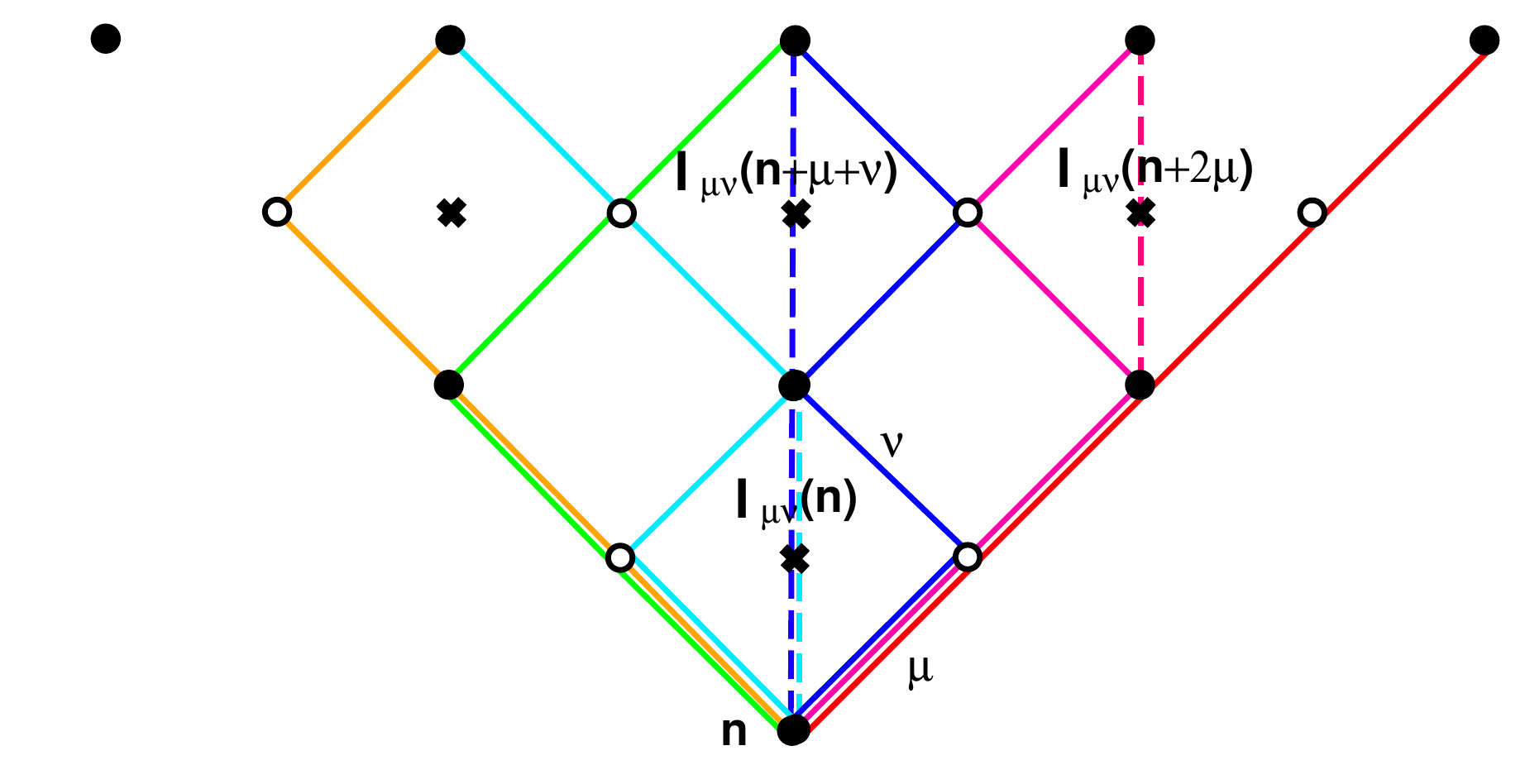}
\caption{\small (Color online) Nodes of the duoverse related to $\bn$ by causal paths in the $\mu\nu$-plane. Several causal paths of depth $L=4$ through active nodes ($\bullet$), passive nodes ($\circ$) and central nodes ($\times$) are shown in different colors. Solid sections of a path do not contribute to its proper time. Dashed  sections of a path through a central node contribute $\ell_{\mu\nu}$ to the proper time, where $\ell_{\mu\nu}>0$ is the spatial separation between the two passive nodes of the same plaquette. In the universe all nodes are active and there is a central node on each plaquette.}
\label{paths}
\end{figure}

Extending \equ{nullpaths} to include paths $\mathfrak{\bar P}_L$ of depth $L$ between two (active) nodes $\bn$ and $\bn'$ that can pass through central nodes, we consider contiguous lists of events of the form,      
\bel{pathspec}
\mathfrak{\bar P}_L=\{\bn_0,\dots,\bn_{j-1},\widehat\bn_j,\bn_{j+1},\dots,\bn_{k-1},\widehat\bn_k,\bn_{k+1},\dots,\bn_L\}\ .
\ee 
Note that any center-event $\widehat\bn_k$ in this list is flanked by two (active) nodes of the same plaquette. We can select the \emph{causal} paths from an (active) node $\bn_0$ to an active node $\bn_L\ge\bn_0$ by demanding that each of its increments either be light- or time-like in the forward direction. A causal path $\mathfrak{C}_L$ thus is a list of $L$ contiguous events of the form\footnote{Every causal path between two active nodes  $\bn$ and $\bn'\ge\bn$ has the \emph{same} depth $L=\sum_\mu h_\mu$\ .},
 \bel{causalpath}
\mathfrak{C}_L=\{\bn_0,\dots,\bn_{j-1},\widehat\bn_j,\bn_{j+1},\dots,\bn_{k-1},\widehat\bn_k,\bn_{k+1},\dots,\bn_L\}\ ,
\ee   
with time- or light-like  positive increments only. Every point of a causal path thus is in the immediate future of the previous one. The adjacent active events determine the plaquette ($\mu\nu$) on which a center-event $\widehat\bn_k$ is located.  The proper time $s(\mathfrak{C}_L)$ of a causal path is the sum of the proper times of its increments,
\bel{propertime}   
s(\mathfrak{C}_L):=\sum_{\widehat \bn_k\in \mathfrak{C}_L}2\sqrt{-\half E_\mu(\bn_{k-1})\cdot E_\nu(\bn_{k-1})}=\sum_{\widehat \bn_k\in \mathfrak{C}_L} |f_{\mu\nu}(\bn_{k-1})|\geq 0\ ,\ \ \text{with } \bn_{k+1}=\bn_{k-1}+\bDelta_\mu+\bDelta_\nu \ .
\ee
The sum is over center-nodes of the path only, since all other increments are null and do not contribute to the proper time.  Note that the proper time defined by \equ{propertime} is an observable composed of basic invariants of the lattice. The consistency condition of \equ{symmc} has to be satisfied, but the backward lat-frames  of the configuration need not be reconstructed to define this proper time.  That \equ{propertime} is a physically sensible definition of the proper time of a causal path can be seen by choosing inertial systems (a gauge), in which the spatially separate events of any traversed plaquette are simultaneous (we proved in \sect{Localization} that this path-dependent choice of gauge is possible and unique up to spatial rotations). In the corresponding inertial systems, the proper time between the two active nodes also gives the spatial extent $\ell_{\mu\nu}(\bn)$ of the plaquette. \equ{propertime} merely defines the proper time of the whole path as the sum of these increments\footnote{Note in this context that for a $34$-plaquette in an inertial system at $\bn$ that satisfies \equ{gauge2}, $|f_{34}(\bn)|=|\xi_3^1|^2+|\xi_3^2|^2=\tau_3=\tau_4$. }.   \fig{paths} depicts some causal paths of depth $L=4$ and their proper times.   

For any particular consistent causal lattice configuration one may select the lattice path with the longest proper time and define the geodesic distance between two causally related nodes as,
\bel{distance}
d(\bn,\bn'):=\max_{\mathfrak{\bar C}_L:\bn\rightarrow\bn'} s(\mathfrak{\bar C}_L)\ ,
\ee
where $L=\sum_\mu h_\mu$ is the depth of all causal paths that relate  $\bn$ to $\bn'$. The number of causal paths that connect a given pair of (active) nodes is finite but can be rather large: the number of causal paths with vanishing proper time already is $L!/(h_1!h_2!h_3!h_4!)$. In general $d(\bn,\bn')=0$ only if $\bn'=\bn+L\bDelta_\mu$, but the geodesic distance between some events may be quite small (of Planck size) even for a large number of steps $L$. From a macroscopic point of view such events essentially are on light-like geodesics.   

These considerations apply to a given causal configuration only. Since the weight $e^{iS_{\rm HP}}$ of the Lorentzian lattice is complex, the weighted proper time of a given path between two causally related nodes in general could be complex even though it is a positive real quantity for any particular configuration. Stationary phase approximation suggests that the average proper time is (almost) real for sufficiently deep paths, but could be complex  for paths of a few steps. The geodesic distance between close-by nodes depends on the configuration and a physical interpretation of correlation functions between two local observables (say $V(\bn)$ and $V(\bn')$) such as 
\bel{correlators}
\vev{V(\bn) V(\bn')}\ ,
\ee
is hardly possible if the geodesic distance between the lattice points $\bn$ and $\bn'$ fluctuates. One perhaps instead should consider \emph{conditional} amplitudes like,
\bel{condcorrel}
\vev{V(\bn) V(\bn')}_{d(\bn,\bn')=\tau} \ .
\ee
Here the correlator is the (complex) weighted average with respect to configurations whose geodesic distance between the two lattice vertices $\bn$ and $\bn'$ is given by $\tau$. One then could ask how this amplitude depends on $\tau>0$.  The conditional amplitude of \equ{condcorrel} thus gives the correlation between the local 4-volumes of a \emph{subset} of all causal configurations. Since the theory is not invariant under translations,  the conditioning of amplitudes in general will have to be more severe. It may in particular be necessary to fix the geodesic distance of an event to the Big Bang, because correlators like that in \equ{condcorrel} almost surely depend on the epoch. 

For nodes $\bn$ and $\bn'$ related by causal paths of great depth $L$, the conditioning in \equ{condcorrel} may not be that important, because the geodesic distance in this case is relatively well defined and should not differ much from the leading semi-classical estimate. Conditional and unconditional correlators coincide in this approximation. One thus perhaps can compute the asymptotic large-distance behavior of amplitudes in the conventional fashion. However, only conditioned correlators may be well defined at short distances on a finite lattice.    

\section{Integration Measure, Regularization and Orientation}
\label{MReg}
To compute expectation values of \SLC-invariant observables, an integration measure for the spinors and transport matrices must be specified. Contrary to ordinary lattice gauge theories with fixed lattice spacing, the short-distance behavior of the present lattice model in addition needs to be regularized in a coordinate invariant geometrical fashion. The proper measure satisfies both objectives and in addition ensures that configurations of this lattice model correspond to oriented triangulated manifolds. 
 
\subsection{Polar Parametrization of Spinors}
\label{PP}
In a "spherical" parametrization of the spinors by their magnitude $\tau$ and (real) angles $0\leq\psi_\mu<2\pi$, $0\leq\varphi_\mu<2\pi$, $0\leq\theta_\mu\leq\pi$,
\bel{paramspinors}
\xi_\mu(\bn)=\sqrt{\tau_\mu(\bn)}e^{i\psi_\mu(\bn)}\widehat\xi_\mu(\bn)=\sqrt{\tau_\mu(\bn)} e^{i\psi_\mu(\bn)}\left(\begin{smallmatrix} e^{-i\varphi_\mu(\bn)/2}\cos(\theta_\mu(\bn)/2)\\e^{i\varphi_\mu(\bn)/2}\ \ \sin(\theta_\mu(\bn)/2) \end{smallmatrix}\right) ,
\ee
$\tau_\mu(\bn)$ is the only non-compact variable,
\bel{magspinor}
0<\tau_\mu(\bn)=\sum_{A=1}^2 |\xi_\mu^A(\bn)|^2=-i\Tr E_\mu(\bn)\ .
\ee

$U^4(1)$-invariance of physical observables implies that they do not depend on the phase $\psi_\mu(\bn)$ of the spinors.  In spherical coordinates the singular anti-hermitian matrix  $E_\mu(\bn)$ thus is parameterized by its temporal extent, $\tau_\mu$, and the three-dimensional unit vector,
\bel{defrhat}
\widehat r_\mu(\bn)=(\sin\theta_\mu(\bn) \cos\varphi_\mu(\bn),\sin\theta_\mu(\bn)\sin\varphi_\mu(\bn),\cos\theta_\mu(\bn))\ ,
\ee
which specifies the direction of the null-ray in a local inertial system at $\bn$.
 
In the parametrization of \equ{paramspinors}, the forward null lat-frames may be written,
\bel{polarE}
E_\mu(\bn)=i\xi_\mu(\bn)\otimes\xi^\dagger_\mu(\bn)=i \tau_\mu(\bn) P_\mu(\bn)\ , 
\ee
where the projector $P_\mu(\bn)$ is the matrix,
\bel{projP}
P_\mu(\bn)=P(\widehat r_\mu(\bn)):=\frac{1}{2}(\one-i\vec\sigma \widehat r_\mu(\bn))= \frac{1}{2}\left(\begin{smallmatrix} 1+\cos(\theta_\mu(\bn)) & e^{-i\varphi_\mu(\bn)}\sin(\theta_\mu(\bn)) \\  e^{i\varphi_\mu(\bn)} \sin(\theta_\mu(\bn)) & 1-\cos(\theta_\mu(\bn))  \\\end{smallmatrix}\right)\ .
\ee 

Using Eqs.~(\ref{polarE})~and~(\ref{projP}) in \equ{Vform}, the 4-volume element $V(\bn)$ of the 5-simplex formed by the four null vectors of a node is given by,
\bel{volpolar}
V(\bn)=\frac{1}{16}\tau_1\tau_2 \tau_3 \tau_4\det\left(\begin{smallmatrix} \widehat r_1 & 1 \\ \widehat r_2 & 1 \\ \widehat r_3 & 1 \\ \widehat r_4 & 1 \\\end{smallmatrix}\right)_{\hspace{-.2em}\bn}=\tau_1(\bn)\tau_2(\bn) \tau_3(\bn) \tau_4(\bn)\widehat V(\bn)\ ,
\ee
with
\bel{tildeV}
\widehat V(\bn)=\frac{1}{16}\det\left(\begin{smallmatrix} \widehat r_1-\widehat r_4 \\ \widehat r_2 -\widehat r_4 \\ \widehat r_3-\widehat r_4\\\end{smallmatrix}\right)_{\hspace{-.2em}\bn}=-\frac{i}{48} \sum_{\mu\nu\rho\sigma}\eps(\mu\nu\rho\sigma)\Tr P_\mu(\bn)\eps P_\nu(\bn)\eps P_\rho(\bn) \eps P_\sigma(\bn) \eps\ .
\ee
The 4-volume thus is proportional to the three-dimensional volume of a tetrahedron with vertices at the unit vectors $\widehat r_1,\widehat r_2,\widehat r_3$ and $\widehat r_4$. It vanishes only when these four points are coplanar and changes sign when a vertex is reflected through the plane of the other three.

In a system where \equ{gauge2} holds, the dependence on $\tau_4$ and $\widehat r_4$ can be eliminated. The 4-volume in this case assumes the form,
\bel{volpolarGF}
V_\text{GF}(\bn)=\frac{1}{16}\tau_1\tau_2 \tau^2_3 \det\left(\begin{smallmatrix}\widehat r_1 & 1 \\ \widehat r_2 & 1 \\ \widehat r_3 & 1 \\ -\widehat r_3 & 1 \\\end{smallmatrix}\right)_\bn=\frac{1}{8}\tau_1(\bn)\tau_2(\bn) \tau^2_3(\bn)\ \left( \widehat r_1(\bn)\times\widehat r_2(\bn)\right)\cdot\widehat r_3(\bn) \ ,
\ee
and is proportional to  the three-volume spanned by the spatial components of $E_1(\bn),\;E_2(\bn)$ and $E_3(\bn)$.     

\subsection{Localized Lattice Integration Measure}
\label{mu}
Since transport matrices and lat-frames are invariants, the diffeomorphism group does not constrain the integration measure. However,  invariance under the \SLC structure group and the local $U^4(1)$ to a large extent dictates the local integration measure of this lattice model.
The invariant integration measure for the \SLC transport matrices is the Haar measure of \SLC. For matrices in the fundamental representation of \SLC of  the form,
\bel{slcrep}
U=\left(\begin{smallmatrix}\alpha&\beta\\\gamma&\delta\end{smallmatrix}\right), \ \text{ with } (\alpha,\beta,\gamma,\delta)\in\mathbb{C}^4 \ \text{ and  } \alpha\delta-\beta\gamma=1\ ,
\ee
the left and right invariant Haar measure is proportional to\cite{Barut:1986},
\bel{Haarslc}
d\mu[U]=\delta^2(1+\beta\gamma-\alpha\delta) d\alpha d\alpha^*d\beta d\beta^* d\gamma d\gamma^* d\delta d\delta^*=d\alpha d\alpha^*d\beta d\beta^* d\gamma d\gamma^*/|\alpha|^2\ .
\ee  

For the purpose of analytic continuation of \SLC to a compact $SU(2)_L\times SU(2)_R$  it may be more convenient to parameterize $U\in\SLC$ in terms of 5 compact Euler angles $0\leq\psi<2\pi,0\leq\theta,\bar\theta<\pi,0\leq\phi,\bar\phi<4\pi$ and a single non-compact variable $0\leq\bar\psi<\infty$,
\bel{EulerSLC}
U=e^{\sigma_3 \psi/2}\; e^{\sigma_2 \theta/2}\;  e^{\sigma_3 \phi/2}\;e^{i\sigma_3 \bar\psi/2}\;e^{\sigma_2 \bar\theta/2}\;e^{\sigma_3 \bar\phi/2}\ ;\ \ U^\dagger= e^{-\sigma_3 \bar\phi/2}\;e^{-\sigma_2 \bar\theta/2}\;e^{i\sigma_3 \bar\psi/2}e^{-\sigma_3 \phi/2}\;e^{-\sigma_2 \theta/2}\;e^{-\sigma_3 \psi/2}\  \ .
\ee
The corresponding \SLC-invariant Haar measure is proportional to,
\bel{Haarslc1}
d\mu[U]=d\psi \sin\theta d\theta d\phi d\bar\psi \sin\bar\theta d\bar\theta d\bar\phi\ ,
\ee
over the entire parameter domain.\footnote{Formally the analytic continuation of \SLC to compact $SU(2)_L\times SU(2)_R$ corresponds to a change of the integration contour for $\bar\psi$ from the positive real axis to the imaginary one, $\bar\psi\rightarrow \pm i\bar\psi$. However, this "Wick rotation" of the high-dimensional integral must be carried out with care\cite{Witten:2010cx} for the contribution from the contour at infinity to vanish. Although important for numerical simulations, this article does not further pursue the analytic continuation of the lattice integrals.}  

Invariance under \SLC transformations of \equ{transformations} furthermore greatly restricts the local integration measure for the null-vectors $E_\nu(\bn)$ and for the corresponding spinors.  For the spherical representation of \equ{polarE} the \SLC-invariant measure at each lattice site is proportional to,
\bel{measureE}
d\mu[E_\nu] \propto d^4E_\nu \ \delta^+(E_\nu\cdot E_\nu)\propto \tau_\nu d^+\tau_\nu\; d\Omega(\widehat r_\nu)\ .
\ee
$d\Omega(\widehat r_\nu)=\sin(\theta_\nu) d\theta_\nu d\varphi_\nu$ here is the $SO(3)$-invariant measure on $S_2$ in spherical coordinates and $d^+\tau_\mu= \Theta[\tau_\mu] d\tau_\mu$. The corresponding \SLC-invariant measure for the spinors includes the integration measure for the phase $\psi_\nu$. Local $U(1)$ invariance requires that,
\bel{measureXi}
d\mu[\xi_\nu] \propto \tau_\nu d^+\tau_\nu d\psi_\nu d\Omega(\widehat r_\nu)\ ,
\ee 
where the domain of the three angles is $0\leq\psi_\nu\leq 2\pi,0\leq\theta_\nu\leq \pi$ and $0\leq\varphi_\nu< 2\pi$.  In the gauge of \equ{gauge2}, the local integration measure for the spinors at each node thus is of the form,
\bel{measureGF}
\int_0^\infty \tau_3^3 d\tau_3 \int_0^\infty \tau_2d\tau_2 \int_0^\infty \tau_1 d\tau_1\int_{S_2} d\Omega_3\int_{S_2}d\Omega_2\int_{S_2} d\Omega_1\int_0^{2\pi} d\psi_3\int_0^{2\pi} d\psi_1\; \rho(V)\ .
\ee
The local $U^4(1)$-invariance of observables was here exploited to set $\psi_2=-\psi_1$ and $\psi_4=-\psi_3$  and the determinant of the Hessian of \equ{Hessian} leads to the weight proportional to $\tau_3^3$ in \equ{measureGF}. The measure of \equ{measureGF} distinguishes $\mu=3$ from $\mu=1,2$, since the partial \SLC gauge-fixing of \equ{gauge2} does not treat the lat-frames equitably. 

Requiring $\SLC$- invariance thus determines the local integration measure up to an invariant density.  $\rho(V)\ge 0$ in \equ{measureGF} is a positive semi-definite function of \emph{local} \SLC-invariants. All these local invariants are constructed from $f_{\mu\nu}(\bn)$ and they include the 4-volume $V(\bn)$ of \equ{volpolar}. Note that $\rho(V)$ in the presnt model generally will depend on the local 4-volume: in the first order formulation of Appendix~\ref{Matteractions}, auxiliary 0-forms are introduced to linearize matter actions. Integrating over such non-dynamical fields leads to a factor of $1/\sqrt{V(\bn)}$ in the lattice measure at each site for each auxiliary degree of freedom. $\rho(V)$ must compensate for the introduction of such non-dynamical and unphysical degrees of freedom.  Note that the dependence of the classical lattice action on the cosmological constant could also be absorbed in $\rho(V)$.  

Assuming that $\rho(V<0)=0$ resolves the sign ambiguity of the volume element and ensures that all lattice configurations correspond to oriented complexes. The next section exploits the fact that the critical limit of the model depends on the asymptotic behavior of $\rho(V)$ for $V\rightarrow 0$ only.    

The partially localized  integration measure of \equ{measureGF} is manifestly invariant under the residual $SU(2)$ structure group. The local rotational invariance of observables therefore can be exploited  to completely localize the residual $SU(2)$ structure group. This localization of a compact group is not compatible \cite{Schaden:1998hz,Neuberger:1986xz} with a BRST-construction, but one evidently can choose lat-frames in which the $\mu=3$ ray has spherical coordinate $\theta_3=0$, with  $\varphi_3$ undetermined, and the $\mu=2$  ray defines the $xz$-plane with $\varphi_2=0$. Renaming $\varphi_1\rightarrow \varphi$,  the local 4-volume of \equ{volpolarGF} in this complete gauge is given by,
\bel{totfixV}
V'_\text{GF}(\bn)=\frac{1}{8}\tau_1(\bn)\tau_2(\bn) \tau^2_3(\bn) \sin{\theta_1(\bn)}\sin{\theta_2(\bn)\sin{\varphi(\bn)}} \ .
\ee 
The local volume element $V$ in this case is positive for $0<\varphi<\pi$ and negative for $\pi<\varphi<2\pi$ (since $0\leq\theta_1,\theta_2<\pi$). One thus can restrict to  \emph{oriented} lattice configurations by the integration domain of $\varphi$ ! For  \emph{oriented} configurations, the integrals of the remaining spinor variables at each active node $\bn$ of the fully localized model are,
\bel{totfixINT}
\int_0^\infty \tau_3^3 d\tau_3 \int_0^\pi d\varphi\int_0^\infty \tau_2d\tau_2 \int_0^\infty \tau_1 d\tau_1\int_0^\pi \sin{\theta_2}d\theta_2\int_0^\pi\sin{\theta_1}d\theta_1 \int_0^{2\pi} d\psi_3\int_0^{2\pi} d\psi_1\; \rho(V'_\text{GF})\ ,
\ee
where the spinors associated with a node are parameterized as,
\bel{GFspinors}
\xi_4=\sqrt{\tau_3}e^{-i\psi_3}\left(\begin{smallmatrix}0\\1\end{smallmatrix}\right)\ ;\ 
\xi_3=\sqrt{\tau_3} e^{i\psi_3}\left(\begin{smallmatrix}1\\0\end{smallmatrix}\right)\  ;\ 
\xi_2=\sqrt{\tau_2} e^{-i\psi_1}\left(\begin{smallmatrix}\cos{(\theta_2/2)}\\ \sin{(\theta_2/2)}\end{smallmatrix}\right)\ ; \ 
\xi_1=\sqrt{\tau_1} e^{i\psi_1}\left(\begin{smallmatrix} e^{-i\varphi/2}\cos(\theta_1/2)\\e^{i\varphi/2}\ \ \sin(\theta_1/2) \end{smallmatrix}\right)\ .
\ee 
Ignoring\footnote{The integrals over $\psi_1$ and $\psi_3$ in \equ{totfixINT} are  absorbed by the TLT of\equ{tildeZ} that enforces the consistency condition. Observables do not depend on these angles.} the remaining spinor angles $\psi_1$ and $\psi_3$, \equ{totfixINT} is an integral over the correct number (6 per site) of \SLC-invariant degrees of freedom. 

\subsection{Invariant Regularization} 
\label{Regularization}
Although discrete, this lattice model is not necessarily regular at short distances and may not possess a critical limit.   Configurations for which the local lattice volume $V(\bn)$ is arbitrary small are not suppressed by triangulating the manifold and may dominate  the lattice integral.   A minimal coarseness must be imposed to avoid the associated UV-instability and obtain a UV-finite lattice model. The UV stability of the model thus is related to the behavior of the density $\rho(V)$ at small $V$. Note that the local 4-volume element $V(\bn)$ is  invariant under reparametrization because the lat-frames are (see \equ{E}).  

Since events are defined by the intersection of light-cones, one cannot demand that the local 4-volume be of fixed size. A viable model requires only that $\rho(V)$ suppress configurations with small local 4-volumes $V(\bn)$. In the critical limit when the number of lattice sites $N$ becomes large, while the total 4-volume of the universe remains constant, only the asymptotic behavior of $\rho(V)$ for small $V$ matters. I will assume that it is power-like
\bel{asymprho}
\rho(V\sim 0)\propto \Theta[V] \,V^\gamma(1+{\cal O}(V))\ ,
\ee 
with an a priori unknown global exponent $\gamma$ that could depend on the number of lattice sites $N$ and on the cosmological constant $\lambda$ (and possibly other lattice couplings). 

In the complete gauge of \equ{gauge2}, the asymptotic UV behavior of the density in \equ{asymprho} leads to lattice integrals at each (active) node of the form,
\bel{fixINTReg}
\int_0^\infty \tau_3^{3+2\gamma} d\tau_3\int_0^\pi (\sin{\varphi})^\gamma d\varphi \int_0^\infty \tau^{1+\gamma}_2 d\tau_2 \int_0^\infty \tau^{1+\gamma}_1 d\tau_1\int_0^\pi (\sin{\theta_2})^{1+\gamma} d\theta_2\int_0^\pi(\sin{\theta_1})^{1+\gamma}d\theta_1 \int_0^{2\pi} d\psi_3\int_0^{2\pi} d\psi_1 \ ,
\ee
where the parametrization of the spinors is given in \equ{GFspinors}. 

For a lattice with $N$ sites, the total 4-volume of the uni- (or duo-)verse will be constant on a curve $\gamma(\lambda;N)$.  The intersection of these curves in the limit $N\rightarrow\infty$ then formally yields the critical points $\gamma_*, \lambda_*$ of the continuum theory (if a critical limit exists).    

\section{The Strong Coupling Limit $\lambda\rightarrow\infty$}
\label{SClimit}
It is interesting to consider the (na{\"i}ve) strong coupling limit of this model in which curvature terms of the action are simply neglected.   The lattice action in this limit is ultra-local and the model has a number of symmetries\footnote{Although the evolution of our universe is believed to be dominated by the cosmological term long after the Big Bang, we here consider rather small lattices that hardly provide a faithful representation of the universe in its later stages.}. These enable us to define a finite lattice model and establish that $\gamma^*=-2$ in this strong coupling limit. At this critical value of the exponent $\gamma$ the 4-volume of the universe collapses for any lattice with a finite number of sites. 

Using \equ{GFspinors} and the definition of \equ{f}, the lattice integrals of  \equ{fixINTReg} at each active node $\bn$ may be written in terms of the invariant lengths $\ell^2_{\mu\nu}(\bn)$ of \equ{lengths2} as, 
\bel{invINT}
\left(\prod_{\mu<\nu}\int_0^\infty d\ell^2_{\mu\nu}\right)V^{\gamma-1}\Upsilon(a,b,c)\ ,
\ee
with $\Upsilon$ defined in \equ{upsilon} and, 
\bel{abc}
a(\bn)=\ell_{12}(\bn)\ell_{34}(\bn),\ \ \ b(\bn)=\ell_{13}(\bn)\ell_{24}(\bn),\ \ \  c(\bn)=\ell_{14}(\bn)\ell_{23}(\bn)\ .
\ee
Expressed by the lengths, the  local invariant volume element of \equ{Vform} is,
\bel{detg}
V(\bn)=\sqrt{-\det(\ell^2_{\mu\nu}(\bn))}=\left(2a^2b^2+2b^2c^2+2c^2a^2-a^4-b^4-c^4\right)_\bn^{1/2}\ .
\ee 
Note that the triangle inequalities imposed by the $\Upsilon$-distribution in \equ{invINT} arise automatically in the change of integration variables. They ensure that the volume in \equ{detg} is real. The fact that the localized measure of \equ{fixINTReg} can be expressed  in terms of \SLC-invariants indirectly confirms that the localization was unique.

Since the cosmological part of the lattice action  $ S_{SC}\sim\sum_\bn V(\bn)$ is ultra local, the generating function of the na{\"i}ve strong coupling limit decomposes into a product of independent generating functions for each site when consistency constraints are ignored. We restrict the configuration space to consistent complexes (that represent triangulated causal manifolds)  by imposing the inequalities of \equ{inequalities}, respectively \equ{dettgineq}. The generating function $Z_{SC}[\lambda;\gamma]$  in the na{\"i}ve  strong coupling limit we are considering formally thus is,
\bel{ZSC}
Z_{SC}[\lambda;\gamma]=\prod_{\bn\in\bar\bLambda}\left[ \left(\prod_{\mu<\nu}\int_0^\infty d\ell^2_{\mu\nu}\right)V^{\gamma-1}e^{-4 i\lambda V}\Upsilon(a,b,c)\Upsilon(\widetilde a,\widetilde b,\widetilde c) \right]_\bn\ ,
\ee
where $\widetilde a, \widetilde b$ and $\widetilde c$ are given by the $\widetilde \ell_{\mu\nu}(\bn)=\ell_{\mu\nu}(\bn-\mu-\nu)$ as in \equ{defabc}.  The $\Upsilon$-distributions  of \equ{upsilon} ensure that the integral in\equ{ZSC} is over configurations $\{\ell_{\mu\nu}(\bn)\}$ that correspond to triangulated causal manifolds only. Note that the triangle inequalities of \equ{upsilon} actually depend on the squares  of the lengths $\ell^2_{\mu\nu}$ only, as might be expected since the continuum action in this limit depends on the metric $g_{ik}$ only. 

For a hypercubic lattice with $N^4$ active sites and periodic boundary conditions, the variables of \equ{abc} and \equ{defabc} are invariant under $3N^3$ continuous scaling  symmetries. These form an Abelian dilation group generated by,
\begin{align}\label{scalingG}
\delta^{(i)}_\bn \ell_{i4}(\bn')=\ell_{i4}(\bn')\ \ \ \delta^{(i)}_\bn \ell_{jk}(\bn')=-\ell_{jk}(\bn'),\ \ \ &\text{for all } \bn'=\bn+q\Delta^{(i)}\ \text{with }q\in\mathbb{Z}\ \text{and } \{i,j,k\}=\{1,2,3\} \nonumber\\
&\text{where } \Delta^{(i)}:=\bDelta_j+\bDelta_k-\bDelta_i-\bDelta_4\ .
\end{align}
This definition of the generators may appear to single out the $4$-th lattice direction, but in fact implies only that $\ell_{\mu\nu}$ scales oppositely to $\ell_{\rho\sigma}$, with $\{\mu,\nu,\rho,\sigma\}=\{1,2,3,4\}$. By definition  $\delta_{\bn}^{(i)}\equiv\delta_{\bn'}^{(i)}$ if $\bn$ and $\bn'$ differ by a multiple of $\Delta^{(i)}$. There thus are only $3N^3$ independent generators on a periodic $N^4$ lattice and the number of these scaling symmetries is proportional to the number of sites on the 3-dimensional surface of a hyper-cubic lattice.  If one ignores  the consistency constraints, the number of independent scaling generators increases to $3N^4$ and is proportional to the number of lattice sites.

One readily verifies that the integration measure $\prod_{\bn,\mu<\nu}\int_0^\infty d\ell^2_{\mu\nu}$ of \equ{ZSC}  is invariant under the dilation group generated by \equ{scalingG}.  The periodic lattice thus is invariant under $3N^3$ non-compact scaling symmetries and the generating function of \equ{ZSC} diverges.

\begin{figure}[h]
\includegraphics[width=4in]{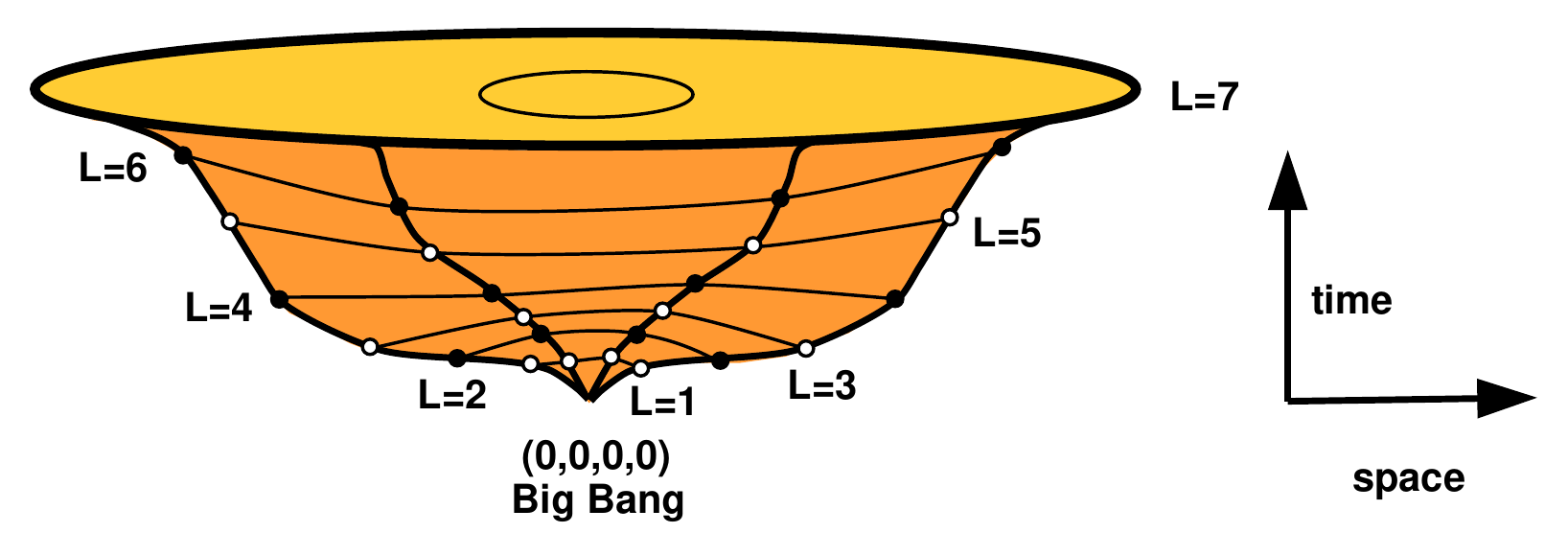}
\caption{\small Conic lattice: an impressionist rendering of the conic part of the 4-dimensional universe that is causally connected to the Big Bang (the event labeled by (0,0,0,0)). [This figure does not represent results of a numerical simulation.] The four "spinal" null rays on the light cone emanating from the Big-Bang event are depicted together with events on them. Spatial hyperplanes labeled by their depth $L$ are indicated schematically. Note that these hyperplanes do \emph{not} correspond to cosmic time because the vertices of the spatial tetrahedrons of a null-simplex generally are not simultaneous in \emph{any} inertial system. The sketch presumes that these hyperplanes rapidly approach $S_3$-hyperspheres with increasing lattice depth. The connected part of the duoverse sketched here is rather small with a maximal depth $L=7$, that is  all its events can be reached in $\leq 7$ light-like steps from the Big Bang. }
\label{conic}
\end{figure}

However, the expectation of \emph{observables} that are invariant under these scaling symmetries, such as any lattice 4-volume, can be computed by localizing the integrals with respect to the scaling group. More interestingly, this dilation group generally is broken by boundary conditions at the "surface" of the lattice. For physical reasons, we would like to model the causally connected part of the universe by a conic four-dimensional lattice like the one sketched in \fig{conic}. This conic section is obtained by standing a hypercubic lattice on one of its corners and cutting it off at the spatial hyperplane with depth $L_\text{max}$; the apex at $\bn=(0,0,0,0)$ of the cone in this simple model represents the Big Bang (BB), all other events on the lattice being causally related to it. 

The quantities $a,b,c$ of \equ{abc} are invariant under the scaling symmetries by construction and the 4-volume of \equ{detg} is as well. $V^2(\bn)$ furthermore is a quadratic form in $a^2,b^2,c^2$  which in addition is invariant under a local SL(2,R) symmetry. Although this local SL(2,R) may be an accidental symmetry of the action and explicitly broken by curvature contributions, it also is a non-compact symmetry of the local integration measure without consistency constraints. If consistency constraints are imposed, only a subgroup of local SL(2,R) transformations at boundary nodes is preserved.

In Appendix~\ref{SLR} the scaling symmetries of \equ{scalingG} and the local SL(2,R) are examined in detail. It is found that these non-compact residual symmetries can be localized by imposing boundary conditions on the cone. Viewing these boundary conditions as a localization by the symmetry, one can adjust the integration measure so that the expectation of invariant observables does not depend on them. The boundary conditions of \equ{bcc} effectively identify parts of faces of the global spatial tetrahedron of given depth $L$.  Note that this identification turns the triangulated  3-dimensional spatial hyperplane of depth $L$ of the hypercubic lattice we started with into a triangulated closed 3-manifold.

Ignoring restriction of the integration domain imposed by $\Upsilon(\widetilde a,\widetilde b,\widetilde c)$, the generating function $Z_{SC}$ of \equ{ZSC} factorizes in generating functions at each site. The local SL(2,R) and scaling symmetries in this case (see Appendix~\ref{SLR}) allow one to deduce that,
\bel{ZSCscaling}
Z_{SC}[\lambda;\gamma]=\left(\frac{z(\gamma)}{\lambda}\right)^{N(\gamma+2)}\ ,
\ee
where $N$ is the number of sites of the lattice and $z(\gamma)$ for $\gamma>-2$ is a  finite integral that does not depend on $\lambda$.  From \equ{ZSCscaling}, the expected  4-volume of the lattice universe at strong coupling thus is given by,
\bel{vevV}
\vev{iV^\text{Univ.}}_{SC}=-\frac{1}{4}\frac{\partial}{\partial \lambda}  \ln Z_{SC}[\lambda;\gamma]=\frac{(\gamma+2)N}{4\lambda}\ ,
\ee
when consistency constraints are ignored. The variance of the total 4-volume in this case is,
\bel{vevV2}
\vev{(i V^\text{Univ.})^2}_{SC}-\vev{i V^\text{Univ.}}^2_{SC}= \frac{\partial^2}{16\partial \lambda^2}  \ln Z_{SC}[\lambda;\gamma]=\frac{(\gamma+2)N}{16\lambda^2}=\frac{\vev{i V^\text{Univ.}}_{SC}}{4 \lambda}\geq 0\ .
\ee

The standard interpretation of $Z_{SC}$ as generating (here depth-) ordered vacuum expectation values  Since the expectation of the variance of an hermitian operator ought to be positive, \equ{vevV2} implies that the 4-volume of Lorentzian space-time should be associated with an anti-hermitian operator $\hat V$ on the Hilbert space and that one measures eigenvalues of  $i \hat V^\text{Univ.}$. This is in keeping with anti-unitary time conjugation in ordinary space-time: although $V(\bn)$ changes sign, $i V(\bn)$ is invariant under time reversal. This assignment is compatible with the fact that the generating function of the strong coupling limit turns into a real probability measure upon analytic continuation  $4i\lambda\rightarrow\beta>0$. 

\begin{figure}[h]
\centering
\begin{minipage}[b]{0.57\textwidth}
\flushleft
\includegraphics[width=0.95\textwidth]{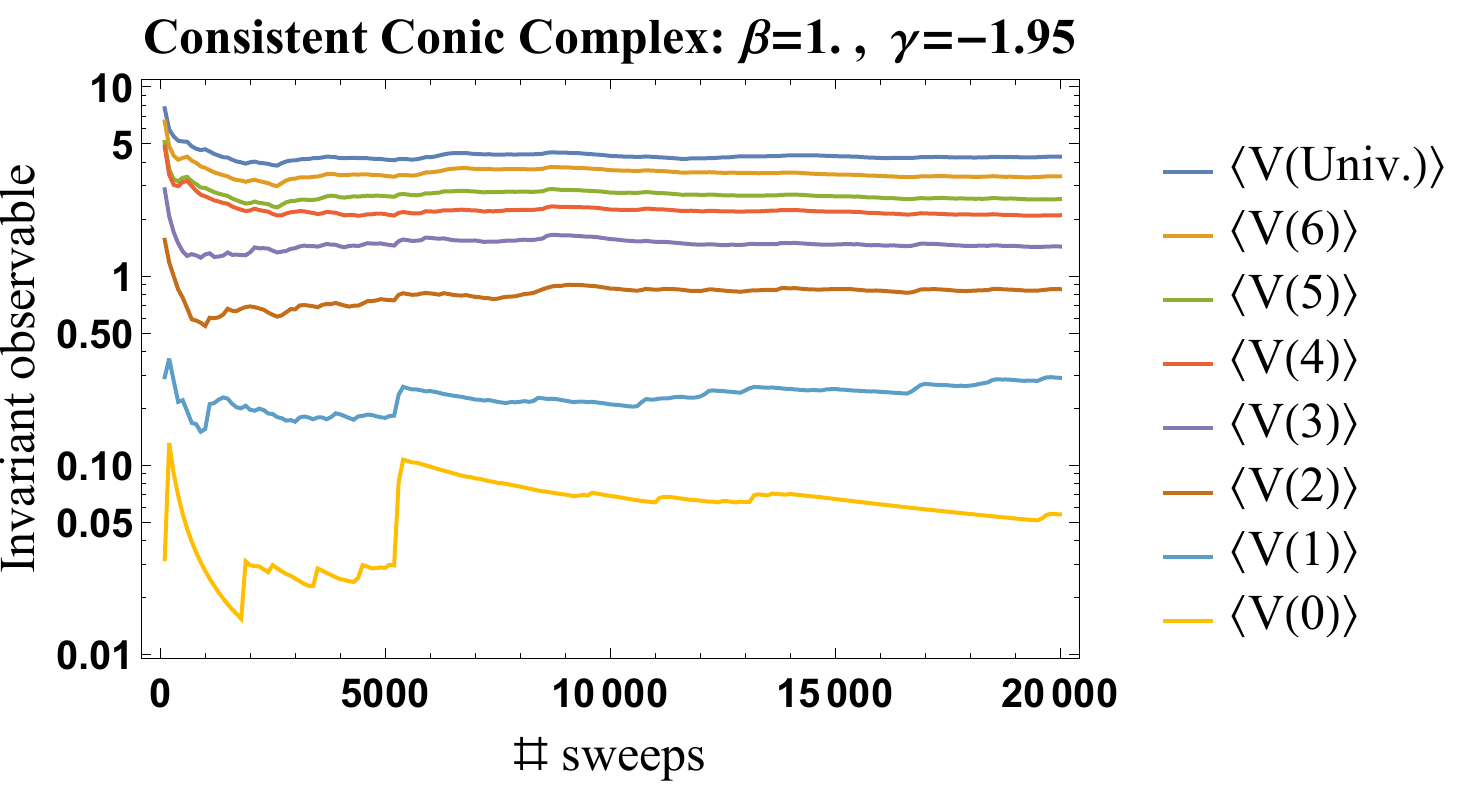}
\hspace{-7em}a)
\label{7a}
\end{minipage}%
\begin{minipage}[b]{0.43\textwidth}
\flushright
\includegraphics[width=1.0\textwidth]{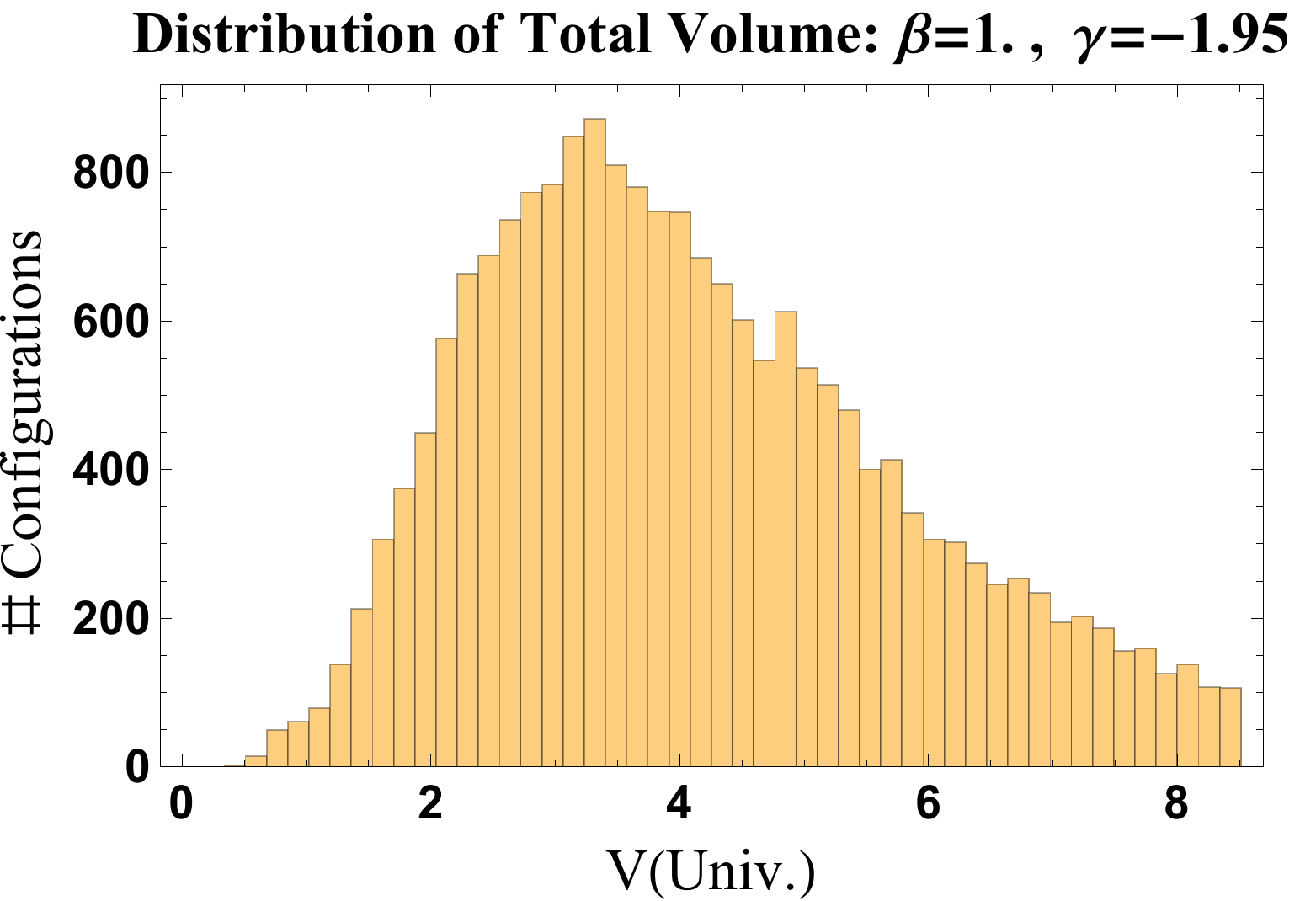}
\hspace{-2em}b)
\label{7b}
\end{minipage}
\begin{minipage}[b]{0.5\textwidth}
\flushleft
\includegraphics[width=0.85\textwidth]{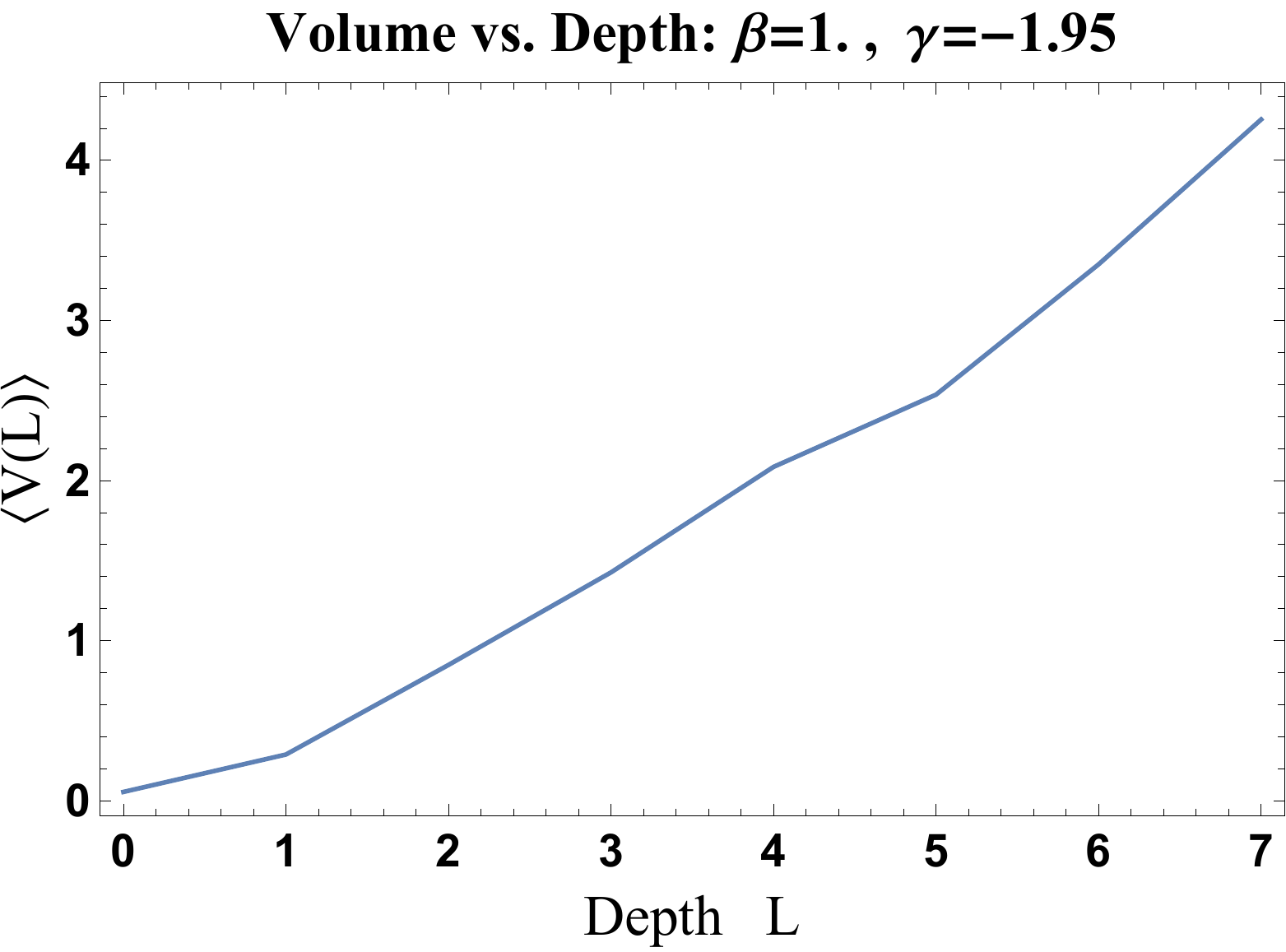}
\hspace{-2em}c)
\label{7c}
\end{minipage}%
\begin{minipage}[b]{0.5\textwidth}
\flushright
\includegraphics[width=0.9\textwidth]{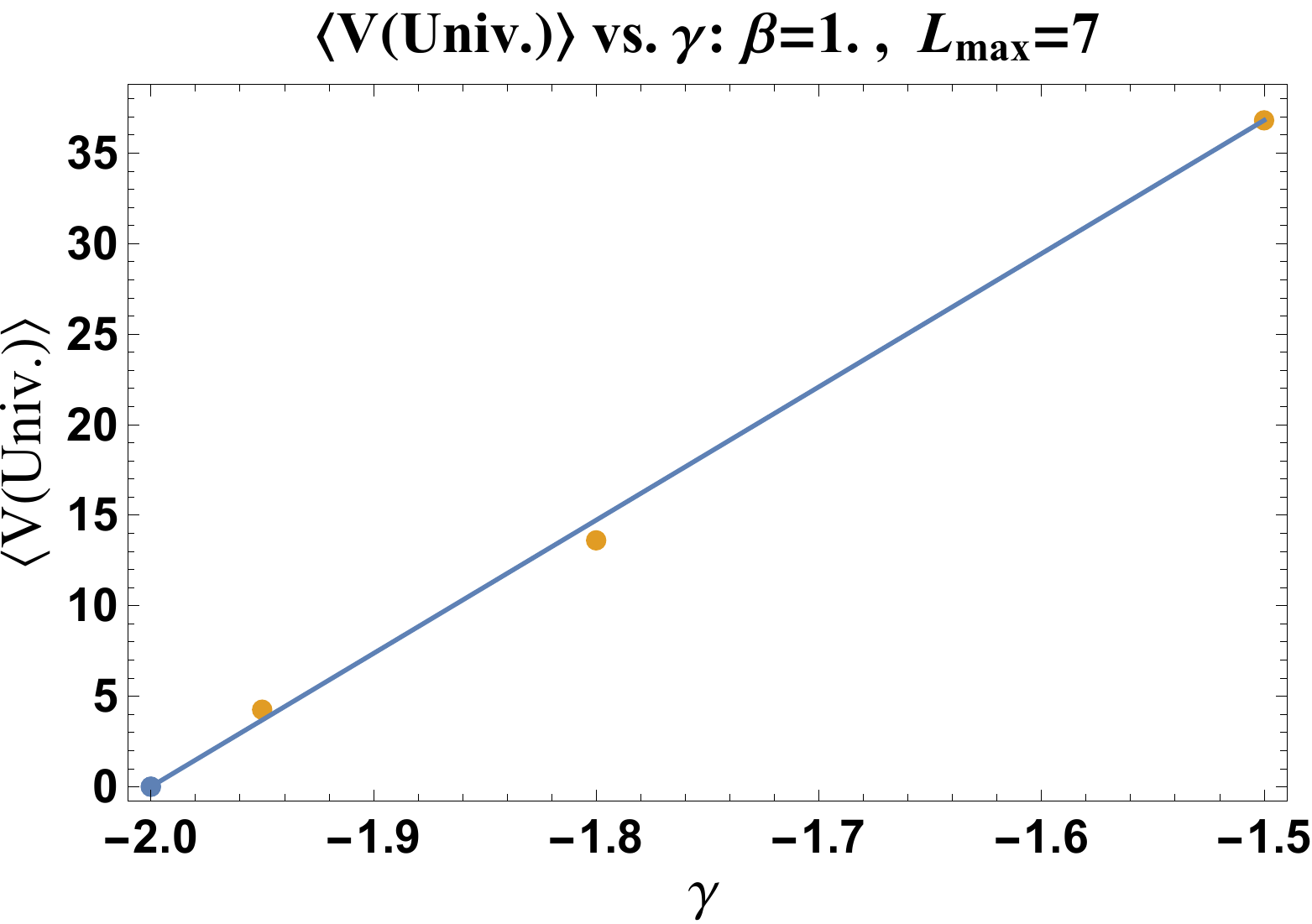}
\hspace{-2em}d)
\label{7d}
\end{minipage}
\caption{\small (Color online) Numerical results for a consistent conic null lattice at strong coupling. The cosmological constant in this numerical study is purely imaginary $\beta=4i\lambda=1$.  $\gamma>-2$  is the exponent of the density $\rho(V)\propto V^\gamma$. a) Running averages of the invariant subvolumes $\vev{V(L)}$ up to depth $L\leq L_\text{max}=7$ of the conic lattice. $\vev{V(\text{Univ.})}=\vev{V(L_\text{max})}$ is the average total 4-volume of the conic null lattice.  b) Distribution of the total 4-volume of the conic null lattice with $L_{max}=7$ for $\beta=1$ and $\gamma=-1.95$.  c) Dependence of the average 4-volume of the null lattice on the depth to the apex of the cone for $\beta=1$ and $\gamma=-1.95$. Note that the 4-volume increases much slower than the number of nodes, $\text{\#nodes}\propto L^4$. d) Total 4-volume of the cone up to depth $L_\text{max}$ for $\beta=1$ but different exponents $\gamma$. The numerical results for $\gamma=-1.5,-1.8,-1.95$ and a constrained linear fit to them are shown. 
The extrapolation to vanishing 4-volume at $\gamma=\gamma^* \sim -2$ is within the numerical accuracy of the simulations.}
\label{simulation}
\end{figure}

Inclusion of the consistency conditions requires numerical simulation even in the na{\"i}ve strong coupling limit. The numerical results for a conic null-lattice with a maximal depth  $L_\text{max}=7$ (corresponding to $330$ nodes) for a purely imaginary cosmological constant $\beta=4 i\lambda=1$ are presented in \fig{simulation}. For these simulations the  residual non-compact "surface" symmetries were all localized in the way described in Appendix~\ref{SLR}. For Im$\lambda<0$, the corresponding generating function is finite and a Metropolis-Hastings algorithm was used to obtain the shown results for $\beta=1$.  \fig{simulation}d) shows that the 4-volume of the cone vanishes for $\gamma\sim -2$. The constraints imposed by the consistency conditions therefore do not change the critical exponent $\gamma^*$. This is not too surprising, since the consistency constraints are irrelevant for the UV-behavior of the model. To see reasonably fast convergence on the finite lattice, we simulated the model for $\gamma\ge -1.95$.  \fig{simulation}a) shows expected volumes $\vev{V(L)}$ of the cone up to depth $L$, $\vev{V(7)}=\vev{V^\text{Univ.}}$ corresponding to the volume of the null lattice cone itself. These are some of the invariants of this model. They all converge even for $\gamma=-1.95\sim\gamma^*$ with the largest fluctuations in $V(0)$, the 4-volume of the simplex whose apex is the Big Bang event. The expected 4-volume of the cone cut at successive slices increases monotonically with depth. \fig{simulation}c) shows that  does not increase as fast as one might expect if each simplex on average had the same 4-volume (as in the unconstrained case):  the number of nodes in each of the sub-volumes is $\text{\#nodes}(L)=1,5,15,35,70,126,210,330$ for $L=0,1,2,3,4,5,6,7$ respectively. Requiring causality thus changes the "shape" of the Universe even at strong coupling. 
\fig{simulation}b) depicts the (unnormalized) distribution of the total 4-volume $V(7)$ of the lattice cone at $\beta=1$ and $\gamma=-1.95$. It resembles a $\Gamma$-distribution.   

Apart from an analytical continuation to real values of $\lambda>0$, the na{\"i}ve strong coupling limit of this model appears to be accessible numerically. Although devoid of curvature, this limit perhaps is not quite as trivial and uninteresting as might be presumed.            

\section{Summary and Discussion}
We considered the triangulation of causal Lorentz manifolds by a null lattice with hypercubic coordination. The nodes of this lattice represent space-time events. Spinors describe the light-like separation to neighboring events in the inertial system of the node at which they are based. Links between nodes are associated with \SLC transport matrices.  The basic \SLC-invariant observables of this model are those of \equ{invariants}. Despite the labeling by a  lattice, this is a geometrical construction based on simplexes with four light-like edges and six spatial ones. It is similar to Regge's\cite{Regge:1961px,Loll:1998aj} approach but retains fixed coordination and is based on an internal \SLC symmetry. As shown in \sect{thelattice}, the triangulation of a causal manifold by light-like signals is physical and coordinate invariant. Macroscopically this method is used to triangulate events by radio signals cemitted  from four GPS satellites. Conceptually the only difference here is that the emission events are themselves triangulated in the same fashion instead of being determined by atomic clocks on known orbits.  

We investigated two topologically different causal Lorentz manifolds: the \emph{universe} with a single causally connected component and the \emph{duoverse} with two. In some ways, the latter is aesthetically more appealing. Its lattice structure topologically is that of a body-centered hypercubic lattice whose two hypercubic sub-lattices are in dual position.The causally disconnected components of the duoverse are described by fields on these mutually dual sub-lattices which are joined by spatial links (on which \SLC transport matrices live) only.  Physical observables in the duoverse are constructed from closed lattice loops and the lattice Hilbert-Palatini action is given by a sum of elementary loops (see \equ{LatHPduo}).  The duoverse could offer a novel accommodation for dark matter that does not interact causally with ordinary matter but does affect the observed local curvature. Depending on global boundary conditions, the two causally disconnected components of the duoverse in fact could be conjugate parts of a single non-orientable connected manifold and thus could lead to novel interpretations of the "arrow of time", the absence of anti-matter in our part of the duoverse and other globally two-sided phenomena.  However,  a universe with a single causally connected component seems to be a viable, if less compelling, alternative. 

Ignoring global considerations, we first addressed local issues of the lattice theory. In \sect{MC} necessary and sufficient conditions for a lattice configuration to represent a triangulated causal manifold were derived.  For null lat-frames the local lattice "metric", $\ell^2_{\mu\nu}=|f_{\mu\nu}(\bn)|^2$, with (complex) amplitudes $f_{\mu\nu}=-f_{\nu\mu}$ defined by \equ{f}, is positive semi-definite, vanishing for $\mu=\nu$. The components of this symmetric matrix of spatial lengths satisfy the triangle inequalities of \equ{tineq} with sides given by \equ{defineabc}. These inequalities are equivalent to demanding that $\det(\ell^2_{\mu\nu})\leq 0$, which is necessary for a real local 4-volume.  The consistency condition of \equ{symmc} requires that the six spatial lengths on the backward light cone of a node satisfy the corresponding  inequalities of \equ{inequalities}, i.e. that the local volume of the corresponding \emph{backward} null-simplex $\widetilde{ch}(\bn)$ is real as well.  In the spinor formulation, this consistency requirement can be imposed by the TLT of \equ{tildeZ}. The corresponding TLT for null-latframes is obtained in Appendix~\ref{coframeMC}.  Equivalently, these inequalities may be enforced by Heaviside distributions.

The \SLC structure group of this lattice theory is not compact and the invariant generating function diverges for any finite number of sites. In \sect{Localization} the \SLC structure group was partially localized to the compact $SU(2)$ subgroup of spatial rotations. This is always possible and a local (equivariant) TFT was constructed that effects this partial gauge fixing.  For simplicity and physical reasons, we mainly considered the constraints of \equ{gauge2}. 

The integration measure of this lattice model is strongly constrained by its \SLC structure group and local $U^4(1)$-invariance of the spinors. The measure for the \SLC transport matrices is the (non-compact) Haar measure. The lattice integration measure for a spherical parametrization of the spinors was derived in \sect{MReg}.  The invariant integration measure is determined up to a density $\rho(V)$ that depends on local \SLC invariants only, and in particular depends on the local 4-volume $V(\bn)$.   We used the gauge condition of \equ{gauge2} to localize this \SLC-invariant integration measure for the spinors. Although not necessary for numerical evaluation, we in addition completely localized the residual compact $SU(2)$-structure group.   The spherical parametrization of the spinors in this case is given in \equ{GFspinors}. This completely localized measure for the spinors in fact can be rewritten in terms of the manifestly \SLC-invariant spatial lengths of \equ{lengths2} as in \equ{invINT}. The restrictions on the length variables arise automatically in the change of variables.  

The \emph{apriori} undetermined density $\rho(V)$ of the integration measure allows for an  invariant regularization of the short-distance behavior of the model that suppresses small local 4-volumes. Assuming the leading dependence of the density $\rho(V)$ for $V\sim0$ to be  power-like as in \equ{asymprho}, we argue that the lattice integration measure in the spinor parametrization of \equ{GFspinors} is \equ{fixINTReg}. It depends on a parameter $\gamma$. For a lattice with a large number $N$ of sites, the dependence of $\gamma(\lambda,N)$ on the cosmological coupling constant may be determined by demanding that the  uni- (or duo-)verse have a given average total observable 4-volume. Critical values  $\gamma^*$ and $\lambda^*$ are points of intersection of these curves as the number of sites $N\rightarrow\infty$.    

\sect{SClimit} examined this model in the na{\"i}ve strong coupling limit, in which  the cosmological term dominates and curvature corrections are ignored. The lattice action in this limit is ultra-local and depends on the local 4-volume only. \equ{detg} gives the local 4-volume in terms of the length variables. Since there is  no dependence on transport matrices, the lattice model simplifies in this unphysical limit and the action and measure can be expressed in terms of the six invariant lengths $\ell^2_{\mu\nu}(\bn)$ at each site.  We considered a conic lattice where the apex of the topologically hypercubic complex represents the Big Bang and the cone terminates at a spatial depth $L_\text{max}$. We systematically found a number of additional non-compact surface symmetries that reflect the fact that there is some freedom in the placement of events on the boundary of the conic manifold: certain choices of events on the boundary can lead to different consistent triangulations of the \emph{same} manifold and we therefore believe that these lattice symmetries are a consequence of residual diffeomorphisms on the boundary of the manifold.    In Appendix~\ref{SLR} this residual group of lattice symmetries is systematically derived. Since the group of residual symmetries is non-compact, it must be localized before attempting a numerical simulation. We find that this is achieved by imposing certain boundary conditions and adjusting the integration measure so that the expectation of invariant  observables does not depend on this localization. In the na{\"i}ve strong coupling limit these observables include all partial lattice 4-volumes.   The boundary conditions on spatial lengths of \equ{bcc} change the topology of the original hyperplane to depth $L$ to that of a closed 3-manifold, much like the identification of the endpoints of a spatial segment to given depth $L$ in the 1+1 dimensional results in a cone with topology $R^+\times S_1$.   

Implementing the additional boundary constraints, numerical simulations on small conic lattices with $L_\text{max}\leq 7$ and purely imaginary cosmological constant $4 i \lambda=\beta=1$ converge rather well in the na{\"i}ve strong coupling limit (see \fig{simulation}. They indicate that the critical exponent  at which the 4-volume of the finite cone vanishes is $\gamma^*\simeq -2$ in the consistent model.  This is to be expected since the consistency constraints are non-local, and $\gamma^*=-2$ when they are relaxed. The numerical results also show that the na{\"i}ve strong coupling limit of the consistent theory is not trivial: we computed the expected partial 4-volumes of the cone at depths $0\leq L\leq L_\text{max}=7$. \fig{simulation}c) shows that they increase much slower than the number of simplexes. When consistency conditions are imposed, the generating functional no longer factorizes into independent generating functions for each simplex and the simulations show that the correlations in fact are substantial.  These initial numerical results are encouraging and perhaps warrant a more detailed investigation of  the model.

 A number of issues have to be resolved before this lattice model becomes numerically accessible. Most importantly, the Lorentzian lattice weight of any configuration is a phase $e^{iS_{HP}}$. Lacking an analytic continuation, it is difficult\footnote{New approaches\cite{Aarts:2009dg,Greensite:2014cxa, Langfeld:2015qoa,Gattringer:2015pea}, originally developed to accommodate a non-vanishing chemical potential in ordinary lattice gauge theories, may help overcome this problem.} to numerically evaluate expectations with complex generating functions. The oscillatory  integrals in fact may be finite (as are Fresnel integrals) but are difficult to control and evaluate numerically.  We so far only verified that the strong coupling limit of this model may be well defined for a purely imaginary cosmological constant. 

The complex weight of the Lorentzian model also makes it difficult to define an "average" proper time between two causally related nodes. \equ{propertime} and \equ{distance} are a reasonable definition of the proper time of a path and of the geodesic distance between two causally related active nodes of any \emph{particular} causal configuration only. For an ensemble of such configurations, the proper time of a path between two particular nodes fluctuates and its average could even be a complex quantity. To address the  general problem of defining distances in fluctuating geometries, one could consider conditioned amplitudes such as \equ{condcorrel}. In ordinary LGT's this conditioning has no effect since distances between lattice nodes are known a priori on a space-time manifold that does not fluctuate.  However, such a conceptual change in the definition of quantum amplitudes may be crucial in the construction of a quantum field theory based on ensembles of causal manifolds.    
   
\begin{acknowledgments}
I would like to thank D. Zwanziger and P. Cooper for encouragement and a first reading of the manuscript and am deeply indebted to T. Jacobson for constructive criticism. Many improvements and clarifications on the original manuscript derive from our Socratic dialogue\cite{communT}.    
\end{acknowledgments}
\appendix
\section{Naturalness of the Cosmological Constant in Semi-Classical Quantum Gravity}
\label{cosconstant}
I argue semi-classically that the incredibly small value of the dimensionless cosmological constant $\lambda =32 \pi \ell_\text{Planck}^2 \Lambda\sim 10^{-120}$ is related to the quantum fluctuations of the 4-volume $V_4$ of the observable universe. A similar argument is given in \cite{Sorkin:2007bd, Barrow:2011zp}, but here I treat the cosmological constant as a coupling constant rather than the expectation of a fluctuating field. This coupling generally will itself depend on the epoch, i.e. on the 4-volume of the observable universe and on the energy density $\rho$ of matter. The infrared free dimensionless coupling $\lambda=32\pi \Lambda \ell^2_P$ tends to vanish at large cosmic scales but $\lambda$ is assumed to remain large compared to $\ell_\text{Planck}^4\rho$ (the cosmological term then dominates the evolution of the universe at later stages).  Conversely, $\lambda$ could be large but dominated by the energy density in the early universe and lead to a strongly coupled theory that may be difficult to describe (semi-)classically. 

Let us therefore just note that fluctuations $\Delta V$ of the 4-volume of the observable universe proportional to the expected 4-volume $\vev{V}$,   
\bel{volumefluc}
(\Delta V)^2:=\vev{V^2}-\vev{V}^2= l^4_c V\ ,
\ee 
semi-classically imply a cosmological constant 
\bel{cosestimate} 
\lambda =\Lambda l_P^2= 32 \pi \Lambda \ell^2_\text{Planck}\sim\frac{32\pi (16\pi^2)\ell_\text{Planck}^4}{ l_c^2 \sqrt{V}}\ .
\ee
 
\equ{volumefluc} follows if the total 4-volume of the universe is composed of a large number, $N$, of volumes $v_i$ drawn independently from a single distribution whose variance is finite\footnote{In this case $l^4_c=(\Delta v)^2/\bar v$, where $\bar v$ is the mean and $(\Delta v)^2$ is the variance of this distribution.}. That the variance of the 4-volume is proportional to the 4-volume itself continues to hold under somewhat weaker assumptions.  Much of the argument in \cite{Sorkin:2007bd} is devoted to deriving \equ{volumefluc} from causal set theory\cite{Bombelli:1987,Carlip:2012wa}. Note that the present model comes close to a concrete realization of this idea: the total  number of nearly independent 4-volumes in the strong coupling regime of \sect{SClimit} is $N=L^4$  at depth $L$ where $L$ is proportional to cosmic time at later stages of the evolution.    

Averaging  over many oscillations of the phase $e^{iS/\hbar}\sim e^{-i\Lambda V_4/(8\pi\ell_\text{Planck}^2)}$ in the functional integral would lead  to  vanishing correlation functions.\footnote{The overall constant phase due to the average 4-volume cancels and is irrelevant in a semi-classical computation of amplitudes.} Fluctuations of the cosmological part of the action $S$ thus cannot greatly exceed $h=2\pi \hbar$.  Neglecting curvature and matter contributions to the action,  the cosmological term  provides the main restriction on volume fluctuations. 

In epochs dominated by the cosmological part of the action, a semi-classical approach to quantum gravity therefore implies,   
\bel{qf}    
\frac{\Lambda}{8\pi \ell^2_\text{Planck}} \Delta V\sim 2\pi\ .
\ee

With the estimate \equ{volumefluc} in \equ{qf}, one then arrives at \equ{cosestimate}.

In \equ{cosestimate} we introduced the Planck length $\ell_\text{Planck}=\sqrt{G\hbar/c^3}$. It is the  length scale associated with semi-classical gravitational phenomena and it would be unnatural  if $l_c\ll \ell_\text{Planck}$. For the purpose of a rough estimate it suffices to assume that $l_c\sim \ell_\text{Planck}$, although it may be a few times larger. 

It remains to determine the average 4-volume of the observable universe.  A somewhat na{\"i}ve estimate in terms of the radius $r\sim 4.4\times 10^{26}m$ of the observable universe and its age $c\tau\sim 1.3\times 10^{26}m$ yields   
\bel{estimate}
 \lambda \lesssim (32\pi)16\pi^2\sqrt{3 \ell^4_\text{Planck}/(4 \pi r^3 c\tau)}\sim 3 \times10^{-120} \ ,
\ee
which is about $1/5$ of the value of $\lambda\sim 1.7\times 10^{-119}$ found experimentally\footnote{The coupling $\lambda=\Lambda l_P^2=32\pi \Lambda \ell^2_\text{Planck}$ used here is about $32\pi\sim 100$ times the cosmological constant in Planck units.}  and about twice $(32 \pi)/t_U^2\sim 1.6 \times 10^{-120}$ where $t_U=8\times 10^{60}$ is the present expansion age of the universe in natural units\cite{Barrow:2011zp}.  Considering the crudeness of the estimates, this is encouraging and an indication that gravity is quantized: $\lambda \gg 10^{-120}$ would imply highly suppressed quantum fluctuations and cast doubt on a semi-classical description of quantum gravity. $\lambda \ll 10^{-120}$ on the other hand would imply an unnaturally large universe. 

Note that consistent with an infrared free coupling, quantum fluctuations of a 4-volume larger than the observable one ought to be larger and the cosmological constant  correspondingly smaller. 

The cosmological constant often is considered an effective representation of the vacuum energy density. Its small value compared to typical (often UV-divergent) vacuum energies has been used as an argument for super-symmetric theories. However, the measured effective cosmological constant we are considering does not represent these vacuum energies. As with other coupling constants, only the \emph{renormalized} cosmological constant can be measured. If quantum gravity is regularized, the measured effective cosmological constant is the difference of a bare cosmological constant and the ground state energy density of matter. It essentially determines the average 4-volume of the observable universe and has little to do with vacuum energies. 

\section{First order Matter Actions}
\label{Matteractions}
The first order Hilbert-Palatini action of \equ{HP} can be written in terms of forms. These diffeomorphism invariant objects retain their geometrical meaning when space-time is discretized. The coupling to matter also can be written in terms of forms and  first order matter actions continue to depend polynomially on the co-frame 1-form (with the same sign ambiguity as the Hilbert-Palatini action). 

In curved space-time, the first order formulation of the invariant action for a gauge field with curvature two-form ${\cal F}(A)=F_{ik}(A)\; dx^i\wedge dx^k$ is,   
\bel{actionV}
S_A= \frac{1}{g^2}\int_M e^a\wedge e^b\wedge [\frac{1}{6} e^c\wedge e^d \tr \tuu{B}{e}{f} \tdd{B}{e}{f}-2\tr\tuu{B}{c}{d}{\cal F}(A) ]\eps_{abcd}\ ,
\ee
where the trace is over internal degrees of freedom and  $B_{ab}=-B_{ba}$ are auxiliary  0-forms. A shift of the auxiliary fields leads to the equivalent and more familiar non-polynomial action,
\bel{actionV2}
S_A=\frac{1}{g^2} \int_M d^4x\;  \text{det}(e)\; \tr [\tdd{F}{i}{k}(A) \tuu{F}{i}{k}(A)-4\tuu{B}{a}{b}\tdd{B}{a}{b} ]\ , \ \text{with}\ \ F^{ik}(A)=g^{il}g^{km}F_{lm}(A) \ ,
\ee
if the co-frame is invertible. The first order action of \equ{actionV} thus differs from the more conventional quadratic action by a local correction to the cosmological constant that depends on  the auxiliary fields. In the lattice formulation the auxiliary fields could be integrated out. This would change the local density of the integration measure in \sect{mu} to by a factor proportional to the local 4-volume $V^{-3}$, $\rho(V)\rightarrow V^{-3}\rho(V)$. Correlators that do not depend on the auxiliary fields thus remain unaltered and the  additional term does not change the dynamics and can be absorbed in a change of the local integration measure. Note that \equ{actionV2} again is proportional to $\text{det}(e)$ rather than $|\text{det}(e)|$.  Contrary to the quadratic action of\equ{actionV2}, the first order form in\equ{actionV} is polynomial in all fields and depends on the co-frame 1-form only.

There are some notable similarities between \equ{actionV} and \equ{HP}. \equ{actionV} in fact reduces to the Hilbert-Palatini action for $\tr\tuu{B}{a}{b}{\cal F}(A)=\frac{g^2}{2 l_P^2}R^{ab}(\om)$. This would be the case for an $so(3,1)$-curvature form in the fundamental representation and an auxiliary field $(B^{ab})_{cd}=\frac{\Lambda}{6}( \delta^a_c\delta^b_d- \delta^a_d\delta^b_c)$ with $g^2=2 \lambda/3$. The action is stationary for this particular auxiliary field only if the curvature, $\tuu{\cal F}{a}{b}=\tuu{R}{a}{b}(\om)-\frac{\Lambda}{3}e^a\wedge e^b=0$ vanishes. The classical Hilbert-Palatini action thus can be interpreted as an $so(3,1)$ gauge theory\cite{MacDowell:1977ma,Mansouri:1977f, Witten:1988wi,Catterall:2009nz} - with pure gauge configurations of $so(4,1)$ as classical solutions. Diffeomorphism invariance in this case is a property of the classical solutions only\cite{Witten:1988wi}.     
  
The first order action for a scalar field in general coordinates is of the form,
\bel{actionScalar}
S_\phi=\int_M  e^a\wedge e^b\wedge e^c\wedge\tr [\half  B^e B_e e^d- B^d d\phi] \eps_{abcd}\ .
\ee
The auxiliary fields $B^a$ again may be integrated out in favor of changing the density of the integration measure of \sect{mu} by $\rho(V)\rightarrow V^{-2}\rho(V)$.

The invariant action of a spinor field may also be written in terms of the co-frame 1-form without explicit reference to frames,
\bel{actionSpin} 
S_\psi= \int_M  e^a \wedge e^b \wedge e^c  \wedge \bar\psi\, \sigma^d {\cal D} \psi\;\eps_{abcd} =\int_M d^4x\; \text{det}(e)\; \bar\psi\, \sigma^a e^k_a {\cal D}_k \psi\ .
\ee
Here $ \bar\psi\, \sigma^a {\cal D} \psi=  \bar\psi\, \sigma^a {\cal D}_k \psi\, dx^k$  is a 1-form and $\{\sigma^a,\;a=1,\dots,4\}$ is a normalized basis of anti-hermitian $2\times 2$ matrices. Note that the spinor action is polynomial only if the $so(3,1)$-connection of the covariant derivative ${\cal D}\psi=d\psi+\om^{ab}\sigma_{ab}\psi$ is assumed to not depend on the (co-)frame. 

Without cosmological term, the Jacobian of the invariant volume may be absorbed in the Hamiltonian formulation\cite{Jacobson:1988yy} of the Hilbert-Palatini action written in terms of frames. This is not possible when a cosmological term or matter interactions are included. In terms of frames, the action of the interacting quantum theory in this case is inherently non-polynomial. A first order formulation of the action using co-frames on the other hand remains polynomial when coupled to matter. 

Another reason for quantizing co-frames rather than frames is that the length scale  $l_P$ can be absorbed in $e^a_k/l_P$. The canonical mass dimensions (with $\hbar=c=1$) of all fields and couplings in this case is non-negative in $d=4$ dimensions.

\begin{center}
\begin{tabular}{|l|c|c|c|c|c|c|c|c|c|}
\hline
field or coupling & $e_k^a/l_P$ & $\omega_k^{ab}$ & $\lambda$ & $A_k$ & $B^{ab}$ & $g$ & $\phi$ & $B^a$ & $\bar\psi \psi$\\
\hline
can. mass dim.&1&1&0&1&0&0&0&0&0\\
\hline
\end{tabular}

Table 1: Canonical dimensions of fields and couplings in $d=4$ dimensions
\end{center}

The first order actions depend on the dimensionality $d$ of space-time in two ways. The totally anti-symmetric Levi-Civita symbol $\eps(abcd)$ with $d$ arguments plays a prominent r{\^o}le in Eqs.~(\ref{HP}-\ref{actionSpin}) and the number of co-frames in the monomials is determined by the order of the form they couple to [all actions may be written as inner products of a $p$-form constructed from the co-frames and a $d-p$ form that does not depend on them].  The number of local $d$-forms one can construct with the given field content to some extent determines the form of the local action, although not as uniquely as desired  since $\phi$, $\bar\psi\psi$ and the auxiliary 0-forms are dimensionless.  
However, the local Hilbert-Palatini action of \equ{HP}  is unique in $d=4$ dimensions. The other three local $4$-forms one can construct with this field content,  $e^a\wedge e^b\wedge\tdd{R}{a}{b}(\om)$, $\tdd{R}{a}{b}(\om)\wedge\tuu{R}{a}{b}(\om)$ and $\tuu{R}{a}{b}(\om)\wedge\tuu{R}{c}{d}(\om) \eps_{abcd}$, are topological densities for the Nieh-Yan, Pontyagin and Euler characteristics of the manifold\cite{Kaul:2011va} that do not alter the continuum dynamics.
 
 In four dimensional space-time, vertices couple up to 6 fields in Eqs.~(\ref{actionV})~(\ref{actionScalar})~and~(\ref{actionSpin}) but at most 4 in \equ{HP}. In $d>3$ dimensions, the first-order actions all lack quadratic terms. The lowest vertices in $d=4$ couple 3 fields and there is no \lq\lq{}natural\rq\rq{} \SLC-invariant background about which a consistent perturbative expansion can be performed. 

\section{Reconstruction of Backward Rays}
\label{Reconstruction}
We here explicitly reconstruct four light-like vectors $\widetilde E_\mu=(\vec r_\mu, - \tau_\mu)$ with $\tau_\mu=|\vec r_\mu|>0$ on the (past) light cone of a node whose inner products,
\bel{Ereconstruct}
-2 \widetilde E_\mu\cdot \widetilde E_\nu=2 \tau_\mu \tau_\nu-2\vec r_\mu\cdot \vec r_\nu=\widetilde\ell^2_{\mu\nu}, \ \ \text{for }\ \mu<\nu\ ,
\ee
are known. The spatial lengths $\{\widetilde\ell_{\mu\nu}=\widetilde\ell_{\nu\mu}>0,1\leq \mu<\nu\leq 4\}$ are assumed to satisfy the inequality \equ{dettgineq}. 

This reconstruction determines the local geometry of a lattice configuration and is found to be unique up to (local) Lorentz transformations. Note that in the universe the lengths $\widetilde\ell_{\mu\nu}$ are the invariant spatial distances between two \emph{active} diagonal nodes of a lattice plaquette whereas they give the displacement between two \emph{passive} or \emph{virtual} nodes in the duoverse. 

As explained in \sect{Localization} we may choose a local inertial system in which,
\bel{Recongauge}
\sum^4_{\mu=1} \vec r_\mu=0\ 
\ee   
Squaring \equ{Recongauge} and using \equ{Ereconstruct} to eliminate the scalar products, one finds that,
\bel{taubar}
\left(\sum^4_{\mu=1} \tau_\mu\right)^2=\sum_{\mu<\nu} \widetilde\ell^2_{\mu\nu}=:16\bar\tau^2\ ,
\ee
where $\bar\tau$ denotes the average temporal separation of the four events from the node. Note that by construction $\bar\tau$ is minimal in the inertial system of \equ{Recongauge}. Using that,
\bel{erecon0}
4 \bar \tau ( \tau_\mu+\tau_\nu-\tau_\rho-\tau_\sigma)=\widetilde\ell^2_{\mu\nu}- \widetilde\ell^2_{\rho\sigma},\ \text{for } \{\mu,\nu,\rho,\sigma\}=\{1,2,3,4\}\ ,
\ee
one obtains the (rotation invariant) temporal separations to the four events,
\begin{align}\label{erecon}
 \tau_\mu&=\bar\tau+\frac{1}{16\bar\tau}\Big(\sum_{\nu\neq\mu} \widetilde\ell^2_{\mu\nu}-\hspace{-1em}\sum_{\mu\neq\rho<\sigma\neq\mu} \hspace{-1em}\widetilde\ell^2_{\rho\sigma}\Big)=\half\left(\sum_{\nu\neq\mu} \widetilde\ell^2_{\mu\nu}\right) \Big/\sqrt{\sum_{\rho<\sigma}\widetilde\ell^2_{\rho\sigma}}\   .
\end{align}

The condition of \equ{Recongauge} is invariant under rotation of the inertial system and we may choose one in which (in spherical coordinates),
\begin{align}\label{Recondirections}
\vec r_1&=\tau_1(\sin\theta_1\cos\varphi_1,\sin\theta_1\sin\varphi_1,\cos\theta_1)\nonumber\\
\vec r_2&=\tau_2(\sin\theta_2\cos\varphi_2,\sin\theta_2\sin\varphi_2,\cos\theta_2)\nonumber\\
\vec r_3&=\tau_3(\sin\theta_3,0,\cos\theta_3)\nonumber\\
\vec r_4&=\tau_4(0,0,1)\ ,
\end{align}
with $0\leq\theta_i\leq\pi$ and $0\leq\varphi_i<2\pi$. In this inertial system, the three angles $\theta_i$  are uniquely determined in terms of the lengths $\widetilde \ell_{14},\widetilde \ell_{24}$ and $\widetilde \ell_{34}$ and the temporal components $\tau_\mu$ obtained in \equ{erecon},
\begin{align}\label{recontheta}
\sin(\theta_i/2)=\frac{\widetilde\ell_{i4}}{2\sqrt{\tau_i \tau_4}}\leq 1,\  \ i=1,2,3\ .
\end{align}
The remaining angles $\varphi_1$ and $\varphi_2$ are found by noting that,
\begin{align}\label{reconvarphi}
\cos\varphi_1&=\frac{2 \tau_1 \tau_3(1-\cos\theta_1\cos\theta_3)-\ell^2_{13}}{2 \tau_1 \tau_3\sin\theta_1\sin\theta_3}\nonumber\\
\cos\varphi_2&=\frac{2 \tau_2 \tau_3(1-\cos\theta_2\cos\theta_3)-\ell^2_{23}}{2 \tau_2 \tau_3\sin\theta_2\sin\theta_3}\ .
\end{align}
Although the right hand sides are known, \equ{reconvarphi} does not determine the angles $\varphi_{1,2}$ uniquely since  $\cos\varphi=\cos(2\pi-\varphi)$. One thus in general has four distinct solutions to \equ{reconvarphi}. The choice is narrowed to (in general) two possibilities by requiring that,
\begin{align}\label{choosephi}
\ell^2_{12}=2 \tau_1 \tau_2(1-\cos\theta_1\cos\theta_2-\sin\theta_1\sin\theta_2\cos(\varphi_1-\varphi_2))\ .
\end{align}
The reconstruction becomes unique for an orientable manifold with positive volume element, since the 4-volume spanned by the lat-frame $\{E_\mu\}$ is positive only if, 
\bel{orientablerecon}
\cos\theta_1\sin\theta_2\sin\theta_3\sin\varphi_2-\cos\theta_2\sin\theta_3\sin\theta_1\sin\varphi_1+ \cos\theta_3\sin\theta_1\sin\theta_2\sin(\varphi_2-\varphi_1)>0 \ .
\ee
Orientability thus selects between the two remaining choices for $\varphi_{1,2}$. 

We here assumed that the solutions to \equ{recontheta}, \equ{reconvarphi} and \equ{choosephi} are consistent, i.e. that all angles are real. As argued in \sect{MC},  this reconstruction is possible if the six spatial lengths  $\{\ell_{\mu\nu}>0,\mu<\nu\}$ satisfy the inequality of \equ{dettgineq}.    Note that we implicitly require this inequality in the  reconstruction when the 4-volume is assumed to be real and positive.  

\section{Consistency in terms of lat-frames}
\label{coframeMC}
We here construct a local TLT that imposes the constraints of \equ{symmc} and leaves the additional \SLC free. This equivariant TLT \cite{Birmingham:1991ty, Schaden:1998hz} is based on the nilpotent  BRST transformation,
\begin{align}\label{BRST}
s \widetilde E^a_\mu(\bn)&=c^a_\mu(\bn)+\omega^a_{\ b}(\bn) \widetilde E^b_\mu(\bn)&s \bar c^{\mu\nu}(\bn)&=g^{\mu\nu}(\bn)=g^{\nu\mu}(\bn)\nonumber\\
s c^a_\mu(\bn) &=\phi^a_{\ b}(\bn) \widetilde E^b_\mu(\bn)+\omega^a_{\ b}(\bn)  c^b_\mu(\bn)& s g^{\mu\nu}(\bn)&=0\nonumber\\
s\phi^a_{\ b}(\bn)&=\omega^a_{\ c}(\bn)\phi^c_{\ b}(\bn)-\phi^a_{\ c}(\bn)\omega^c_{\ b}(\bn)&s b^{\mu\nu}(\bn)&=\bar\omega^{\mu\nu}(\bn)=-\bar\omega^{\nu\mu}(\bn)\nonumber\\
s\omega^a_{\ b}(\bn)&=\omega^a_{\ c}(\bn)\omega^c_{\ b}(\bn)-\phi^a_{\ b}(\bn)&s\bar\omega^{\mu\nu}(\bn)&=0
\end{align}
and the BRST-exact action,
\begin{align}\label{TLT}
S_{TLT}&=s\sum_{\mu,\nu,\bn}[\bar c^{\mu\nu}(\bn) (E_\mu(\bn-\mu-\nu)\cdot E_\nu(\bn-\mu-\nu)-\widetilde E_\mu(\bn)\cdot \widetilde E_\nu(\bn))+ 2 b^{\mu\nu}(\bn)\widetilde E_\mu(\bn)\cdot c_\nu(\bn) ]\nonumber\\
&=\sum_{\mu,\nu,\bn} g^{\mu\nu}(\bn) [E_\mu(\bn-\mu-\nu)\cdot E_\nu(\bn-\mu-\nu)-\widetilde E_\mu(\bn)\cdot \widetilde E_\nu(\bn)]+\nonumber\\
&\qquad\qquad+2 [\bar c^{\mu\nu}(\bn)+\bar\omega^{\mu\nu}(\bn)]\widetilde E_\mu(\bn)\cdot c_\nu(\bn) +2 b^{\mu\nu}(\bn)[\widetilde E_\mu(\bn)\cdot\phi(\bn)\cdot\widetilde E_\nu(\bn)+c_\mu(\bn)\cdot c_\nu(\bn)]\ .
\end{align}
Here $c_\mu^a(\bn)$ are the ghosts associated with the \emph{backward} lat-frames $\widetilde E_\mu^a(\bn)$. The ghosts $\omega^{ab}=-\omega^{ba}$ generate infinitesimal local $SO^+(3,1)$ transformations of the fields in \equ{BRST}.  $\phi^{ab}=-\phi^{ba}$ are the commuting topological ghosts\cite{Birmingham:1991ty}  that maintain the nilpotency of $s$. The remaining fields impose constraints and transform as doublets under this BRST. $g^{\mu\nu}(\bn)$ is a symmetric  bosonic Lagrange multiplier field enforcing the constraints~(\ref{symmc}). $\bar c^{\mu\nu}(\bn)=\bar c^{\nu\mu}(\bn)$ are the corresponding 10 anti-ghosts at each site. The $b^{\mu\nu}(\bn)=-b^{\nu\mu}(\bn)$ are 6 bosonic fields with ghost number~$-2$ that constrain the 16 ghosts $c^a_\mu(\bn)$. The $\bar\omega^{\mu\nu}(\bn)=-\bar\omega^{\nu\mu}(\bn)$ are their BRST-variations (with ghost number $-1$). The TLT-action of \equ{TLT} by construction  is $SO^+(3,1)$ gauge invariant and does not depend on the ghost $\omega^{ab}$ that generates proper Lorentz transformations. 

Changing variables of the topological ghost to  $\phi_{\mu\nu}=s(\widetilde E_\mu\cdot c_\nu)=\widetilde E^a_\mu\phi_{ab}\widetilde E^b_\nu+c_\mu\cdot c_\nu$, the bosonic integrals over $b^{\mu\nu}$ and $\phi^{\mu\nu}$ lead\footnote{In $d$ space-time dimensions, the Jacobian for this change of variables is $\det{M}\propto (\det{\widetilde E})^{d-1}$ where $M$ is the $d(d-1)/2\times d(d-1)/2$ matrix $M^{[ab]}_{[\mu\nu]}(\bn)=\frac{\partial \phi_{\mu\nu}(\bn)}{\partial \phi_{ab}(\bn)}\propto \widetilde E^a_\mu(\bn) \widetilde E^b_\nu(\bn)-\widetilde E^a_\nu(\bn) \widetilde E^b_\mu(\bn)$  constructed from the $d\times d$-matrix $\widetilde E(\bn)$ at each site.} to a factor $(\det{\widetilde E(\bn)})^{-3}$ at each lattice site. Combining symmetric and antisymmetric anti-ghosts to  $\bar C^{\mu\nu}(\bn)=\bar c^{\mu\nu}(\bn)+\bar\omega^{\mu\nu}(\bn)$, there are altogether 16 anti-ghosts per site and a corresponding number of ghost fields $c_\mu^a$. The  integration over ghosts and anti-ghosts yields a factor $(\det{\widetilde E_\mu^a(\bn)})^4$. In any space-time dimension, this TLT imposes altogether $d(d+1)/2$ local constraints and leads to a factor $ \det{\widetilde E_\mu^a(\bn)}$ in the measure at each site. Note that this determinant arises due to the quadratic dependence of the constraint on $\widetilde E_\mu^a$ -- we could have deduced this factor in the measure by a Faddeev-Popov like procedure. For an orientable manifold one must choose the orientation of the $\widetilde E_\mu$ so that $\det{\widetilde E_\mu^a(\bn)}\ge 0$ at every site.    

In the TLT one simply drops (or saturates) the integral over the ghosts $\omega^a_b(\bn)$. Equivalently one may eliminate these fields in\equ{BRST} from the outset and define an equivariant BRST transformation $s_e$ that is not nil-potent,
\begin{align}\label{eBRST1}
s_e \widetilde E^a_\mu(\bn)&=c^a_\mu(\bn)&s_e \bar c^{\mu\nu}(\bn)&=g^{\mu\nu}(\bn)=g^{\nu\mu}(\bn)\nonumber\\
s_e c^a_\mu(\bn)  &=\phi^a_{\ b}(\bn) \widetilde E^b_\mu(\bn)& s_e g^{\mu\nu}(\bn)&=0\nonumber\\
s\phi^a_{\ b}(\bn)&=0&s_e b^{\mu\nu}(\bn)&=\bar\omega^{\mu\nu}(\bn)=-\bar\omega^{\nu\mu}(\bn)\nonumber\\
&&s_e\bar\omega^{\mu\nu}(\bn)&=0\ .
\end{align}
$s_e^2$ is readily seen to generate a local $SO^+(3,1)$ transformation with parameters $\phi^a_{\ b}(\bn)$.  

Upon integration of the $\phi_{\mu\nu}, b^{\mu\nu}, \bar C^{\mu\nu}$ and $c^a_\mu$-fields at each site we thus formally arrive at the \emph{local} effective TLT action,

\bel{STLTeff}
S_\text{eff}^{TLT}= \sum_{\mu,\nu,\bn}g^{\mu\nu}(\bn) (E_\mu(\bn-\mu-\nu)\cdot E_\nu(\bn-\mu-\nu)-\widetilde E_\mu(\bn)\cdot \widetilde E_\nu(\bn))+\bar c^\mu(\bn) \widetilde E^a_\mu(\bn) c_a(\bn)\ ,
\ee
where the integral over the new 4-component ghosts $c_a(\bn)$ and anti-ghosts $\bar c^\mu(\bn)$ yields the factor $\det{ \widetilde E_\mu^a(\bn)}$ in the measure at each site.  It is interesting to note that integrating some of the ghosts and Lagrange multiplier fields of the TLT action in\equ{TLT} led to the effective local action of \equ{STLTeff} whose eBRST-symmetry is no longer manifest. Integration of the (anti-)ghosts $\bar c^\mu(\bn)$, $c_a(\bn)$ and of the Lagrange multiplier $g^{\mu\nu}(\bn)$ gives the TLT partition function,
\begin{align}
\label{ZTLT}
Z_{TLT}[E]&\propto \int \prod_\bn \det\widetilde E(\bn)\Theta[\det\widetilde E(\bn)]\prod_\mu d^4\widetilde E_\mu(\bn)\ \Theta[-\widetilde E^4_\mu(\bn)]\prod_{\nu\leq\mu}\delta(E_\nu(\bn-\mu-\nu)\cdot E_\mu(\bn-\mu-\nu)-\widetilde E_\mu(\bn)\cdot \widetilde E_\nu(\bn)) \nonumber\\
&\propto \prod_\bn \Upsilon(\widetilde a(\bn),\widetilde b(\bn),\widetilde c(\bn)) \left(\prod_{\mu<\nu}\int_0^\infty d\beta\;\delta(E_\nu(\bn-\mu-\nu)\cdot E_\mu(\bn-\mu-\nu) -\beta)\right) \nonumber\\
&\propto \prod_\bn \Upsilon(\widetilde a(\bn),\widetilde b(\bn),\widetilde c(\bn))\ ,
\end{align}
where $\widetilde a(\bn),\widetilde b(\bn),\widetilde c(\bn)$ are given in terms of $\widetilde\ell^2(\bn)=-2E_\nu(\bn-\mu-\nu)\cdot E_\mu(\bn-\mu-\nu)=\ell^2(\bn-\mu-\nu)\ge 0$ as in \equ{defabc} and the distribution $\Upsilon(a,b,c)$, defined in \equ{upsilon}, imposes the non-local inequalities of \equ{inequalities}. To arrive at the second line of \equ{ZTLT} we fixed the \SLC-invariance of the backward lat-frames as was previously done for the forward lat-frames. We then changed integration variables from backward lat-frames to the corresponding \SLC-invariant lengths. The resulting integration measure for the lengths is given by the analog of \equ{invINT} (with  $\gamma=0$). The $\det{\widetilde E}$ factor cancels against the  inverse local volume $V^{-1}$ of \equ{invINT}. Due to the $\delta$-distributions,  the restrictions on the domain of the backward lat-frames become the non-local inequality constraints on the forward latframes imposed by $\Upsilon(\widetilde a(\bn),\widetilde b(\bn),\widetilde c(\bn))$.       

The inequalities of \equ{inequalities} thus may be imposed by a TLT with the local action of \equ{STLTeff}.

\section{Boundary Conditions and Symmetries of a Conic Lattice}
\label{SLR}
\subsection{Na{\"i}ve strong coupling without consistency constraints}
\label{NoCC}
Ignoring the consistency constraints, the integration measure of \equ{invINT} at each site $\bn$ may be written, 
\bel{invINT1}
\left(\prod_{\mu<\nu}\int_0^\infty d\ell^2_{\mu\nu}\right)(-g)^{(\gamma-1)/2}\Theta(-g)\ ,
\ee 
where $g$ is a a quaratic form in $a^2,b^2,c^2$,
\bel{gform}
g=\det{( \ell^2_{\mu\nu})}=a^4+b^4+c^4-2 a^2b^2-2b^2c^2-2c^2a^2=(a^2,b^2,c^2)
\left(\begin{smallmatrix}\hfill1&\hfill-1&\hfill -1\\\hfill -1&\hfill 1&\hfill -1\\\hfill -1&\hfill -1&\hfill1\end{smallmatrix}\right)
\left(\begin{smallmatrix}a^2\\b^2\\c^2\end{smallmatrix}\right)\ ,
\ee
with $a,b,c$ given in terms of the lengths $\ell_{\mu\nu}(\bn)$ as in \equ{abc}.  
$g$ depends only on the squares of spatial separations and the SL(2,R) isometry that leaves $g<0$ invariant is generated by the real matrices,
\bel{SL2R}  
T^1=\frac{1}{2}\left(\begin{smallmatrix}\hfill1&\hfill 1&-1\\\hfill1&\hfill 1&-1\\\hfill0&\hfill 0&-2\end{smallmatrix}\right)\ ,\ 
T^2=\frac{1}{2\sqrt{3}}\left(\begin{smallmatrix}\hfill 3& -1&\hfill 1&\\ \hfill 1&-3 &-1\\\hfill 2&-2&\hfill 0\end{smallmatrix}\right)\ ,\
T^3=\frac{1}{\sqrt{3}}\left(\begin{smallmatrix}\hfill 0&\hfill 1&\hfill-1&\\ \hfill -1&\hfill 0 &\hfill 1\\\hfill 1&\hfill -1&\hfill 0\end{smallmatrix}\right)\ . 
\ee
They commute to: $[T^2,T^3]=T^1,\ [T^3,T^1]=T^2,\ [T^1,T^2]=-T^3$. 

The action of this SL(2,R) on the 5-dimensional space,
\bel{lengthspace}
Q=\{(\ell^2_{14},\ell^2_{24},\ell^2_{34},\ell^2_{23},\ell^2_{13},\ell^2_{12});  \ \text{with}\ \det{(\ell^2_{\mu\nu})=-1}\}\ ,
\ee
is induced by (for instance) considering $Q$ as a fibre bundle over the 3-dimensional base $B=\{ v:=(\ell^2_{23},\ell^2_{13},\ell^2_{12})\}$, with 2-dimensional fibers $F_v:=\{f=(\ell^2_{14},\ell^2_{24},\ell^2_{34});g_v(f)=1\}$ where $g_v(f)$ is the quadratic form of \equ{gform} restricted to a base point,
\bel{gvform}
g_v(f):=f^T M(v) f\ \ \text{ with  } M(v)=\left(\begin{smallmatrix}-\ell^4_{23}&\hfill \ell^2_{23}\ell^2_{13}&\hfill \ell^2_{23}\ell^2_{12}\\\hfill\ell^2_{13}\ell^2_{23}& -\ell^4_{13}&\hfill \ell^2_{13}\ell^2_{12}\\\hfill\ell^2_{12}\ell^2_{23}&\hfill \ell^2_{12}\ell^1_{13}&-\ell^4_{12}\end{smallmatrix}\right)
\ee
The SL(2,R)  acts on the fibers $F_v$ of this bundle with generators,
\bel{indSLR}
  \widetilde T^{(i)}(v)=\text{diag}(\ell^{-2}_{23},\ell^{-2}_{13},\ell^{-2}_{12})\  T^i \ \text{diag}(\ell^2_{23},\ell^2_{13},\ell^2_{12})\ ,
\ee
and the corresponding variation of spatial separations is,
\bel{trafolength}
\delta^{(i)} \ell^2_{j4}= \widetilde T^{(i)k}_{\;j}(v) \ell^2_{k4}\ \text{with } \delta^{(i)} \ell^2_{23}=\delta^{(i)}\ell^2_{13}=\delta^{(i)}\ell^2_{12}=0 \ .
\ee
To geometrically interpret the action of this SL(2,R), note that it changes only three lengths with a common vertex of the spatial tetrahedral face in a manner that preserves the 4-volume of the light-like 5-simplex. We chose a particular SL(2,R) subgroup of all 4-volume preserving transformations that leaves $v:=(\ell^2_{23}(\bn),\ell^2_{13}(\bn),\ell^2_{12}(\bn))$ invariant and acts on the three lengths $f$ with common 4$^\text{th}$ vertex. We could just as well have considered isomorphic SL(2,R) groups that act on another vertex and do not change the lengths of the corresponding triangular base of the tetrahedron. We label these SL(2,R) groups by the vertex they change, as SL$^{(\mu)}$. They are related because $Q$ is invariant under 3 continuous dilation symmetries that do not transform $a,b$ or $c$. These dilations $\{D^i,i=1,2,3\}$ commute with the SL$^{(\mu)}$'s and transform the lengths as,
\begin{align}\label{scalingsym}
\ell_{14}\xrightarrow{D^1} e^{\alpha_1} \ell_{14}\;,\ \ell_{23}\xrightarrow{D^1} e^{-\alpha_1} \ell_{23}\ ;\ 
\ell_{24}\xrightarrow{D^2} e^{\alpha_2} \ell_{24}\;,\ \ell_{13}\xrightarrow{D^2} e^{-\alpha_2} \ell_{13}\ ;\ \ell_{34}\xrightarrow{D^3} e^{\alpha_3} \ell_{34}\;,\ \ell_{12}\xrightarrow{D^3} e^{-\alpha_3} \ell_{12}\ .
\end{align}
It is readily verified that any SL$^{(i)}$ transformation is the composition of an SL$^{(4)}$ transformation with two dilations  $D^j$ and $D^k$, where $\{i,j,k\}\in\{1,2,3\}$. 

The volume preserving group of transformations of a null-simplex $ch(\bn)$ therefore is isomorphic to 
\bel{diffeoG} 
{\cal G}=\text{SL}^{(4)}\times D^1\times D^2\times D^3\ ,
\ee 
and the remaining symmetry of the unconstrained lattice integration measure with $N$ sites is ${\cal G}^N$.

The expectation of quantities invariant  under ${\cal G}^N$ can be evaluated by localizing the invariant measure of \equ{invINT1}. One for instance can demand at every node that,  
\bel{localization}
a^2=b^2=c^2   \ \text{ and }\  \ell^2_{14}=\ell^2_{23},\ \ell^2_{24}=\ell^2_{13},\  \ell^2_{34}=\ell^2_{12}\ ,
\ee
which is the same as requiring that the tetrahedron of the  null-simplex be regular\cite{RindlerPenroseBook} with,
\bel{regtetra}
 \sqrt{3}\ell^4_{\mu\nu}(\bn)=\sqrt{-g(\bn)}=V(\bn),\  \text{for all}\ 1\leq\mu<\nu\leq 4\ .
\ee 
The localization conditions of \equ{localization} can be (uniquely) imposed. The last 3 conditions of \equ{localization} evidently are reached by the dilation symmetries of \equ{scalingsym}. To show that $a^2=b^2=c^2$ can always be satisfied by an SL(2,R) transformation, consider the Morse function,  
\bel{morseV} 
{\cal M}(h)=a^4(h)+b^4(h)+c^4(h)\ge 0
\ee
as a function of $h\in$SL(2,R)/SO(2). ${\cal M}(h)$ is invariant under the compact  SO(2) subgroup of SL(2,R). The critical points of ${\cal M}(h)$ on SL(2,R)/SO(2) satisfy,
\bel{critpM}
0=(a^2+b^2-2 c^2)(a^2+b^2+c^2)=(a^2-b^2)(a^2+b^2+c^2)\ .
\ee
For $a^2+b^2+c^2>0$, ${\cal M}(h)$ is minimal when $a^2=b^2=c^2$ and the SL(2,R) transformation that reaches this minimum is unique up to (compact) $SO(2)$ rotations.  Invariants of ${\cal G}$ thus are functions of the local 4-volume $V=-\sqrt{-\det \ell^2_{\mu\nu}}=\sqrt{-g}$ only. For the purpose of computing the expectation of an invariant, $I(V)$, one thus may localize the  6-dimensional integral of \equ{invINT1} to a one-dimensional one over the 4-volume $V$ of each null-simplex,
\bel{locint1}   
\left(\prod_{\mu<\nu}\int_0^\infty d\ell^2_{\mu\nu}\right)(-g)^{(\gamma-1)/2}\Theta(-g) I(V)\ ,\ \longrightarrow \int_0^\infty dV\; V^{\gamma+1}\, I(V)\ .
\ee 
Without the consistency constraints imposed by $\Upsilon(\widetilde a,\widetilde b,\widetilde c)$ in \equ{ZSC}, the generating function in the na{\"i}ve strong coupling limit would localize to, 
\bel{ZSCloc1}
Z_{SC}[\lambda;\gamma]\propto\prod_{\bn\in\bar\bLambda}\left[ \int_0^\infty dV(\bn)\,  V(\bn)^{\gamma+1}e^{-4 i\lambda V(\bn)} \right]\propto \lambda^{-N(\gamma+2)}\ ,
\ee
which is \equ{ZSCscaling}. 

\subsection{Consistent na{\"i}ve strong coupling limit}
The factors $\Upsilon(\widetilde a,\widetilde b,\widetilde c)=\Theta(-\det{\widetilde\ell^2_{\mu\nu}(\bn)})$ in \equ{ZSC} enforce consistency constraints of the complex which ensure that every configuration represents a triangulated causal manifold. These constraints destroy many of the symmetries of the measure we have in their absence and the generating function no longer decomposes into independent factors for each node. Consistency thus imposes some correlations that prevent us from analytically calculating the generating function. In the end we have to resort to numerical simulations in the consistent na{\"ive} strong coupling limit.

However, a non-compact subgroup of symmetries ${\cal S}\subset{\cal G}^N$ survives the consistency constraints.  The generating function $Z_{SC}$ diverges and an efficient numerical simulation of the consistent lattice model at strong coupling demands that these non-compact symmetries be localized.
 
${\cal S}$ is systematically found by considering the action of subgroups of ${\cal G}$ on null-simplexes $ch(\bn\in \partial C)$ on the boundary of the hypercubic cone $C$ and extending their action in a manner that preserves the 4-volumes of all forward \emph{and} backward null-simplexes of the lattice. We determine the full set of symmetries in this way because any spatial length $\ell_{\mu\nu}(\bn)$ between two nodes, $\bn+\mu$ and $\bn+\nu$, on a (cubic) face of the boundary, $\partial C$, is \emph{not} constrained by consistency. The finite hypercubic cone $C$ here is the collection of nodes,
\bel{conedef}
C=\{\bn=\sum_{\mu=1}^4 n_\mu \bDelta_\mu; n_\mu\in \mathbb{N}\ \text{with } L=\sum_{\mu=1}^4 n_\mu\leq L_\text{max}, \mu=1,\dots,4\}\ .
\ee
At least one of the $n_\mu$ vanishes for nodes $\bn\in \partial C$ on the boundary of the cone $C$ . There are four generic types of nodes to consider on $\partial C$:
\begin{enumerate}
\item[1)] $0=n_1=n_2=n_3=n_4$.\newline
The apex of this null-simplex denotes the Big-Bang event. The diagonal links $[\bn+\mu,\bn+\nu]$ of $ch(0,0,0,0)$ are all on faces of $\partial C$ and their spatial lengths therefore are not constrained by consistency. The full group ${\cal G}$ is a symmetry of $ch(0,0,0,0)$.
\item[2)]$0=n_1=n_2=n_3, n_4>0$.\newline
This 1-dimensional subspace of nodes is the 4$^\text{th}$ spine of the cone $C$. The diagonal links $[\bn+\mu,\bn+\nu]$ of spinal null-simplexes are all on faces of $\partial C$ and their spatial lengths are not constrained by consistency.  A spinal simplex thus is invariant under the full group ${\cal G}$.
\item[3)]$0=n_1=n_2, n_3>0,n_4>0$.\newline
This is a 2-dimensional subspace of nodes on the surface of $C$. All diagonal links except $[\bn+\bDelta_1, \bn+\bDelta_2]$ are on a face of the tetrahedron and therefore are not constrained by consistency. The corresponding length $\ell_{12}(\bn)$ is invariant under  $SL^{(4)}\times D^1\times D^2\subset {\cal G}$.  Note that $D^1$ and $D^2$ each act on a pair of lengths on faces of $\partial C$.  The action of $D^3$ rescales the lengths $\ell_{34}(\bn)$ (which is on a face) and $\ell_{12}(\bn)=\widetilde\ell_{12}(\bn+\bDelta_1+\bDelta_2)$ inversely. To preserve the volume of the (adjacent)  \emph{backward} null-simplex $\widetilde{ch}(\bn+\bDelta_1+\bDelta_2)$, this dilation also has to rescale $\widetilde\ell_{34}(\bn+\bDelta_1+\bDelta_2)=\ell_{34}(\bn+\bDelta_1+\bDelta_2-\bDelta_3-\bDelta_4)$. It therefore also has to rescale $\ell_{12}(\bn+\bDelta_1+\bDelta_2-\bDelta_3-\bDelta_4)$, etc. As discussed below, this continues until the length $\ell_{12}(\bn')$ between two nodes on another face is rescaled.  The length $\ell_{12}(\bn')$ scales inversely to $\ell_{34}(\bn)$ on the original face. The action of $D^3$ on $ch(\bn)$ thus can be extended to a non-local symmetry that involves a string of cells.  
\item[4)]$n_1=0, n_2>0, n_3>0, n_4>0$.\newline
The diagonal links $[\bn+\bDelta_2,\bn+\bDelta_3]$, $[\bn+\bDelta_2,\bn+\bDelta_4]$ and $[\bn+\bDelta_3,\bn+\bDelta_4]$ of length $\ell_{23}(\bn), \ell_{24}(\bn)$ and $\ell_{34}(\bn)$ are on faces of $\partial C$ but do not have a common vertex.  The action of the SL$^{(1)}$ on this null-simplex can be extended to a symmetry. Although it changes the lengths of $\ell_{12}(\bn),\ell_{13}(\bn)$ and $\ell_{14}(\bn)$ which this cell shares with backward null-simplexes, the  SL$^{(1)}$ transformation may be followed by appropriate dilations that restore these lengths and instead rescale the separations  $\ell_{23}(\bn), \ell_{24}(\bn), \ell_{34}(\bn)$ between nodes on faces. As in case 3), the action  of $D^{1}, D^{2}$ and $D^{3}$ dilations on this cell can be extended to non-local symmetries that also rescale lengths of neighboring cells.    
\end{enumerate}
Note that these four cases actually represent  classes of surface nodes with four, three, two or one of the $n_\mu$ vanishing. They correspond to nodes at the intersection of 4, 3, 2 or 1 face(s) of the hypercubic cone. 

The remnant ${\cal S}\subset{\cal G}^N$ symmetry group that survives the consistency constraints thus includes a set of non-local dilation symmetries $D_{\bn}^{(i)}$, associated with surface nodes $\bn\in\partial C$ of the cone. Their generators  $\{\delta_{\bn}^{(i)}, i=1,2,3\}$ are given in \equ{scalingG}. For $\{i,j,k\}\in\{1,2,3\}$, $\delta_{\bn}^{(i)}$ generates a symmetry that scales $\ell_{i4}(\bn')$ oppositely to $\ell_{jk}(\bn')$, just as $D^i$ in \equ{scalingsym} does, but it scales these lengths on a one-dimensional string of lattice sites $\bn'$ separated from $\bn$ by an integer multiple  $0\leq p\leq q=\min[n_i,n_4]$ of $\bDelta_j+\bDelta_k-\bDelta_i-\bDelta_4:=\Delta^{(i)}$. These sites are all on the same depth $L$ as $\bn$ and in cases 1) and 2), $q=0$ for all three $D^{(i)}$, whereas $q>0$ for one dilation in case 3) and all dilations in case 4).  Note that there are exactly half as many independent dilations as there are diagonal links on the surface of the hypercubic cone. 

An SL(2,R) symmetry associated with each surface node also survives. T. Jacobson suggested\cite{communT} that the residual group ${\cal S}$ of 4-volume preserving symmetries on the conic null-lattice is due to diffeomorphisms that relate triangulations of one and the same manifold. Even if interior nodes of the triangulation are determined by the intersection of light cones and form a null-lattice, some choice in the triangulation of nodes on the 3-dimensional light-like surface of a conic manifold remains. The residual group ${\cal S}$ relates equivalent null-lattice complexes with the same local 4-volumes. The last restriction implies that the symmetry group of equivalent null lattice configurations may be larger than ${\cal S}$. However, for Im$(\lambda)<0$ the 4-volume of a null-simplex effectively is bounded and it suffices to consider only equivalence classes of configurations under the non-compact group  $\cal S$ to define a finite generating function in the na{\"i}ve strong coupling limit.         

The residual scaling symmetries can be localized by imposing the boundary conditions,
\bel{bcc}
\ell^2_{i4}(\bn)=\ell^2_{jk}(\bn')\ ,
\ee
where the diagonal links $[\bn+\bDelta_i,\bn+\bDelta_4]$ and $[\bn'+\bDelta_j, \bn'+\bDelta_k]$ are on faces of the hypercubic cone and $\bn'=\bn+\min[n_i,n_4]\Delta^{(i)}$, with $\Delta^{(i)}$ given in \equ{scalingG}.

The residual scaling symmetries allow the boundary condition of \equ{bcc} to always be satisfied. By viewing \equ{bcc} as a (unique) localization of the residual scaling symmetries, one can ensure that the expectation of invariants formally does not depend on the boundary condition of \equ{bcc} by inserting, 
\bel{scaleloc}
1=\prod_{[\bn+\bDelta_i,\bn+\bDelta_4]\in \partial C} \left(4 \int_{-\infty}^\infty d\alpha e^{2\alpha} \ell^2_{i4}(\bn) \delta(e^{2\alpha} \ell^2_{i4}(\bn)- e^{-2\alpha}\ell^2_{jk}(\bn'))\right)
\ee
in the generating function $Z_{SC}$ of \equ{ZSC}. Invariance of the measure (and of the strong coupling action and of observables), implies that the divergent volume of the residual scaling groups factorizes and may be absorbed in the normalization of the generating function. The product of integrals in \equ{scaleloc} extends over all sites $\bn$ for which the diagonal link $[\bn+\bDelta_i,\bn+\bDelta_4]$ is on a  face of, $\partial C$.

Upon dropping the (divergent) volume of the residual scaling groups, one effectively has inserted,
\bel{fixscaling}
\prod_{[\bn+\bDelta_i,\bn+\bDelta_4]\in \partial C} \left(\ell^2_{i4}(\bn) \delta(\ell^2_{jk}(\bn+\min[n_i,n_4]\Delta^{(i)})-\ell^2_{i4}(\bn))\right)
\ee
in the integration measure of \equ{ZSC}.  

We thus have imposed the boundary condition of \equ{bcc} while correcting the integration measure so that the expectation of invariant observables remain unaffected. Observables that are invariant under the scaling group in general are not local. They construction include any partial 4-volume of the cone and in particular the 4-volume of a slice of the lattice at a given depth $L$ of the cone. 

As in \sect{NoCC} the infinite volume of the $SL(2,R)$ symmetry at each surface node also has to  be factorized and absorbed in the normalization before attempting a numerical simulation. As shown in \sect{NoCC}, SL(2,R) invariance can be used to achieve that,
\bel{spinal}
\ell_{14}(\bn)\ell_{23}(\bn)=\ell_{24}(\bn)\ell_{13}(\bn)=\ell_{34}(\bn)\ell_{12}(\bn)\ ,
\ee
for any null-simplex $ch(\bn)$ with $\bn\in \partial C$.

As with the scaling symmetries, these conditions can be enforced without affecting the expectation of invariants by viewing \equ{spinal} as an equivariant localization of the coset $SL(2,R)/SO(2)$. It can be shown that inserting,
\bel{fixSLR}
\prod_{\bn\in \partial C}\left(\ell^4_{14}\ell^4_{23}\delta( \ell^2_{14}\ell^2_{23}-\ell^2_{14}\ell_{23})
\delta(\ell^2_{24}\ell^2_{13}-\ell^2_{34}\ell^2_{12})\right)_{\bn}\ ,
\ee
in the integration measure of \equ{ZSC} achieves this.

We localized the finite lattice model  by inserting  \equ{fixSLR} and \equ{fixscaling} in the integration measure of \equ{ZSC} and used the $\delta$-distributions to evaluate the integrals of certain lengths of boundary nodes. The remaining lattice integral of the na{\"i}ve but now consistent strong coupling limit is finite and exhibits the correct scaling property: within numerical accuracy, the total expected lattice 4-volume is finite for $\gamma>\gamma^*=-2$ for any cosmological constant with Im$(\lambda)<0$  and vanishes in the limit $\gamma\rightarrow\gamma^*=2$. Numerical results for several invariants on a small lattice are presented in \fig{simulation} of \sect{SClimit}.     

\bibliography{%
              biblio/b1-gravity,%
              }
 \end{document}